\documentclass[preprint,aps,prc,showpacs,nofootinbib,secnumarabic,floatfix]{revtex4-1}

\usepackage{graphicx}
\usepackage{bm}
\usepackage{color}
\usepackage{gensymb}
\usepackage{amsmath}
\usepackage{slashed}
\usepackage{appendix}
\usepackage{hyperref}

\newcommand{\ltsim}{\protect\raisebox{-0.5ex}{$\:\stackrel{\textstyle <}
	{\sim}\:$}}
\newcommand{\gtsim}{\protect\raisebox{-0.5ex}{$\:\stackrel{\textstyle >}
	{\sim}\:$}}

\newcommand{\bvec}[1]{\ensuremath{\boldsymbol{#1}}}

\begin{document}

\title{Multiple rescattering effects in the hard knockout reaction $\mathbf{^2}$H$\bvec{(p,2p)n}$}  

\author{A.B. Larionov}
\email{e-mail: larionov@theor.jinr.ru}
\affiliation{Bogoliubov Laboratory of Theoretical Physics, Joint Institute for Nuclear Research, 141980 Dubna, Russia}

\begin{abstract}
  The interaction of a proton with a deuteron is the simplest nuclear reaction. However, it allows the study of precursors of nuclear medium effects
  such as initial-state/final-state interactions (ISI/FSI). In case of hard proton knockout, the deviation of ISI/FSI from the 'standard'
  values may carry a signal of color transparency. In this regard, it is important to define the 'standard' as precisely as possible.
  This work continues previous studies within the framework of the Generalized Eikonal Approximation (GEA). The focus is on processes where
  the participating protons experience multiple soft rescattering on the spectator neutron. It is shown that correct treatment of
  deviations of the trajectories of outgoing protons from the longitudinal direction leads to a significant modification of partial amplitudes
  with soft rescattering of two outgoing protons and non-vanishing amplitudes with rescattering of incoming and outgoing protons.
  The new treatment of multiple rescattering is important in kinematics with a forward spectator neutron.
\end{abstract}

\maketitle

\section{Introduction}
\label{intro}

The quantum-mechanical description of multiple scattering processes is an extremely complex theoretical problem. For high-energy projectiles, the Glauber approximation
is applicable \cite{Glauber,Glauber:1970jm}, which has proven to be quite effective. The diagrammatic formulation of the multiple scattering theory
was given by Gribov \cite{Gribov:1968gs} and Bertocchi \cite{Bertocchi:1972cj}. At moderately relativistic energies, this formulation allowed the derivation
of the Glauber formalism and, in addition, a more accurate theoretical approach called the Generalized Eikonal Approximation (GEA)
\cite{Frankfurt:1996xx,Frankfurt:1996uz,Sargsian:2001ax}.
The GEA formalism was applied to the hard $A(e,e^\prime N)(A-1)^*$ and $d(e,e^\prime p)n$ reactions, when the energy transfer to the nucleus $\gtsim$ several GeV 
\cite{Frankfurt:1996xx,Sargsian:2001ax,Boeglin:2024spd}, and the hard $d(p,2p)n$ \cite{Frankfurt:1996uz,Larionov:2022gvn}, $A(\bar p, \chi_c)(A-1)^*$ \cite{Larionov:2013nga},    
$d(\bar p, J/\psi)n$ and $d(\bar p, \psi^\prime)n$ \cite{Larionov:2019mwa}, and $d(\bar p, \pi^-\pi^0)p$ \cite{Larionov:2019xdn,Larionov:2022uri}
reactions induced by incoming (anti)protons at similar energies. In such reactions, particles produced in a hard collision with a target nucleon are
emitted at small polar angles in the rest frame of the target nucleus, which allows to significantly simplify calculations.

The authors of Ref.~\cite{Frankfurt:1996uz} proposed a GEA-based model for the large-angle $d(p,2p)n$ process and evaluated the effects of color transparency (CT)
on the nuclear transparency ratio at $p_{\rm lab}=6-20$ GeV/c. The model of Ref.~\cite{Frankfurt:1996uz} includes single and double soft rescattering diagrams.
The transverse momentum transfers in soft elastic $pn$ amplitudes were calculated with respect to the incoming proton beam direction $z$,
which allowed the authors to linearize the fast particles propagators with respect to the momentum transfers along $z$.
In coordinate space, this resulted in the ordering of the $z$-coordinates of the proton and neutron in the deuteron according to the order
of the scattering processes in a given partial amplitude. (This leads, in particular, to the disappearance of partial amplitudes with rescattering of the incoming
and outgoing protons, see Fig. 1e,f in Ref.~\cite{Frankfurt:1996uz}.). At high beam momenta this is a natural assumption.
However, as the beam momentum decreases, its validity becomes questionable due to the large polar scattering angles in the laboratory system.

In Ref.~\cite{Larionov:2022gvn}, the process $d(p,2p)n$ was considered in a broader beam momentum range $p_{\rm lab}=6-75$ GeV/c.
The soft momentum transfers in the single rescattering diagrams were calculated relative to the actual directions of the asymptotic momenta of the particles.
Accordingly, the positions of the proton and neutron in the deuteron were also ordered along the same directions.
However, the calculation of double soft rescattering diagrams was done in the same way as in Ref.~\cite{Frankfurt:1996uz}.

In Ref.~\cite{Larionov:2025equ}, the uncertainties due to the choice of the deuteron wave function (DWF) were studied.
It has been shown that at least up to the transverse momentum of the spectator neutron of about 0.4 GeV/c these uncertainties do not sensitively change
the nuclear transparency ratio and the tensor analyzing power to influence conclusions on CT.

Another source of uncertainty is the way in which the amplitudes of multiple soft rescattering are treated.
Thus, the purpose of the present work is to rederive rescattering amplitudes in the GEA by treating the soft momentum transfers with respect to the actual
directions of the asymptotic particle momenta. This leads not only to corrected double rescattering amplitudes when both outgoing protons
experience soft rescattering on the spectator neutron, but also to finite amplitudes with rescattering of incoming and outgoing protons,
i.e. double- and triple rescattering amplitudes. It is shown that the proper treatment of the
geometry of the rescattering process is important for a forward moving spectator, i.e. when the center-of-mass (c.m.) velocity of the colliding
$pp$-system is small.

In addition, it is shown that multiple rescattering amplitudes of any order are possible when two outgoing protons undergo successive rescattering
on a slow spectator neutron.
The corresponding infinite sums are evaluated and included in the form of the ``renormalized''
double rescattering amplitudes.

The work is organized as follows.
In sec. \ref{form}, the GEA model is explained with an accent on new features related to the more precise treatment of the outgoing proton trajectories.
The calculation of the so-called $\Delta$-factors, which appear in GEA in contrast to the Glauber model, has been revised and improved compared
to previous versions \cite{Frankfurt:1996uz,Larionov:2022gvn}. The infinite series of the soft rescattering amplitudes of the two
outgoing protons are calculated.
The nuclear transparency ratio obtained with the old and new treatment of the multiple scattering is presented in sec. \ref{numerics}.
The summary is given in sec. \ref{summary}.
Appendix \ref{pdForw} includes the derivation of the forward elastic $pd$ scattering amplitude as a consistency check
of the GEA.

\section{Basic formalism}
\label{form}

\begin{figure}
  \begin{center}
    \begin{tabular}{cc}
  \includegraphics[scale = 0.4]{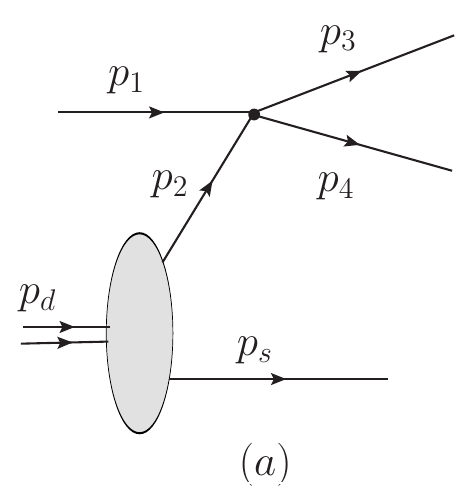} &
  \includegraphics[scale = 0.4]{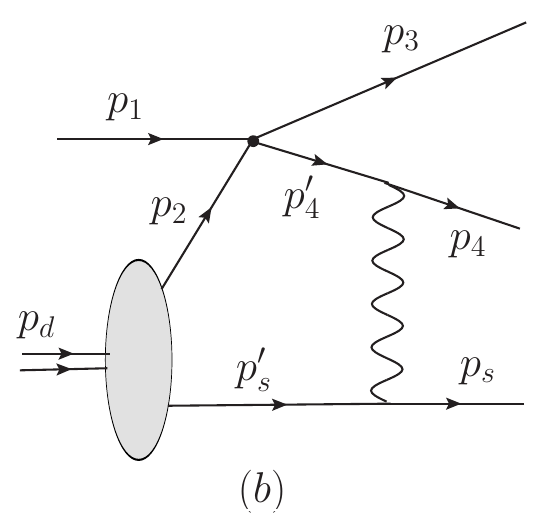}\\
  \includegraphics[scale = 0.4]{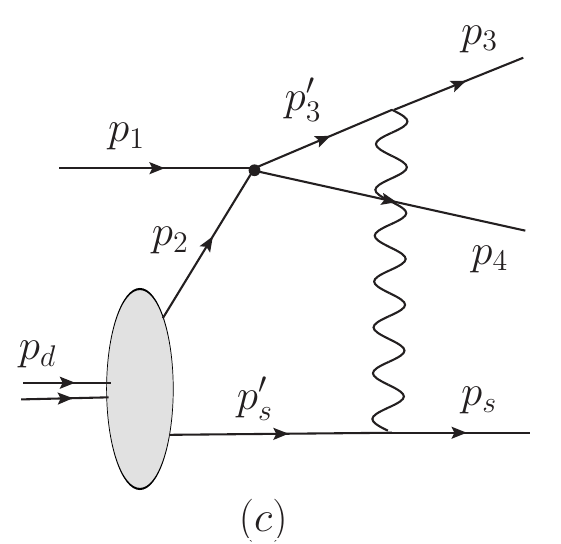} &
  \includegraphics[scale = 0.4]{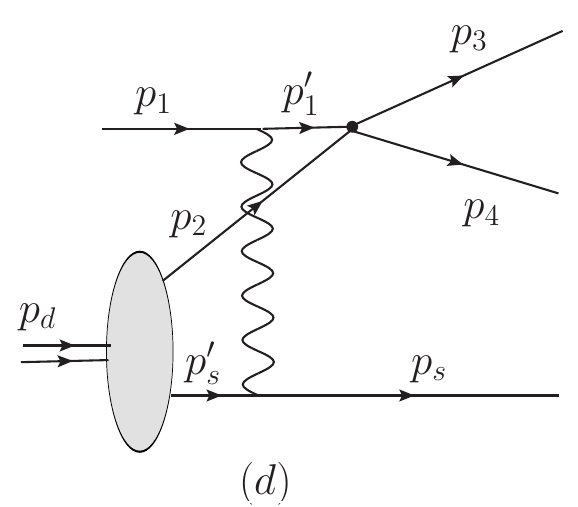}\\
  \includegraphics[scale = 0.4]{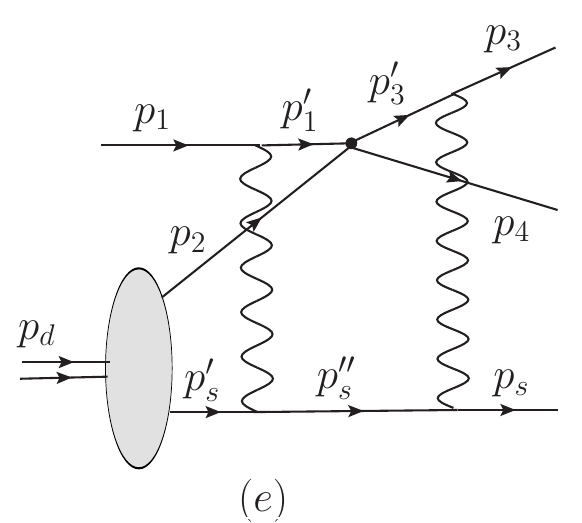} &
  \includegraphics[scale = 0.4]{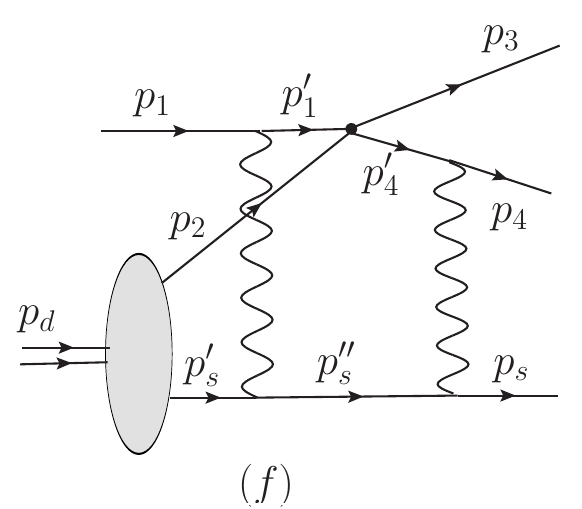} \\
  \includegraphics[scale = 0.4]{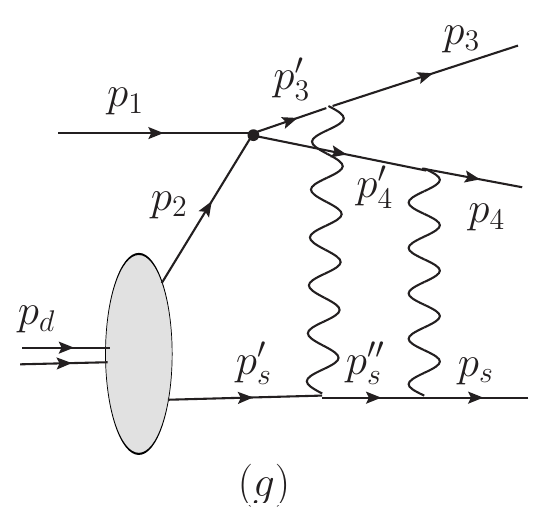} &
  \includegraphics[scale = 0.4]{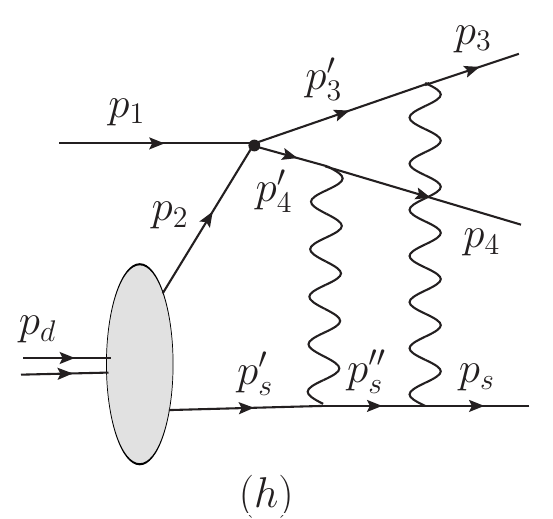}\\
  \includegraphics[scale = 0.4]{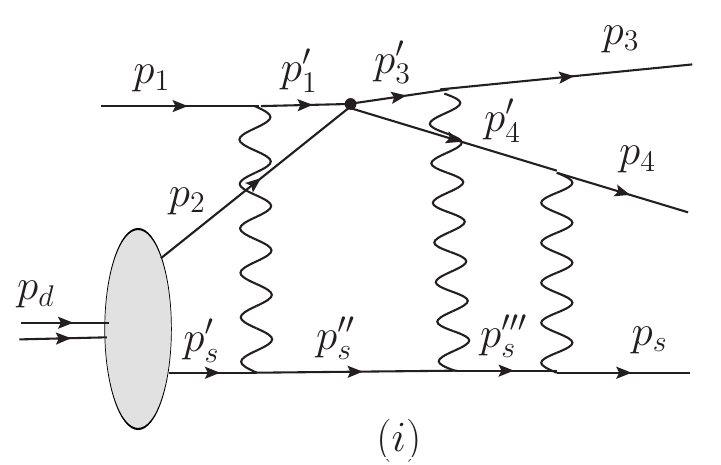} &
  \includegraphics[scale = 0.4]{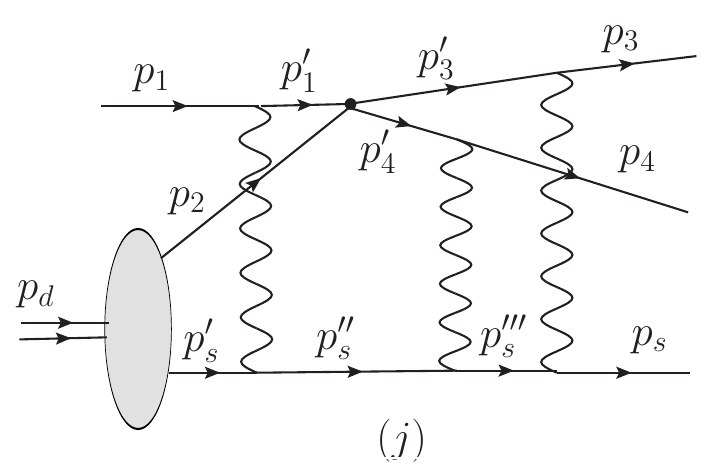}\\
    \end{tabular}
    \end{center}
\caption{\label{fig:diagr} Feynman diagrams for the process $p d  \to p p n$ with a slow spectator neutron.
  The wavy lines denote soft elastic scattering amplitudes.
  The four-momenta of the deuteron, neutron, beam proton, struck proton, and outgoing protons are denoted as
  $p_d$, $p_s$, $p_1$, $p_2$, $p_3$, $p_4$, respectively. The primed quantities denote the four-momenta of
  intermediate particles in the amplitudes with soft rescattering. }
\end{figure}
Figure \ref{fig:diagr} shows the partial amplitudes which include the impulse approximation (IA) amplitude
and all possible amplitudes with single, double, and triple soft rescattering of the incoming and outgoing protons.
The IA amplitude (a) and the amplitudes with single rescattering (b),(c) and (d) have been already calculated
in the GEA (see Eqs.~(7),(16) and (22) in Ref.~\cite{Larionov:2022gvn}) and, thus, we start with the double rescattering amplitude (e):
\begin{eqnarray}
  M^{(e)} &=&   M_{\rm hard}(p_3,p_4,p_1) \int \frac{d^4p_s^{\prime\prime}}{(2\pi)^{4}} \int \frac{d^4p_s^{\prime}}{(2\pi)^{4}}
              iM_{\rm el}(p_3,p_s,p_3^\prime) \frac{i}{p_3^{\prime 2}-m^2+i\epsilon} \nonumber \\
              && \times \frac{i}{p_1^{\prime 2}-m^2+i\epsilon} \frac{i}{p_s^{\prime\prime 2}-m^2+i\epsilon} iM_{\rm el}(p_1^\prime,p_s^{\prime\prime},p_1)
              \frac{i}{p_s^{\prime 2}-m^2+i\epsilon} \frac{i\Gamma_{d \to pn}(p_d,p_s^\prime)}{p_2^2-m^2+i\epsilon}~,  \label{M^(e)}
\end{eqnarray}
where $M_{\rm hard}$ and $M_{\rm el}$ are the invariant amplitudes of the hard and soft elastic scattering, respectively,
$m$ is the nucleon mass, and $\Gamma_{d \to pn}$ is the deuteron virtual decay vertex.
The hard amplitude is taken out of integrations
over internal four-momenta, since this amplitude changes only weakly on the momentum scale of soft rescattering.
Following Ref.~\cite{Frankfurt:1996uz},
we neglect the dependence of the soft rescattering amplitudes on the particle energy (on-shell approximation), 
and integrate over the time components of the four-momenta $p_s^{\prime}$ and $p_s^{\prime\prime}$, closing the integration contours
in the lower part of the complex plane, where only the particle poles of the propagators of intermediate spectator contribute,
\begin{equation}
  \int \frac{dp^0}{2\pi} \frac{i}{p^2-m^2+i\epsilon} = \frac{1}{2E},~~~E=\sqrt{\bvec{p}^2+m^2}~. \label{contIntegral}
\end{equation}
This corresponds to fixing the time ordering from left to right
in the diagram of Fig.~\ref{fig:diagr} (e) and leads to the following expression:
\begin{eqnarray}
  M^{(e)} &=&   M_{\rm hard}(p_3,p_4,p_1) \int \frac{d^3k^{\prime\prime}}{(2\pi)^{3}} \int \frac{d^3k^{\prime}}{(2\pi)^{3}}
              iM_{\rm el}(k^{\prime\prime}) \frac{i}{p_3^{\prime 2}-m^2+i\epsilon} \nonumber \\
              && \times \frac{i}{p_1^{\prime 2}-m^2+i\epsilon} \frac{1}{2E_s^{\prime\prime}} iM_{\rm el}(k^{\prime})
              \frac{1}{2E_s^{\prime}} \left(\frac{2E_s^\prime m_d}{p_2^0}\right)^{1/2} (2\pi)^{3/2} \phi(-\bvec{p}_s^\prime)~,  \label{M^(e)_3d}
\end{eqnarray}
where the integrations over $d^3p_s^{\prime\prime}d^3p_s^{\prime}$ have been equivalently replaced by integrations $d^3k^{\prime\prime}d^3k^\prime$
with $k^{\prime}=p_s^{\prime\prime}-p_s^{\prime}$, $k^{\prime\prime}=p_s-p_s^{\prime\prime}$ being sequential four-momentum transfers 
to the spectator; $E_s^{\prime}=\sqrt{\bvec{p}_s^{\prime 2} + m^2}$ and $E_s^{\prime\prime}=\sqrt{\bvec{p}_s^{\prime\prime 2} + m^2}$
are the on-shell energies of the intermediate spectator. In Eq.(\ref{M^(e)_3d}), we used the relation between the deuteron virtual decay vertex
and the DWF, which is applicable in the deuteron rest frame when the spectator four-momentum is on the mass shell
(see Ref.~\cite{Larionov:2018lpk} for derivation):
\begin{equation} 
 \frac{i\Gamma_{d \to pn}(p_d,p_s^\prime)}{p_2^2-m^2+i\epsilon} 
 = \left(\frac{2E_s^\prime m_d}{p_2^0}\right)^{1/2} (2\pi)^{3/2} \phi(-\bvec{p}_s^\prime)~,     \label{Gamma_d}
\end{equation}
where $m_d$ is the deuteron mass.

\begin{figure}
  \begin{center}
     \includegraphics[scale = 0.50]{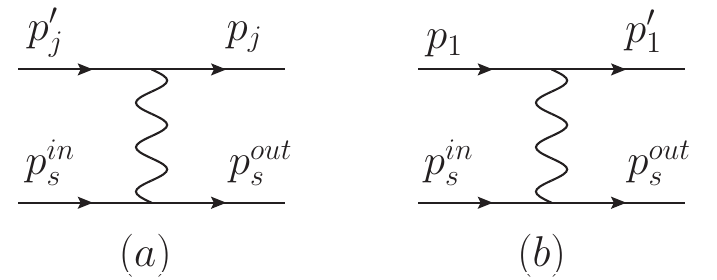}
  \end{center}
  \caption{\label{fig:deltas} Soft elastic transition amplitudes between the intermediate and asymptotic states
    for the outgoing protons $j=3,4$ (a) and for the incoming proton 1 (b).}
\end{figure}

In the diagrams of Fig.~\ref{fig:diagr}, all possible combinations of the intermediate propagators with soft rescattering amplitudes
are either of type (a) or type (b) in Fig.~\ref{fig:deltas}.
The inverse propagator of intermediate proton related to the transition amplitude of the type (a) in Fig.~\ref{fig:deltas}
can be written as
\begin{equation}
  p_j^{\prime 2}-m^2+i\epsilon = 2|\bvec{p}_j| (p_s^{in,\tilde z} - p_s^{out, \tilde z} + \Delta_j + i\epsilon)~,~~~j=3,4~,       \label{j_inv_prop}
\end{equation}
where the $\tilde z$-axis is directed along $\bvec{p}_j$ and
\begin{equation}
  \Delta_j \equiv \frac{E_j(E_s^{out}-E_s^{in})}{|\bvec{p}_j|} + \frac{(p_s^{out}-p_s^{in})^2}{2|\bvec{p}_j|}
  \simeq \frac{(E_j-E_s^{in})(E_s^{out}-E_s^{in})}{|\bvec{p}_j|}~,~~~j=3,4~.              \label{Delta_j}
\end{equation}
In the case of transition amplitude of type (b) in Fig.~\ref{fig:deltas}, the inverse propagator of the intermediate proton
is expressed as
\begin{equation}
  p_1^{\prime 2}-m^2+i\epsilon = 2|\bvec{p}_1| (p_s^{out,z} - p_s^{in,z} - \Delta_1 + i\epsilon)~,       \label{1_inv_prop}
\end{equation}
where the $z$-axis is directed along the beam momentum $\bvec{p}_1$ and
\begin{equation}
  \Delta_1 \equiv \frac{E_1(E_s^{out}-E_s^{in})}{|\bvec{p}_1|} - \frac{(p_s^{out}-p_s^{in})^2}{2|\bvec{p}_1|}
  \simeq \frac{(E_1+E_s^{in})(E_s^{out}-E_s^{in})}{|\bvec{p}_1|}~.               \label{Delta_1}
\end{equation}
In the second step in Eqs.(\ref{Delta_j}),(\ref{Delta_1}), the term $\pm\bvec{p}_s^{in}\bvec{k}/|\bvec{p}_j|$ ($\bvec{k}=\bvec{p}_s^{out}-\bvec{p}_s^{in}$)
that changes sign when taking integral over $d^3k$ was neglected.

The Glauber approximation can be obtained if one sets $\Delta=0$. This is equivalent to the assumption that the nucleons-scatterers (the spectator neutron in our case)
inside the nucleus remain static and implies approximate conservation of the longitudinal momenta of the particles in small-angle
elastic scattering at high energies, as can be seen from the zeros of the inverse propagators (\ref{j_inv_prop}),(\ref{1_inv_prop}).
Including finite $\Delta$'s becomes important at recoil momentum of the spectator neutron of the order of or more than the Fermi momentum, i.e. when the amplitudes
with rescattering become the leading ones while the IA amplitude is small. Finite $\Delta$'s are also needed for a proper account of the relativistic kinematics
which leads to the conservation of the $p^{-}=\sqrt{m^2+\bvec{p}^2} - p^z$ components of the light-cone momenta \cite{Frankfurt:1996xx}
as can be seen from Eqs.(\ref{j_inv_prop}),(\ref{1_inv_prop}) in the ultra-relativistic limit $E_j=|\bvec{p}_j| \to +\infty$.

Below we propose a modified calculation of $\Delta$'s compared to the calculations from Refs.~\cite{Frankfurt:1996uz,Larionov:2022gvn}.
The most important changes are for the case when at least one of the energies, $E_s^{in}$, $E_s^{out}$, is related to the intermediate
spectator neutron which already suffered at least one rescattering (i.e. it is in the middle of rescattering chain).

Eqs.(\ref{Delta_j}),(\ref{Delta_1}) include the energies of the incoming and outgoing spectator,  $E_s^{in}=\sqrt{(\bvec{p}_s^{in})^2+m^2}$ and $E_s^{out}=\sqrt{(\bvec{p}_s^{out})^2+m^2}$,
which depend on the momenta. The momentum $\bvec{p}_s^{in}$ is not fixed  by the kinematics of the process and is always an integration variable.
So is also $\bvec{p}_s^{out}$ unless it is the asymptotic momentum of the spectator. In order to further simplify Eq.(\ref{M^(e)_3d})
and similar formulas for other partial amplitudes of Fig.~\ref{fig:diagr} it is necessary to set
some fixed  'effective' values of the $\Delta$'s in Eqs.~(\ref{j_inv_prop}),(\ref{1_inv_prop}). This can be done as follows.
Suppose, the energy of spectator before rescattering, $E_s^{in}$, is known. If the spectator after rescattering is in the asymptotic state then
its energy is also known, i.e. $E_s^{out}=E_s$, and the corresponding $\Delta$ is fully determined by Eq.(\ref{Delta_j}) or Eq.(\ref{Delta_1}).
If the spectator after rescattering is still in the intermediate state then its energy can be estimated as
\begin{equation}
  E_s^{out} = \sqrt{(E_s^{in})^2 + \langle k_t^2 \rangle + \Delta_j^2}~,~~~j=1,3,4~,            \label{E_s^out} 
\end{equation}
where $\langle k_t^2 \rangle = B_{pn}^{-1}(|\bvec{p}_j|)$ is the average transverse momentum transfer squared
and $B_{pn}(|\bvec{p}_j|)$ is the slope parameter of the $pn$ elastic amplitude (see Eq.~\ref{M_el} below).
Eq.(\ref{E_s^out}) is obtained neglecting any correlation between momentum of the spectator before rescattering and momentum transfer
and setting the longitudinal momentum transfer to satisfy the on-shell condition for the intermediate state of the proton $j$. 
By substituting Eq.(\ref{E_s^out}) in Eqs.(\ref{Delta_j}),(\ref{Delta_1}) and solving the resulting quadratic equation with respect
to $\Delta_j$ one obtains the following formula:
\begin{equation}
  \Delta_j = \frac{1}{1-\beta^2} \left(E_s^{in}\beta - \sqrt{(E_s^{in}\beta)^2 - \langle k_t^2 \rangle (1-\beta^2)} \right)~,   \label{Delta_j_fin}
\end{equation}
where
\begin{equation}
  \beta = \left\{ \begin{array}{ll}   
                    |\bvec{p}_1|/(E_1+E_s^{in}) & \mbox{~~for~~} j=1 \\
                    |\bvec{p}_j|/(E_j-E_s^{in}) & \mbox{~~for~~} j=3,4~.
                 \end{array}
           \right.     \label{beta}
\end{equation}
(The choice of the '-' sign at the square root in Eq.(\ref{Delta_j_fin}) ensures that $\Delta_j$ is finite in the high-energy limit $\beta \to 1$.)
The procedure is repeated for the next soft rescattering from left to right in the amplitude by setting the new $E_s^{in}$ equal to the previously determined $E_s^{out}$.
For the first rescattering, $E_s^{in}=m$ is set since in this case the spectator momentum is strongly suppressed by the DWF.

Having fixed $\Delta$'s, we can now proceed to simplify Eq.(\ref{M^(e)_3d}). The following relations are used to transfer the propagators and wave functions
to the coordinate representation:
\begin{eqnarray}
  && \frac{i}{p+i\epsilon} =\int dz^0 \Theta(z^0) \mbox{e}^{i p z^0}~,   \label{Dcoord} \\
  && (2\pi)^{3/2} \phi(-\bvec{p}_s^\prime) = \int d^3 r \mbox{e}^{i\bvec{p}_s^\prime\bvec{r}} 
                   \phi(\bvec{r})~, ~~~\bvec{r}=\bvec{r}_2-\bvec{r}_s~,     \label{phi(p_2)}
\end{eqnarray}
where $\Theta(x)$ is the Heaviside step function.
By using relations (\ref{j_inv_prop}),(\ref{1_inv_prop}),(\ref{Dcoord}) the propagators can be written as
\begin{eqnarray}
  && \frac{i}{p_1^{\prime 2}-m^2+i\epsilon} = \frac{i}{2|\bvec{p}_1| (k^{\prime z}  - \Delta_1 + i\epsilon)}
     = \frac{1}{2|\bvec{p}_1|} \int dz^0 \Theta(z^0) \mbox{e}^{i(k^{\prime z}  - \Delta_1)z^0}~,       \label{1_prop_(e)}\\
  && \frac{i}{p_3^{\prime 2}-m^2+i\epsilon} = \frac{i}{2|\bvec{p}_3| (-k^{\prime\prime \tilde z} + \Delta_3 + i\epsilon)}
     = \frac{1}{2|\bvec{p}_3|} \int d\tilde z^0 \Theta(\tilde z^0) \mbox{e}^{i(-k^{\prime\prime \tilde z} + \Delta_3)\tilde z^0}~. \label{3_prop_(e)}
\end{eqnarray}
Thus, it is convenient to perform the integrations in the cylindrical coordinates:
\begin{eqnarray}
  && d^3k^{\prime} = d^2k_t^{\prime} dk^{\prime z}~,       \label{d^3k^p}\\
  && d^3k^{\prime\prime} = d^2k_t^{\prime\prime} dk^{\prime\prime \tilde z}~,   \label{d^3k^pp}
\end{eqnarray}
where $\bvec{k}_t^{\prime} \perp \bvec{e}_z \equiv \bvec{p}_1/|\bvec{p}_1|$, $\bvec{k}_t^{\prime\prime} \perp \bvec{e}_{\tilde z} \equiv \bvec{p}_3/|\bvec{p}_3|$.
The phase of the integrand in Eq.(\ref{phi(p_2)}) can be rewritten as
\begin{equation}
  \bvec{p}_s^\prime\bvec{r} = (\bvec{p}_s - \bvec{k}^{\prime\prime} - \bvec{k}^{\prime})\bvec{r}
  = \bvec{p}_s\bvec{r} - \bvec{k}_t^{\prime\prime}\bvec{\tilde b} - k^{\prime\prime \tilde z} \tilde z
  - \bvec{k}_t^{\prime} \bvec{b} - k^{\prime z}z~,         \label{phaseDWF}
\end{equation}
where $\tilde z = \bvec{r}\bvec{e}_{\tilde z}$ and $\bvec{\tilde b} = \bvec{r} - \tilde z \bvec{e}_{\tilde z}$
are, respectively, the relative longitudinal and transverse displacements of the nucleons in the deuteron with respect to the $\bvec{e}_{\tilde z}$ direction,
while $z= \bvec{r}\bvec{e}_z$ and $\bvec{b} = \bvec{r} - z \bvec{e}_z$
are those with respect to the $\bvec{e}_{z}$ direction.

Substituting Eqs.(\ref{phi(p_2)})-(\ref{phaseDWF}) in Eq.(\ref{M^(e)_3d}) we have:
\begin{eqnarray}
  M^{(e)} &=&  2m^{1/2} M_{\rm hard}(p_3,p_4,p_1) \int \frac{d^2k_t^{\prime\prime} dk^{\prime\prime \tilde z}}{(2\pi)^{3}} \int \frac{d^2k_t^{\prime} dk^{\prime z}}{(2\pi)^{3}}
              \frac{iM_{\rm el}(|\bvec{p}_3|,k_t^{\prime\prime})}{4|\bvec{p}_3|m}
              \int d\tilde z^0 \Theta(\tilde z^0) \mbox{e}^{i(-k^{\prime\prime \tilde z} + \Delta_3)\tilde z^0} \nonumber \\
           && \times \frac{iM_{\rm el}(|\bvec{p}_1|,k_t^{\prime})}{4|\bvec{p}_1|m}
                   \int dz^0 \Theta(z^0) \mbox{e}^{i(k^{\prime z}  - \Delta_1)z^0}
             \int d^3 r \mbox{e}^{i(\bvec{p}_s\bvec{r} - \bvec{k}_t^{\prime\prime}\bvec{\tilde b} - k^{\prime\prime \tilde z} \tilde z
                                        - \bvec{k}_t^{\prime} \bvec{b} - k^{\prime z}z)} 
                   \phi(\bvec{r})~.                      \label{M^(e)_3d_decomp}
\end{eqnarray}
In this equation, we replaced $m_d \to 2m$ since the deuteron binding energy is negligibly small on the energy scale of the considered processes,
and $E_s^{\prime},E_s^{\prime\prime},p_2^0 \to m$ since the Fermi motion is non-relativistic in the deuteron rest frame. 
In the spirit of Glauber theory, the soft elastic $pn$ rescattering amplitudes were assumed to depend on the transverse momentum transfers only
(apart from the dependence on the momenta of the scattered protons which is made explicit).

The integrations over longitudinal momentum transfers in Eq.(\ref{M^(e)_3d_decomp}) can be easily performed giving
\begin{equation}
   \int \frac{dk^{\prime\prime \tilde z}}{2\pi} \int \frac{dk^{\prime z}}{2\pi} \mbox{e}^{i(k^{\prime z}  - \Delta_1)z^0 + i(-k^{\prime\prime \tilde z} + \Delta_3)\tilde z^0
    -ik^{\prime\prime \tilde z} \tilde z -ik^{\prime z}z}
  = \delta(z^0-z) \delta(\tilde z^0+\tilde z) \mbox{e}^{-i\Delta_1z -i\Delta_3\tilde z}~.     \label{longMomInt}
\end{equation}
Integrating over $dz^0$ and $d\tilde z^0$ removes the $\delta$-functions which finally gives
\begin{equation}
  M^{(e)} = 2m^{1/2} M_{\rm hard}(p_3,p_4,p_1) \int d^3r \Theta(z) \Theta(-\tilde z)
  \mbox{e}^{i\bvec{p}_s\bvec{r}-i\Delta_1z-i\Delta_3\tilde z} \phi(\bvec{r})
            \Gamma_1(b) \Gamma_3(\tilde b)~, \label{M^(e)_GEA}
\end{equation}
where
\begin{equation}
    \Gamma_i(b) = -\frac{i}{4 |\bvec{p}_i| m} \int \frac{d^2k_t}{(2\pi)^2} \mbox{e}^{-i\bvec{k}_t\bvec{b}} M_{\rm el}(|\bvec{p}_i|,k_t)~,
                        \label{Gamma_i}
\end{equation}
with $\bvec{k}_t \perp \bvec{p}_i$, is the profile function. Note that the order of particle interactions according to the time ordering of Fig.~\ref{fig:diagr}e
from left to right is expressed by the $\Theta$-functions in Eq.(\ref{M^(e)_GEA}).
The integration over azimuthal angle of the transverse momentum transfer $\bvec{k}_t$ can be performed analytically
which gives
\begin{equation}
    \Gamma_i(b) = -\frac{i}{8 \pi |\bvec{p}_i| m} \int\limits_0^{+\infty} dk_t k_t M_{\rm el}(|\bvec{p}_i|,k_t) J_0(k_tb)~,    \label{Gamma_i_fin}
\end{equation}
where
\begin{equation}
  J_0(x)=\frac{1}{2\pi}\int\limits_0^{2\pi} d\phi\, \mbox{e}^{-ix\cos \phi}    \label{J_0}
\end{equation}
is the Bessel function of the first kind. In the case of the elastic $pn$ amplitude in the standard high-energy form (c.f. Ref.~\cite{Glauber:1970jm}),
\begin{equation}
    M_{\rm el}(p_{\rm lab},k_t)= 2 p_{\rm lab} m \sigma_{p n}^{\rm tot}
     (i+\rho_{\rm p n}) \mbox{e}^{-B_{\rm p n}k_t^2/2}~,           \label{M_el}
\end{equation}
where $\sigma_{p n}^{\rm tot}$ is the total $pn$ cross section,
$\rho_{\rm p n}=\mbox{Re}M_{\rm el}(p_{\rm lab},0)/\mbox{Im}M_{\rm el}(p_{\rm lab},0)$
is the ratio of the real part of the forward $pn$ scattering amplitude to the imaginary one,
and $B_{\rm p n}$ is the slope of the dependence on the transverse momentum transfer,
the profile function can be expressed as
\begin{equation}
    \Gamma_i(b) = -\frac{iM_{\rm el}(|\bvec{p}_i|,0)}{8 \pi |\bvec{p}_i| m B_{\rm p n}} \mbox{e}^{-b^2/2B_{\rm p n}}~.   \label{Gamma_i_Glauber}
\end{equation}
Equations (\ref{Gamma_i}),(\ref{Gamma_i_fin}) are applicable also if the CT effects are included via the dependence of the effective
$pn$ cross section on the longitudinal distance between proton and neutron in the deuteron (see Eq.(41) in Ref.~\cite{Larionov:2022gvn}).
\footnote{Eq.(\ref{Gamma_i_Glauber}) can not be used with CT effects since in that case the proton form factor is modified,
see Eq.(43) in Ref.~\cite{Larionov:2022gvn}.}

The amplitude (f) is obtained from Eq.(\ref{M^(e)_GEA}) by replacing $3 \to 4$ in all multipliers except the hard amplitude.
The latter is asymmetric with respect to the exchange of quantum numbers of the protons 3 and 4,
which provides the same asymmetry of the total amplitude.

The amplitude (g) can also be obtained in the GEA by explicitly taking into account the different directions of the momenta of incoming and outgoing protons.
We will not repeat the derivation which is much similar to that of Eq.(\ref{M^(e)_GEA}). The resulting expression is
\begin{equation}
  M^{(g)} =  2m^{1/2} M_{\rm hard}(p_3,p_4,p_1) \int d^3r \Theta(-\tilde z) \Theta(-\tilde{\tilde  z}) 
                  \mbox{e}^{i\bvec{p}_s\bvec{r}-i\Delta_3\tilde z-i\Delta_4\tilde{\tilde  z}} \phi(\bvec{r})
             \Gamma_3(\tilde b) \Gamma_4(\tilde{\tilde b})~.    \label{M^(g)_GEA}
\end{equation}
The directions of $\tilde z$ and $\tilde{\tilde z}$ axes are chosen along the momenta $\bvec{p}_3$ and $\bvec{p}_4$, respectively.

It is also straightforward to calculate the triple rescattering amplitude (i):
\begin{eqnarray}
   M^{(i)} &=& -2m^{1/2} M_{\rm hard}(p_3,p_4,p_1) \int d^3r \Theta(z) \Theta(-\tilde z) \Theta(-\tilde{\tilde  z}) 
                    \mbox{e}^{i\bvec{p}_s\bvec{r} -i\Delta_1z -i\Delta_3\tilde z -i\Delta_4\tilde{\tilde  z}} \phi(\bvec{r})  \nonumber \\
   && \times \Gamma_1(b) \Gamma_3(\tilde b) \Gamma_4(\tilde{\tilde b})~.   \label{M^(i)_GEA}
\end{eqnarray}
The directions of the $\tilde z$ and $\tilde{\tilde z}$ axes are defined as in Eq.(\ref{M^(g)_GEA}).
The longitudinal momentum transfers $\Delta_1,\Delta_3$ and $\Delta_4$ are calculated
using Eqs.(\ref{Delta_j}),(\ref{Delta_1}),(\ref{Delta_j_fin}).
The amplitudes (h) and (j) are given by Eq.(\ref{M^(g)_GEA}) and Eq.(\ref{M^(i)_GEA}), respectively,
but with another values of $\Delta_3$ and $\Delta_4$, determined by using the procedure described above
(see Eqs.(\ref{Delta_j}),(\ref{Delta_j_fin}),(\ref{beta})).

Some comments are in order here. In the limit $\Delta_i=0$, Eqs.(\ref{M^(e)_GEA}),(\ref{M^(g)_GEA}) and (\ref{M^(i)_GEA}) become the standard double and triple rescattering terms
of Glauber theory, see, e.g. Ref.~\cite{Frankfurt:1996xx}. However, one new feature appears due to the presence of the two outgoing particles.
In Ref.~\cite{Sargsian:2001ax}, the reduction theorem was proved that states that ``high energy particles propagating in the nuclear medium cannot interact with the same bound nucleon
a second time after interacting with another bound nucleon''. This theorem allows to reduce the infinite series of the rescattering diagrams to the limited set of the scatterings,
with no more than one scattering on each bound nucleon. The proof given in Ref.~\cite{Sargsian:2001ax} was made for {\it one} high energy particle.
This theorem explains, in particular, why it is sufficient to take into account only the contributions of the single and double scattering amplitudes when calculating the elastic
$pd$ amplitude \cite{Glauber:1955qq,Bassel:1968stc,Platonova:2010wjt,Uzikov:2016lsc}.
Along these lines, Appendix \ref{pdForw} contains the derivation of the forward elastic $pd$ scattering amplitude
based on Feynman diagrams. It is clearly shown there that triple scattering amplitudes do not contribute.
However, the situation gets different if {\it two} high energy particles propagate in the nucleus.

\begin{figure}
  \begin{center}
    \includegraphics[scale = 0.4]{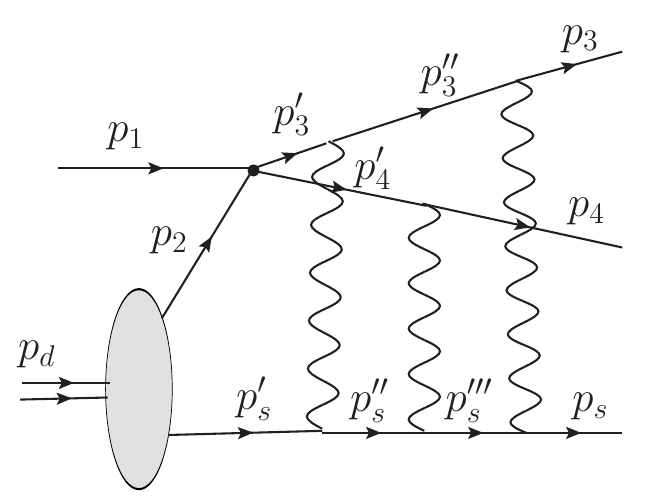}
  \end{center}
  \caption{\label{fig:p3p4p3} Feynman diagram for the process $p d  \to p p n$ with triple rescattering
    of outgoing protons on  
    the slow spectator neutron. Notations are the same as in  Fig.~\ref{fig:diagr}.}
\end{figure}
Let us consider the diagram of Fig.~\ref{fig:p3p4p3} that is obtained by adding the rescattering of the proton 3 on the r.h.s of the diagram (g) of Fig.~\ref{fig:diagr}.
The corresponding invariant amplitude (the notation ``343'' indicates the order of rescatterings) is expressed as
\begin{eqnarray}
  M^{(343)} &=& M_{\rm hard}(p_3,p_4,p_1) \int \frac{d^4p_s^{\prime\prime\prime}}{(2\pi)^{4}} \int \frac{d^4p_s^{\prime\prime}}{(2\pi)^{4}} \int \frac{d^4p_s^{\prime}}{(2\pi)^{4}} 
    iM_{\rm el}(p_3,p_s,p_3^{\prime\prime}) \frac{i}{p_3^{\prime\prime 2}-m^2+i\epsilon} \nonumber \\
    && \times \frac{i}{p_s^{\prime\prime\prime 2}-m^2+i\epsilon} iM_{\rm el}(p_4,p_s^{\prime\prime\prime},p_4^\prime)  \frac{i}{p_4^{\prime 2}-m^2+i\epsilon}
                \frac{i}{p_s^{\prime\prime 2}-m^2+i\epsilon} \nonumber \\
              && \times iM_{\rm el}(p_3^{\prime\prime},p_s^{\prime\prime},p_3^{\prime}) \frac{i}{p_3^{\prime 2}-m^2+i\epsilon}
              \frac{i}{p_s^{\prime 2}-m^2+i\epsilon} \frac{i\Gamma_{d \to pn}(p_d,p_s^\prime)}{p_2^2-m^2+i\epsilon}~.  \label{M^(343)}
\end{eqnarray}
After integrations over time components of the four-momenta  of the intermediate spectator this amplitude takes the following form:
\begin{eqnarray}
  \lefteqn{M^{(343)} = M_{\rm hard}(p_3,p_4,p_1) \int \frac{d^3k^{\prime\prime\prime}}{(2\pi)^{3}} \int \frac{d^3k^{\prime\prime}}{(2\pi)^{3}} \int \frac{d^3k^{\prime}}{(2\pi)^{3}} 
    iM_{\rm el}(k^{\prime\prime\prime}) \frac{i}{p_3^{\prime\prime 2}-m^2+i\epsilon} \frac{1}{2E_s^{\prime\prime\prime}} iM_{\rm el}(k^{\prime\prime})} && \nonumber \\
    &\times&    \frac{i}{p_4^{\prime 2}-m^2+i\epsilon}
                \frac{1}{2E_s^{\prime\prime}} iM_{\rm el}(k^{\prime}) \frac{i}{p_3^{\prime 2}-m^2+i\epsilon}
              \frac{1}{2E_s^{\prime}} \left(\frac{2E_s^\prime m_d}{p_2^0}\right)^{1/2} (2\pi)^{3/2} \phi(-\bvec{p}_s^\prime)~,  \label{M^(343)_3d}
\end{eqnarray}
where $k^{\prime}=p_s^{\prime\prime}-p_s^{\prime}$, $k^{\prime\prime}=p_s^{\prime\prime\prime}-p_s^{\prime\prime}$,
and $k^{\prime\prime\prime}=p_s-p_s^{\prime\prime\prime}$ are sequential four-momentum transfers to the spectator;
$E_s^{\prime}=\sqrt{\bvec{p}_s^{\prime 2} + m^2}$, $E_s^{\prime\prime}=\sqrt{\bvec{p}_s^{\prime\prime 2} + m^2}$, and $E_s^{\prime\prime\prime}=\sqrt{\bvec{p}_s^{\prime\prime\prime 2} + m^2}$
are the on-shell spectator energies; the $d \to pn$ vertex was expressed via the DWF according to Eq.(\ref{Gamma_d}).  
The first two inverse propagators in Eq.(\ref{M^(343)_3d}) can be expressed by using Eq.(\ref{j_inv_prop}) as
\begin{eqnarray}
  p_3^{\prime\prime 2}-m^2+i\epsilon &=& 2|\bvec{p}_3| (-k^{\prime\prime\prime \tilde z} + \tilde\Delta_3 + i\epsilon)~, \\
  p_4^{\prime 2}-m^2+i\epsilon &=& 2|\bvec{p}_4| (-k^{\prime\prime \tilde{\tilde z}} + \Delta_4 + i\epsilon)~,
\end{eqnarray}
with $\bvec{e}_{\tilde z} \equiv \bvec{p}_3/|\bvec{p}_3|$, $\bvec{e}_{\tilde{\tilde z}} \equiv \bvec{p}_4/|\bvec{p}_4|$.
For the third inverse propagator we can write:
\begin{eqnarray}
  p_3^{\prime 2}-m^2+i\epsilon &=& (p_3+k^{\prime\prime\prime}+k^{\prime})^2 -m^2 +i\epsilon \nonumber \\
  &=& 2 |\bvec{p}_3| (-k^{\prime\prime\prime \tilde z}-k^{\prime \tilde z} + \Delta_3 + i\epsilon)~.
\end{eqnarray}
We can now rewrite the amplitude (\ref{M^(343)_3d}) in the following form:
\begin{eqnarray}
  \lefteqn{M^{(343)} = 2m^{1/2} M_{\rm hard}(p_3,p_4,p_1) \int \frac{d^3k^{\prime\prime\prime}}{(2\pi)^{3}}
    \int \frac{d^3k^{\prime\prime}}{(2\pi)^{3}} \int \frac{d^3k^{\prime}}{(2\pi)^{3}}
\frac{iM_{\rm el}(|\bvec{p}_3|,k_t^{\prime\prime\prime})}{4|\bvec{p}_3|m}
    \frac{i}{-k^{\prime\prime\prime \tilde z} + \tilde\Delta_3 + i\epsilon}} \nonumber \\
&\times&  \frac{iM_{\rm el}(|\bvec{p}_4|,k_t^{\prime\prime})}{4|\bvec{p}_4|m}
          \frac{i}{-k^{\prime\prime \tilde{\tilde z}} + \Delta_4 + i\epsilon}
          \frac{iM_{\rm el}(|\bvec{p}_3|,k_t^{\prime})}{4|\bvec{p}_3|m}
          \frac{i}{-k^{\prime\prime\prime \tilde z}-k^{\prime \tilde z} + \Delta_3 + i\epsilon}    \nonumber \\
&\times& (2\pi)^{3/2} \phi(-\bvec{p}_s+\bvec{k}^{\prime\prime\prime}+\bvec{k}^{\prime\prime}+\bvec{k}^\prime)  \nonumber \\
&=& 2m^{1/2} M_{\rm hard}(p_3,p_4,p_1) \int \frac{d^2k_t^{\prime\prime\prime}dk^{\prime\prime\prime \tilde z} }{(2\pi)^{3}}
    \int \frac{d^2k_t^{\prime\prime}dk^{\prime\prime \tilde{\tilde z}}}{(2\pi)^{3}}
    \int \frac{d^2k_t^{\prime}dk^{\prime \tilde z}}{(2\pi)^{3}}    \nonumber \\
  &\times&  \frac{iM_{\rm el}(|\bvec{p}_3|,k_t^{\prime\prime\prime})}{4|\bvec{p}_3|m}
    \frac{i}{-k^{\prime\prime\prime \tilde z} + \tilde\Delta_3 + i\epsilon}
    \frac{iM_{\rm el}(|\bvec{p}_4|,k_t^{\prime\prime})}{4|\bvec{p}_4|m}
   \int d\tilde{\tilde z}^0 \Theta(\tilde{\tilde z}^0) \mbox{e}^{i(-k^{\prime\prime \tilde{\tilde z}} + \Delta_4)\tilde{\tilde z}^0}   \nonumber \\
  &\times&   \frac{iM_{\rm el}(|\bvec{p}_3|,k_t^{\prime})}{4|\bvec{p}_3|m}
           \int d\tilde z^0 \Theta(\tilde z^0) \mbox{e}^{i(-k^{\prime\prime\prime \tilde z}-k^{\prime \tilde z} + \Delta_3)\tilde z^0}  \nonumber \\
   &\times&   \int d^3 r \mbox{e}^{ i[\bvec{p}_s\bvec{r} - (k^{\prime\prime\prime \tilde z}+k^{\prime \tilde z})\tilde z - (\bvec{k}_t^{\prime\prime\prime}+\bvec{k}_t^{\prime})\bvec{\tilde b}
                                   -k^{\prime\prime \tilde{\tilde z}}\tilde{\tilde z} - \bvec{k}_t^{\prime\prime}\bvec{\tilde{\tilde b}}] } 
                   \phi(\bvec{r})~.  \label{M^(343)_3d_decomp}
\end{eqnarray}
Note that, in Eq.(\ref{M^(343)_3d_decomp}), the DWF depends on {\it all three} partial momentum transfers, $\bvec{k}^{\prime\prime\prime},\bvec{k}^{\prime\prime}$ and $\bvec{k}^\prime$.
This fact does not allow to factorize-out the product of propagators of proton 3 which would otherwise produce zero after integrating over $dk^{\prime\prime\prime \tilde z}$,
similar to the triple scattering term in the $pd$ elastic scattering (see Eq.(\ref{M_{pd}^(e)_fin}) of Appendix \ref{pdForw}).
Instead, the integration over longitudinal momentum transfer $k^{\prime \tilde z}$ can be equivalently replaced by integration over 
$q^{\tilde z}=k^{\prime \tilde z}+k^{\prime\prime\prime \tilde z}$. This immediately leads to the factorizing-out the integral
\begin{equation}
  \int \frac{dk^{\prime\prime\prime \tilde z}}{2\pi} \frac{i}{-k^{\prime\prime\prime \tilde z} + \tilde\Delta_3 + i\epsilon} = \frac{1}{2}~,  \label{LongMomInt0.5}
\end{equation}
which is calculated by using the rule
\begin{equation}
  \frac{1}{x+i\epsilon} = {\cal P} \frac{1}{x} -i\pi\delta(x)~,    \label{LandauRule}
\end{equation}
where ${\cal P}$ denotes the principal value integration.

It should be stressed, that there is no any other term containing the longitudinal momentum transfer from the last scattered proton
(here 3) left in Eq.(\ref{M^(343)_3d_decomp}) after introducing the accumulated momentum transfer $q^{\tilde z}$.
Remaining longitudinal momentum integrations in Eq.(\ref{M^(343)_3d_decomp}) are as follows:
\begin{equation}
 \int \frac{dk^{\prime\prime \tilde{\tilde z}}}{2\pi} \int \frac{dq^{\tilde z}}{2\pi}  \mbox{e}^{i(-k^{\prime\prime \tilde{\tilde z}} + \Delta_4)\tilde{\tilde z}^0}
   \mbox{e}^{i(-q^{\tilde z} + \Delta_3)\tilde z^0}  \mbox{e}^{ -iq^{\tilde z}\tilde z -ik^{\prime\prime \tilde{\tilde z}}\tilde{\tilde z} } 
 = \delta(\tilde z^0 + \tilde z) \delta(\tilde{\tilde z}^0 + \tilde{\tilde z}) \mbox{e}^{-i\Delta_3\tilde z -i\Delta_4\tilde{\tilde z}}~.  \label{longMomInt1}
\end{equation}
Substituting Eqs.(\ref{LongMomInt0.5}),(\ref{longMomInt1}) in Eq.(\ref{M^(343)_3d_decomp}) and using the definition of the profile function, Eq.(\ref{Gamma_i}),
leads to the following formula:
\begin{equation}
  M^{(343)} = -m^{1/2} M_{\rm hard}(p_3,p_4,p_1) \int d^3 r \Theta(-\tilde z) \Theta(-\tilde{\tilde z}) \phi(\bvec{r})
    \mbox{e}^{i(\bvec{p}_s\bvec{r} -\Delta_3\tilde z -\Delta_4\tilde{\tilde z})}\, \Gamma_3^2(\tilde b)  \Gamma_4(\tilde{\tilde b})~.  \label{M^(343)_GEA}
\end{equation}
By comparing this formula with Eq.(\ref{M^(g)_GEA}) for the amplitude $M^{(g)} \equiv M^{(34)}$, one can see that an additional rescattering of the proton 3 brought a factor of
$-\Gamma_3(\tilde b)/2$ in the integrand of Eq.(\ref{M^(g)_GEA}).
It is straightforward to see now, that adding one more rescattering of the proton 4 on the right side of the diagram of Fig.~\ref{fig:p3p4p3} will
bring a factor of $-\Gamma_4(\tilde{\tilde b})/2$ in the integrand of Eq.(\ref{M^(343)_GEA}) etc. In other words, all odd-numbered
soft longitudinal momentum transfers (from proton 3) can be summed and all even-numbered ones (from proton 4) can be summed too.
The DWF depends on these two sums only, and not on other soft partial longitudinal momentum transfers. Therefore, integrations over all but first two (counting from left to right in Fig.~\ref{fig:p3p4p3})
soft longitudinal momentum transfers will produce factors of 1/2 instead of $\Theta$-functions of relative longitudinal coordinates.
As a consequence, every new rescattering of the proton 3 or 4 brings, respectively, a factor of $-\Gamma_3(\tilde b)/2$ or $-\Gamma_4(\tilde{\tilde b})/2$
in Eq.(\ref{M^(g)_GEA}).
Hence, summing up all ladder
\footnote{The ladder diagrams are not exhausting the full set of possible diagrams, since the cross-diagrams are not included.
For example, one can consider the diagram obtained from Fig.~\ref{fig:p3p4p3} by exchanging the
upper rescattering vertices of the proton 3. However, such diagrams can not be treated as ISI/FSI and, thus, we will not include them.}
rescattering diagrams results in the following expression for the ``renormalized'' amplitude (g)
\begin{eqnarray}
  \lefteqn{\tilde M^{(g)} \equiv M^{(34)}+M^{(343)}+M^{(3434)}+M^{(34343)}+ \ldots} \nonumber \\
  &=& 2m^{1/2} M_{\rm hard}(p_3,p_4,p_1) \int d^3 r  \Theta(-\tilde z) \Theta(-\tilde{\tilde z}) \phi(\bvec{r})
 \mbox{e}^{i(\bvec{p}_s\bvec{r} -\Delta_3\tilde z -\Delta_4\tilde{\tilde z})}\, \Gamma_3(\tilde b)  \Gamma_4(\tilde{\tilde b}) \nonumber \\
 && \times  [1-\frac{1}{2}\Gamma_3(\tilde b)+\frac{1}{4}\Gamma_3(\tilde b)\Gamma_4(\tilde{\tilde b})-\frac{1}{8}\Gamma_3(\tilde b)\Gamma_4(\tilde{\tilde b})\Gamma_3(\tilde b) + \ldots ]
 \nonumber \\
 &=& 2m^{1/2} M_{\rm hard}(p_3,p_4,p_1) \int d^3 r \Theta(-\tilde z) \Theta(-\tilde{\tilde z})  \phi(\bvec{r})
 \mbox{e}^{i(\bvec{p}_s\bvec{r} -\Delta_3\tilde z -\Delta_4\tilde{\tilde z})}\, \Gamma_3(\tilde b)  \Gamma_4(\tilde{\tilde b}) {\cal F}~,  \label{M^(g)_GEA_ren}
\end{eqnarray}
where
\begin{equation}
  {\cal F} \equiv  \frac{1-\Gamma_3(\tilde b)/2}{1-\Gamma_3(\tilde b)  \Gamma_4(\tilde{\tilde b})/4} \label{renFac}
\end{equation}
is the renormalization factor.
In a similar way, the renormalized triple rescattering amplitude (i) is obtained by 
multiplying the integrand of Eq.(\ref{M^(i)_GEA}) by the factor ${\cal F}$ of Eq.(\ref{renFac}).
(In the case of the amplitudes (h) and (j) the replacement $\Gamma_3(\tilde b) \leftrightarrow \Gamma_4(\tilde{\tilde b})$ should be performed.)
The convergence condition of the two geometrical series contributing the factor in square brackets in Eq.(\ref{M^(g)_GEA_ren}) is $|\frac{1}{4}\Gamma_3(\tilde b)\Gamma_4(\tilde{\tilde b})|<1$.
We checked in numerical calculations that this condition is well satisfied.
 
In the high-energy limit, when all particles propagate almost parallel to the beam direction $\bvec{e}_z$, $\Theta$-functions in Eq.(\ref{M^(e)_GEA})
become mutually excluding and, thus, $M^{(e)}=M^{(f)}=0$ in agreement with Ref.~\cite{Frankfurt:1996uz}.
Same is of course true for the amplitudes with triple rescattering, i.e. $M^{(i)}=M^{(j)}=0$, as follows from Eq.(\ref{M^(i)_GEA}).
However, the amplitude (\ref{M^(g)_GEA}) does not disappear at high energies as the two $\Theta$-functions simply merge in a single one
(see Eq.(32) in Ref.~\cite{Larionov:2022gvn}).
Thus, this amplitude is expected to be the leading rescattering amplitude, in addition to the single rescattering amplitudes.

At moderate energies, when the outgoing protons have some finite transverse momenta,
there is a certain region in $\bvec{r}$-space where the products of $\Theta$-functions do not disappear and the
amplitudes (\ref{M^(e)_GEA}),(\ref{M^(i)_GEA}) may become finite.
However, in the numerical analysis we did not find a significant contribution of the triple rescattering amplitude (\ref{M^(i)_GEA})
in all kinematics, except for the cases when one of the outgoing protons moves backward in the deuteron rest frame, which is possible only
for a fast forward moving spectator ($\alpha_s \ltsim 0.5$ see Eq.(\ref{alpha_s}) below). Since such extreme kinematics requires relativistic corrections
to the DWF, we will not discuss it.

\section{Model inputs}
\label{inputs}

The rescattering amplitudes discussed in the previous section depend on the following three inputs.

(i) The DWF that determines the deuteron virtual decay vertex, Eq.(\ref{Gamma_d}), is taken in the Paris potential model \cite{Lacombe:1981eg}.
  
(ii) The hard elastic $pp$ scattering amplitude is represented according to Ref.~\cite{Ralston:1988rb} as
  a sum of the quark counting (QC) and Landshoff (L) parts
  \begin{equation}
    M_{\rm hard} = M_{\rm QC} + M_{\rm L}~.     \label{M_hard}
  \end{equation}
  The QC part selects small-size quark configurations in the initial
  and final state. In the simplest approximation, this part of the hard amplitude can be expressed as
 \begin{equation}
     M_{\rm QC} = \left(16\pi(s-4m^2)s \frac{d\sigma_{pp}^{\rm QC}}{dt}\right)^{1/2}~,      \label{M_QC}
 \end{equation}
 where
\begin{equation}
  \frac{d\sigma_{pp}^{\rm QC}}{dt} = 45 \frac{\mu{\rm b}}{{\rm GeV}^2} \left(\frac{10\,{\rm GeV}^2}{s}\right)^{10}
                                  \left(\frac{4m^2-s}{2t}\right)^{4\gamma}~,                     \label{dsigma_pp^QC/dt_par}
\end{equation}
with $\gamma=1.6$, is the corresponding part of the phenomenological parameterization of the large-angle elastic $pp$ cross section from Ref.~\cite{Frankfurt:1994nw}.
\footnote{The spin dependent phase multiplier consistent with antisymmetry of the hard amplitude with respect to the interchange of the quantum numbers between initial
or final protons is implicit in Eq.(\ref{M_QC}). Provided that the non-central part of the soft elastic amplitude $M_{\rm el}$ is neglected, the concrete form of the
multiplier does not influence results of the present work for the unpolarized incoming proton beam.}

The Landshoff part is dominated by large-size configurations. Following Refs.~\cite{Ralston:1988rb,Frankfurt:1994nw}, this part of the amplitude
can be written as
\begin{equation}
  M_{\rm L} = M_{\rm QC} \frac{\rho_1\sqrt{s}}{2} \mbox{e}^{\pm i(\phi(s)+\delta_1)}~, \label{M_L}
\end{equation}
with $\rho_1=0.08$ GeV$^{-1}$, $\delta_1=-2$ and
\begin{equation}
  \phi(s) = \frac{\pi}{0.06} \log\left[\log\left(\frac{s}{0.01~\mbox{GeV}^2}\right)\right]~.     \label{phi}
\end{equation}
The separation of small- and large size configurations in Eq.(\ref{M_hard}) implies that
only a part of every rescattering amplitude proportional to $M_{\rm QC}$ is influenced by CT which leads to the oscillations
of the nuclear transparency ratio with incident momentum (see Refs.~\cite{Ralston:1988rb,VanOvermeire:2006tk,Frankfurt:1996uz,Larionov:2022gvn} for detailed discussion).
In this work, the effects of CT are not taken into account. Thus, the entire hard amplitude factorizes out from the rescattering amplitudes
and cancels out in the nuclear transparency ratio (see Eq.~(\ref{T}) below).

(iii) The soft elastic $pn$ scattering amplitude, $M_{\rm el}(p_{\rm lab},k_t)$, is parameterized according to Eq.(\ref{M_el}).  
  Spin dependence as well as spin-flip transitions are neglected. Thus, Eq.(\ref{M_el}) describes the central part of the
  $pn$ amplitude only. 
  The total $pn$ cross section, $\sigma_{p n}^{\rm tot}$, at $p_{\rm lab} < 3.5$ GeV/c is taken
  in the parameterization of Ref.~\cite{Cugnon:1996kh} while at higher $p_{\rm lab}$ the Regge-Gribov fit
  from Ref.~\cite{Patrignani:2016xqp} is used.
  The ratio of the real and imaginary parts of the forward $pn$ scattering amplitude, $\rho_{\rm p n}$,
  in agreement with experimental data \cite{ParticleDataGroup:2022pth},
  at  $p_{\rm lab} < 4.33$ GeV/c is set equal to -0.5 while at higher $p_{\rm lab}$ the Regge-Gribov fit
  from Ref.~\cite{Patrignani:2016xqp} is used.
  The slope parameter, $B_{\rm p n}$, at $p_{\rm lab} < 9.1$ GeV/c is taken from Ref.~\cite{Cugnon:1996kh} while at higher $p_{\rm lab}$
  the PYTHIA parameterization for elastic $pp$  scattering from Ref.~\cite{Falter:2004uc} is used:
  \begin{equation}
     B_{\rm p n} =  [5.0 + 4(s/{\rm GeV}^2)^{0.0808}] {\rm GeV}^{-2}~.             \label{B_pn}
  \end{equation}
    
\section{Numerical analysis}
\label{numerics}

To evaluate the effects of soft rescattering, we will calculate the nuclear transparency ratio which is defined as
\begin{equation}
    T = \frac{\overline{|M^{(a)}+M^{(b)}+M^{(c)}+M^{(d)}+M^{(e)}+M^{(f)}+M^{(g)}+M^{(h)}|^2}}{\overline{|M^{(a)}|^2}}~,       \label{T}
\end{equation}
where the numerator is the modulus squared of the sum of the IA (a), single rescattering (b),(c),(d), and double rescattering (e),(f),(g),(h) amplitudes of Fig.~\ref{fig:diagr},
and the denominator is the modulus squared of the IA amplitude. The overline means averaging over spin projections of incoming particles
and summation over spin projections of the outgoing particles.
We checked that including the triple rescattering amplitudes (i) and (j) in the numerator of Eq.(\ref{T}) results in less than $0.2\%$ change
of the transparency ratio in the considered kinematic ranges.
Thus, only the contributions of the single and double soft rescattering amplitudes will be discussed below.
The GEA calculation that includes the partial amplitudes (a)-(h) and takes into account the renormalization of the amplitudes (g),(h) will be called a 'full calculation'.

The partial amplitudes were calculated in the deuteron rest frame with the $z$-axis along the proton momentum.
The kinematic variables characterizing the particles in the final state are as follows:
\begin{equation}
  t=(p_1-p_3)^2      \label{t}
\end{equation}
-- the square of four-momentum transfer between the incoming proton and one of the two scattered protons;
\begin{equation}
  \phi=\phi_3-\phi_s   \label{phi}
\end{equation}
-- the azimuthal angle between the transverse momentum of the scattered proton for which $t$ is measured
and that of the spectator neutron; 
\begin{equation}
  \alpha_s = \frac{2(E_s-p_s^z)}{m_d}                 \label{alpha_s}
\end{equation}
-- the light cone variable defined such that, in the infinite momentum frame with the deuteron moving fast backward,
$\alpha_s/2$ is the fraction of the deuteron momentum carried by the spectator neutron;
$p_{st}$ -- the transverse momentum of the neutron. Since, in the absence of particle polarizations, the rotational
symmetry about $z$ axis is satisfied, four variables fully characterize the event.

In order to see the influence of the different treatments of the multiple rescattering effects, the $\phi$-dependence
of the transparency ratio $T$ was studied at several fixed values of $t$, $\alpha_s$ and $p_{st}$, for the proton beam momentum
$p_{\rm lab}=4$ and 15 GeV/c.

\begin{figure}
  \begin{center} 
    \begin{tabular}{ccc}
   \vspace{-1cm}   
   \includegraphics[scale = 0.3]{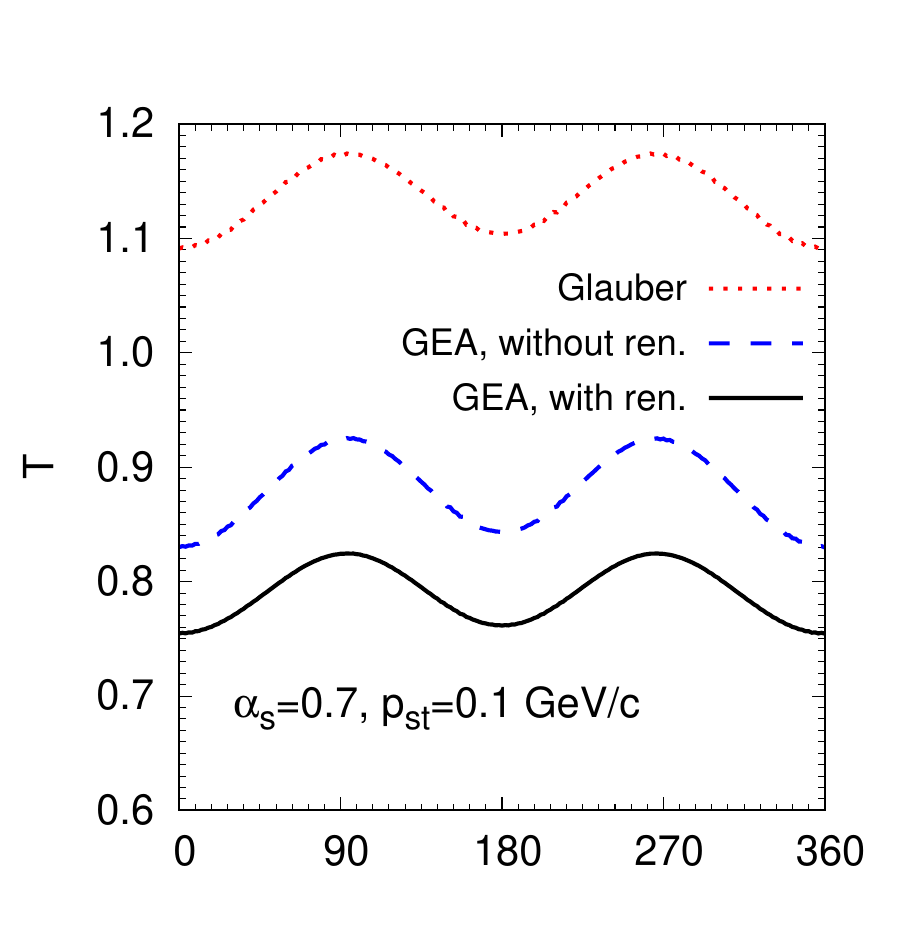} &
   \includegraphics[scale = 0.3]{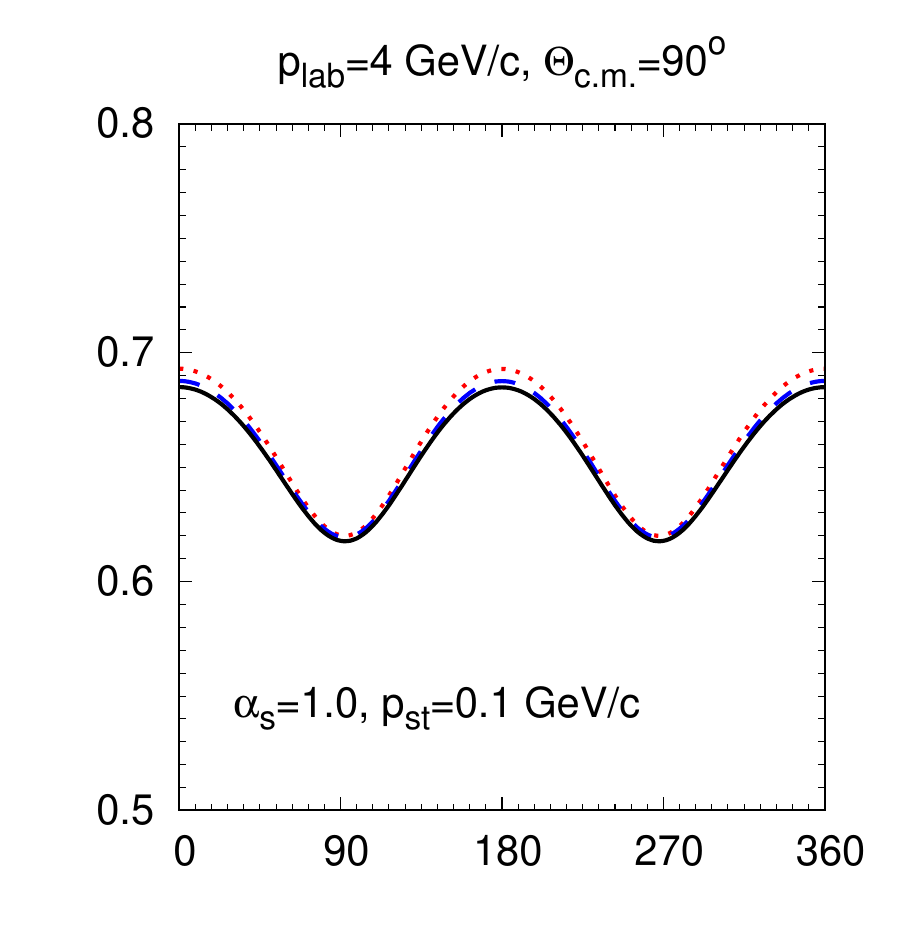} &
   \includegraphics[scale = 0.3]{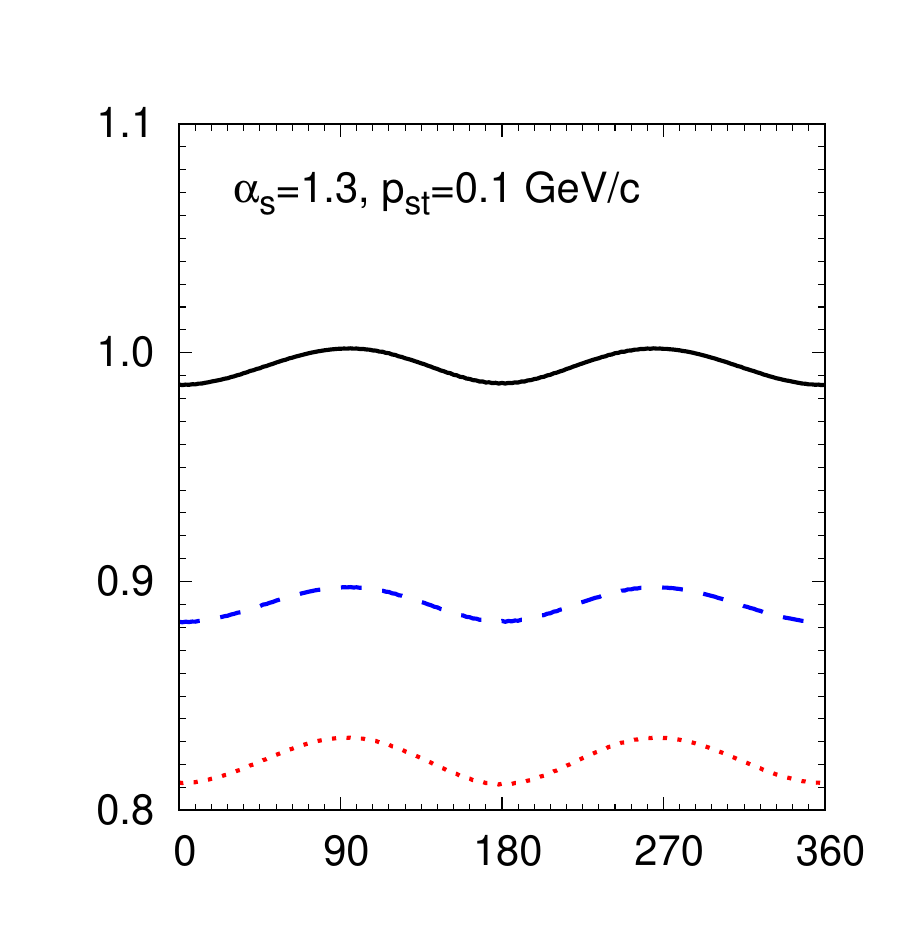} \\
   \vspace{-1cm}
   \includegraphics[scale = 0.3]{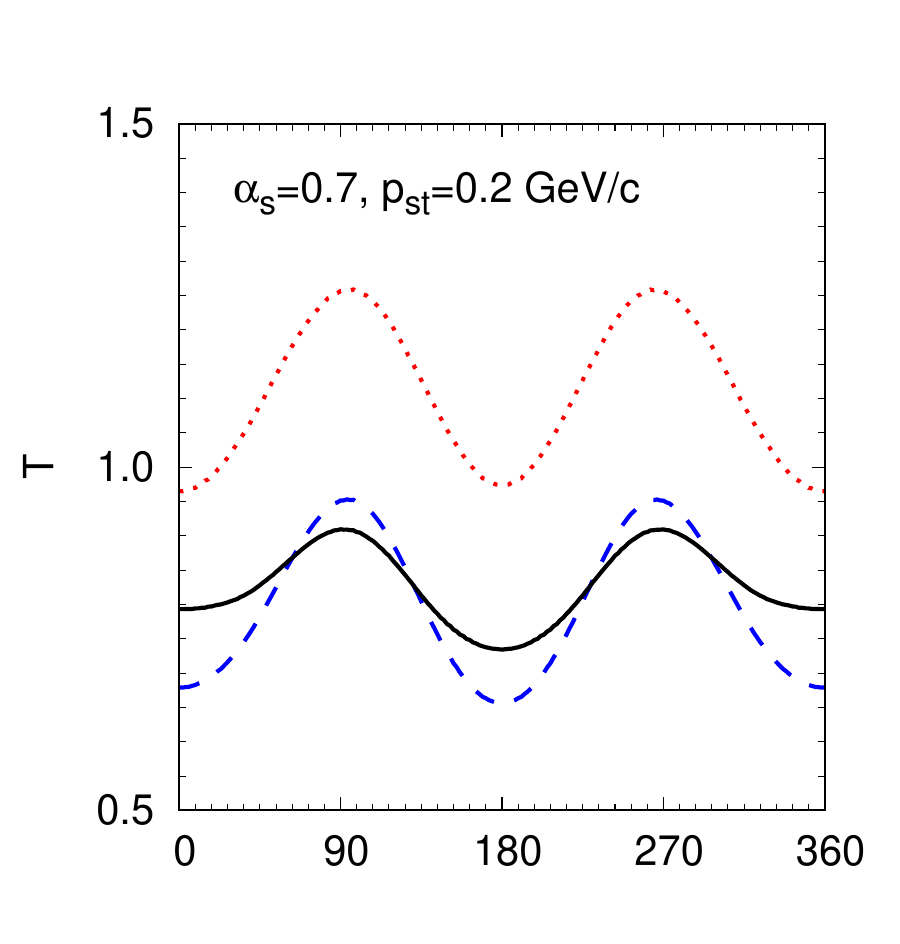} &
   \includegraphics[scale = 0.3]{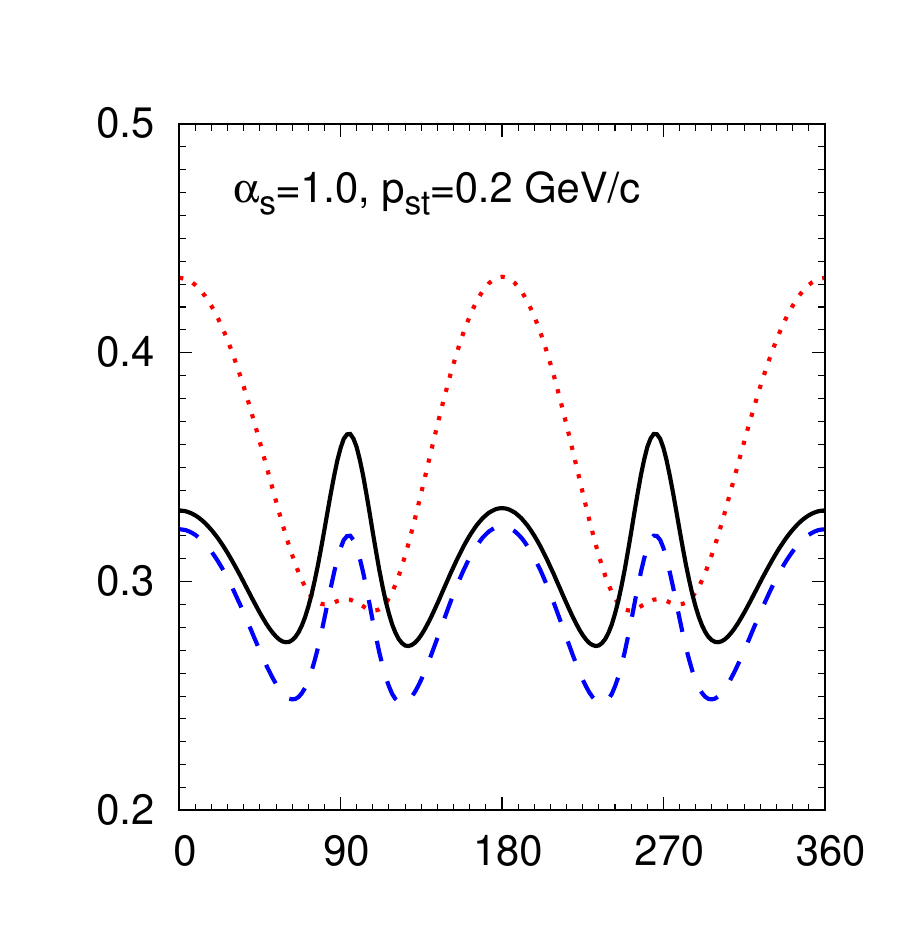} &
   \includegraphics[scale = 0.3]{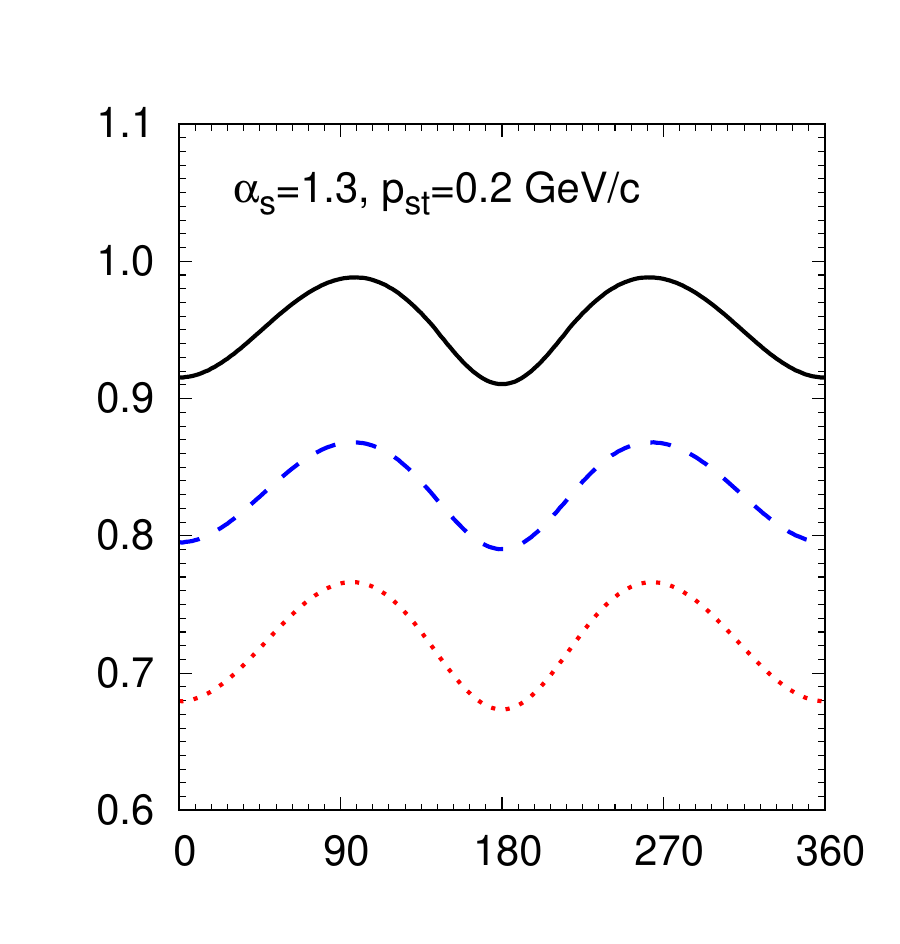} \\
   \vspace{-1cm}
   \includegraphics[scale = 0.3]{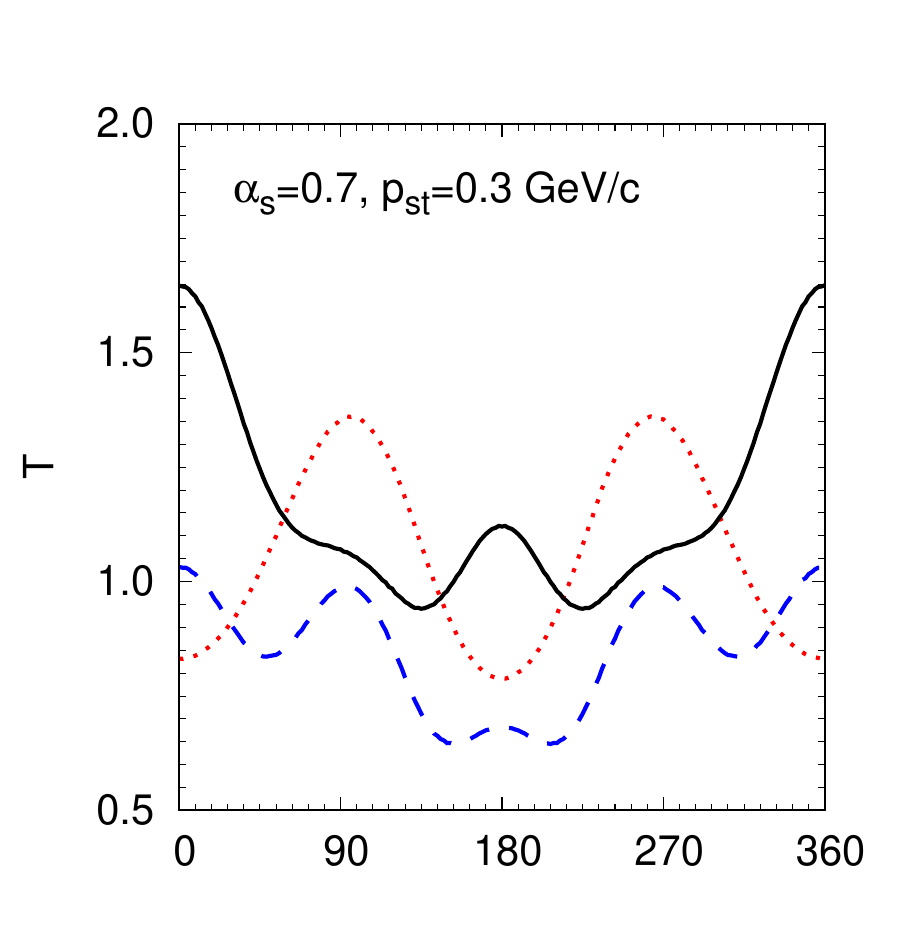} &
   \includegraphics[scale = 0.3]{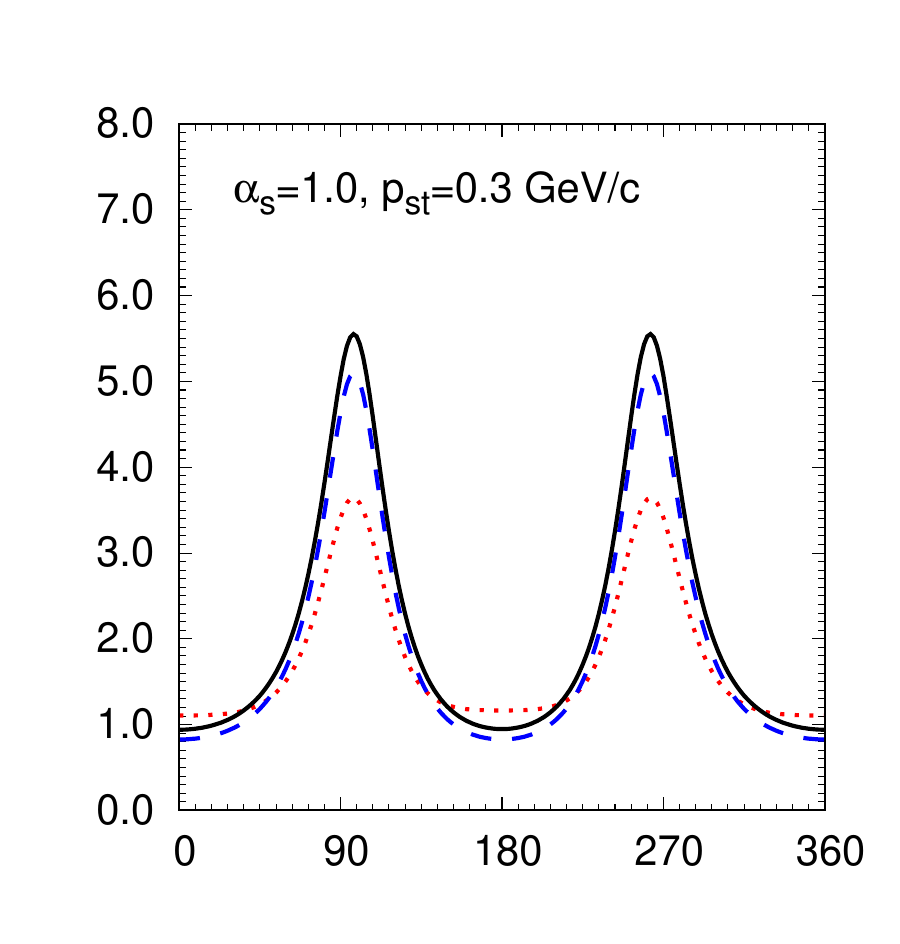} &
   \includegraphics[scale = 0.3]{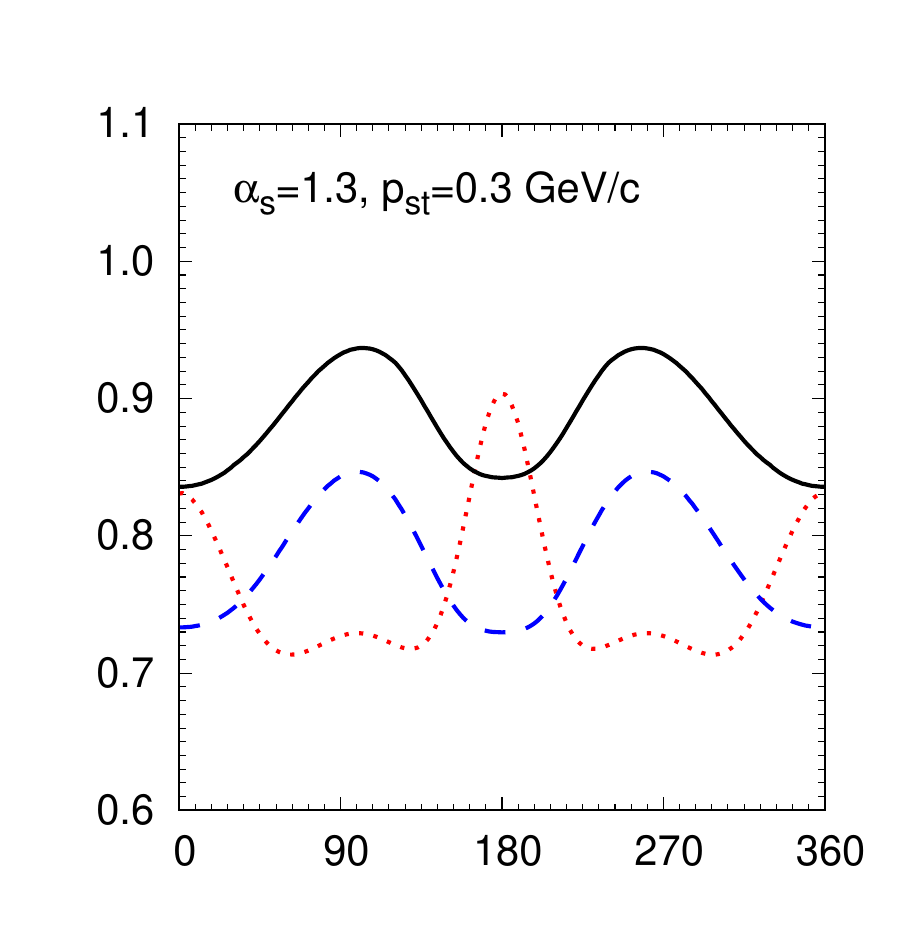} \\
   \includegraphics[scale = 0.3]{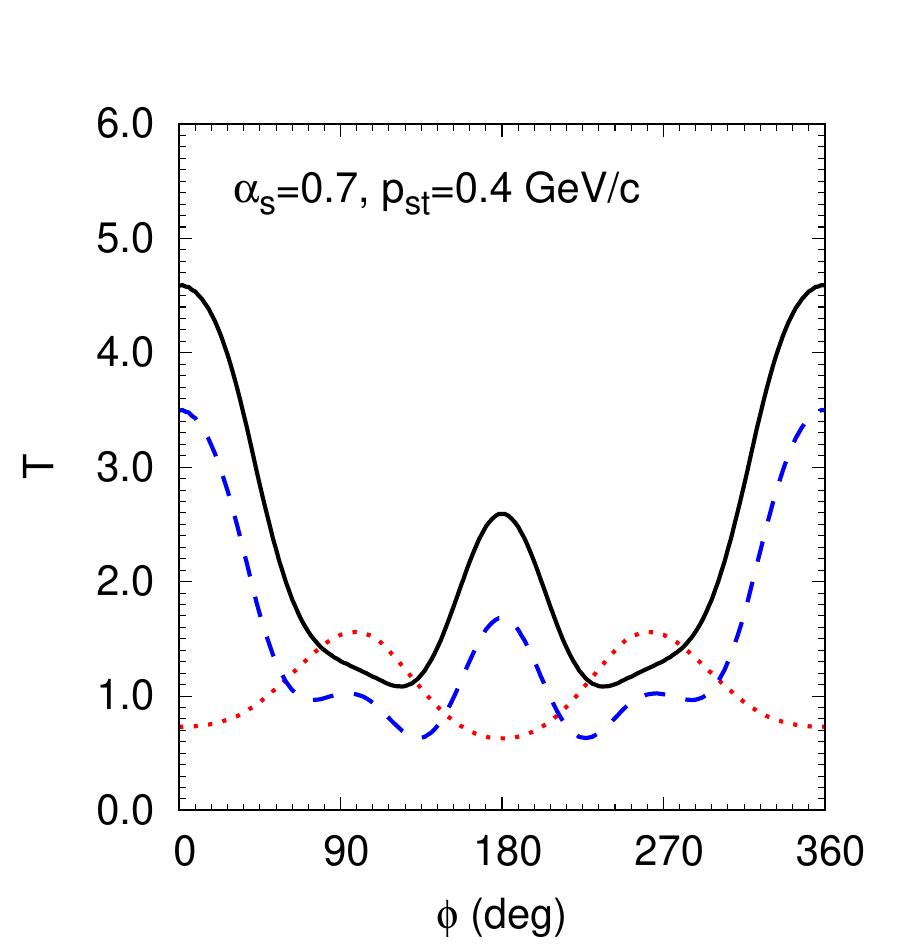} &
   \includegraphics[scale = 0.3]{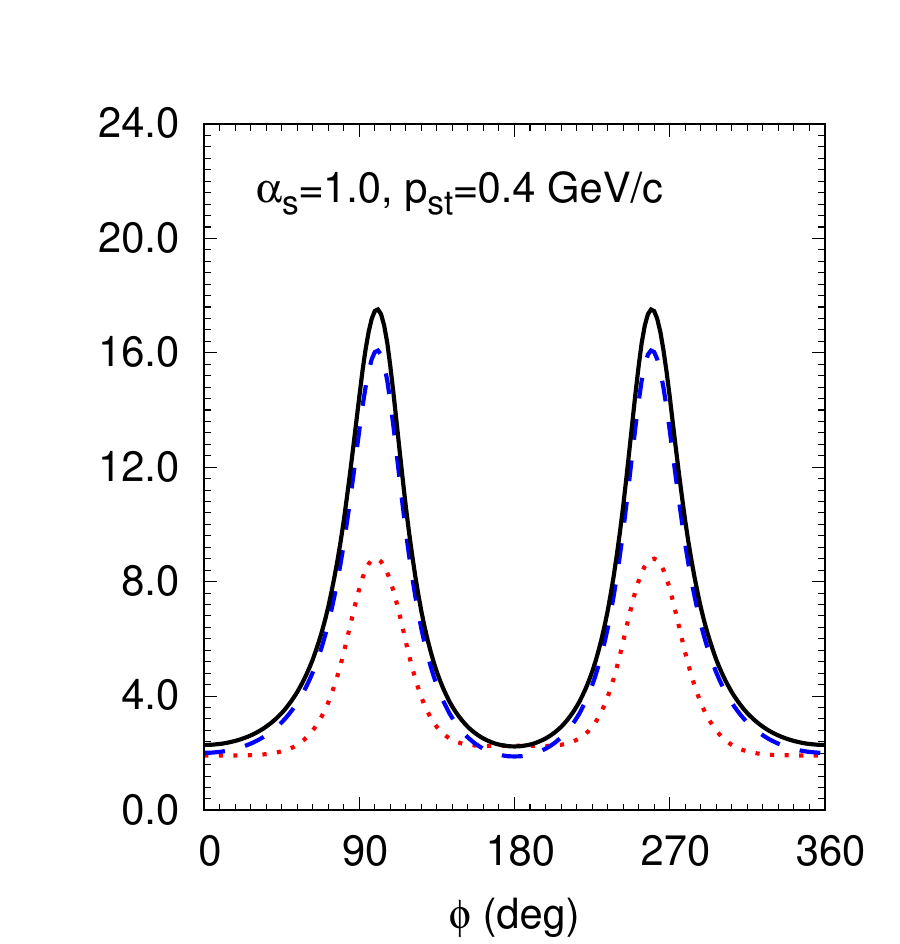} &
   \includegraphics[scale = 0.3]{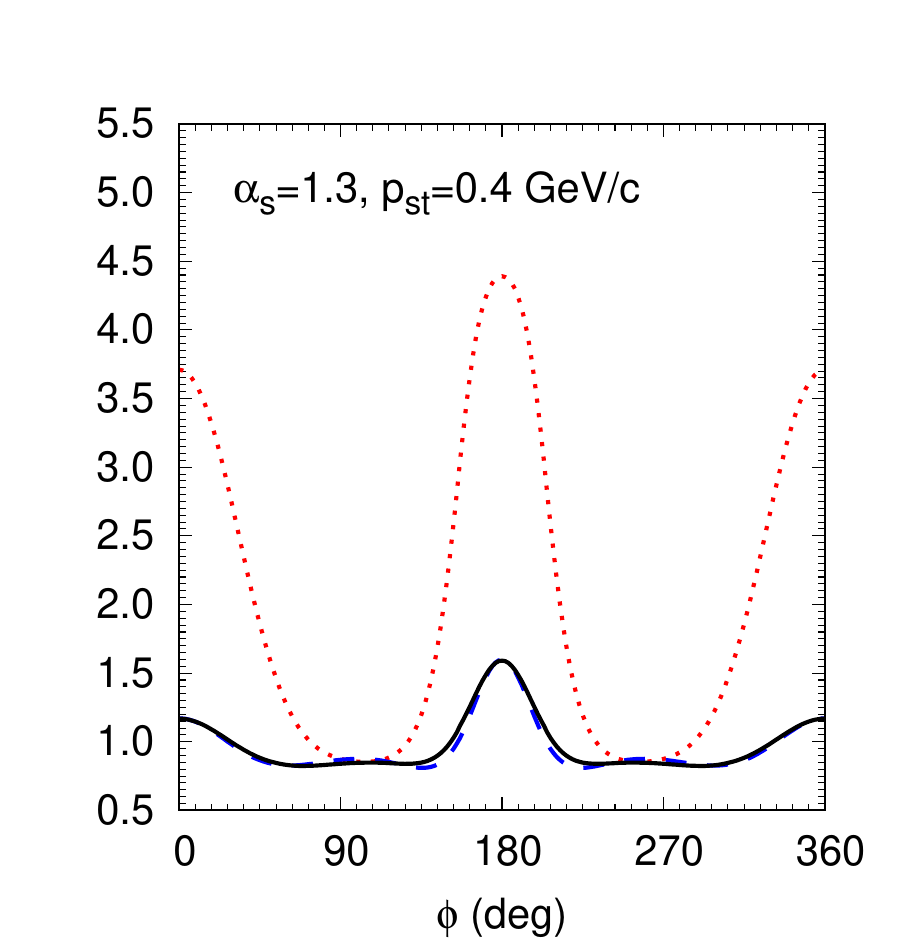} \\
   \end{tabular}
 \end{center}   
  \caption{\label{fig:T_4gevc_t_fix_tst} The azimuthal angle dependence of the transparency ratio $T$ for $pd \to ppn$ at $p_{\rm lab}=4$ GeV/c
    and $\Theta_{\rm c.m.}=90\degree$.
    Different panels correspond to different values of $\alpha_s$ and $p_{st}$ as indicated.
    Shown are the results of calculations with amplitudes (a)-(h) of Fig.~\ref{fig:diagr}.
    Dotted (red) lines -- Glauber model.
    Dashed (blue) lines -- GEA without renormalization.
    Solid (black) lines -- GEA with renormalization of the amplitudes (g) and (h) of the double rescattering
    of the outgoing protons (full calculation).
  }
\end{figure}

It is instructive, first, to see how the results of the GEA are compared to those of the pure Glauber model.
Fig.~\ref{fig:T_4gevc_t_fix_tst} shows the transparency ratio at $p_{\rm lab}=4$ GeV/c and $t=(4m^2-s)/2$, where $s=(p_3+p_4)^2$, that corresponds to
the $pp$ c.m. scattering angle $\Theta_{\rm c.m.}=90\degree$.
In the case of transverse spectator kinematics
($\alpha_s=1$), the transition from the absorptive regime at small $p_{st}$ to the rescattering regime at large $p_{st}$ is clearly visible, in-particular,
at $\phi=90\degree$ and $\phi=270\degree$, where the amplitudes with rescattering are enhanced (see discussion in Ref.~\cite{Larionov:2022gvn}).
The effect of the GEA corrections (i.e. finite $\Delta$'s) to the Glauber model is quite strong
in all kinematics except the one with the smallest recoil momentum of the spectator ($\alpha_s=1, p_{st}=0.1$ GeV/c)
and even qualitatively changes the azimuthal dependence in the kinematics with large recoil momentum of the spectator.
The effect of renormalization on the transparency ratio does not exceed $10-20\%$ in transverse ($\alpha_s=1$) and backward ($\alpha_s=1.3$) kinematics
and becomes larger, reaching $\sim 50\%$ in forward ($\alpha_s=0.7$) kinematics at large transverse momenta.

Below, {\it only} GEA results will be discussed. 
The calculations were performed with and without including the double rescattering amplitudes, i.e. those of Fig.~\ref{fig:diagr}e,f,g,h.
The renormalization factor of the double rescattering amplitudes, Eq.(\ref{renFac}), was taken into account. 
For comparison, the results with the older treatment of the double rescattering amplitudes, see Refs.~\cite{Frankfurt:1996uz,Larionov:2022gvn}
are also shown.

\begin{figure}
  \begin{center} 
    \begin{tabular}{ccc}
   \vspace{-1cm}   
   \includegraphics[scale = 0.3]{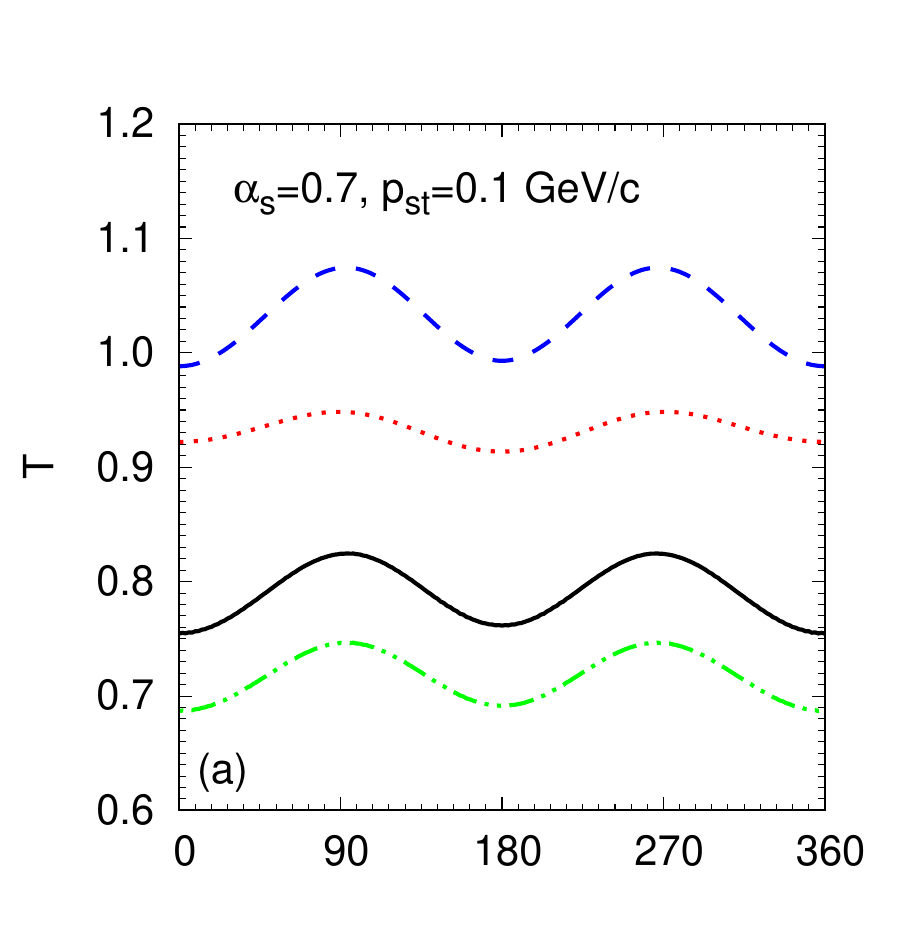} &
   \includegraphics[scale = 0.3]{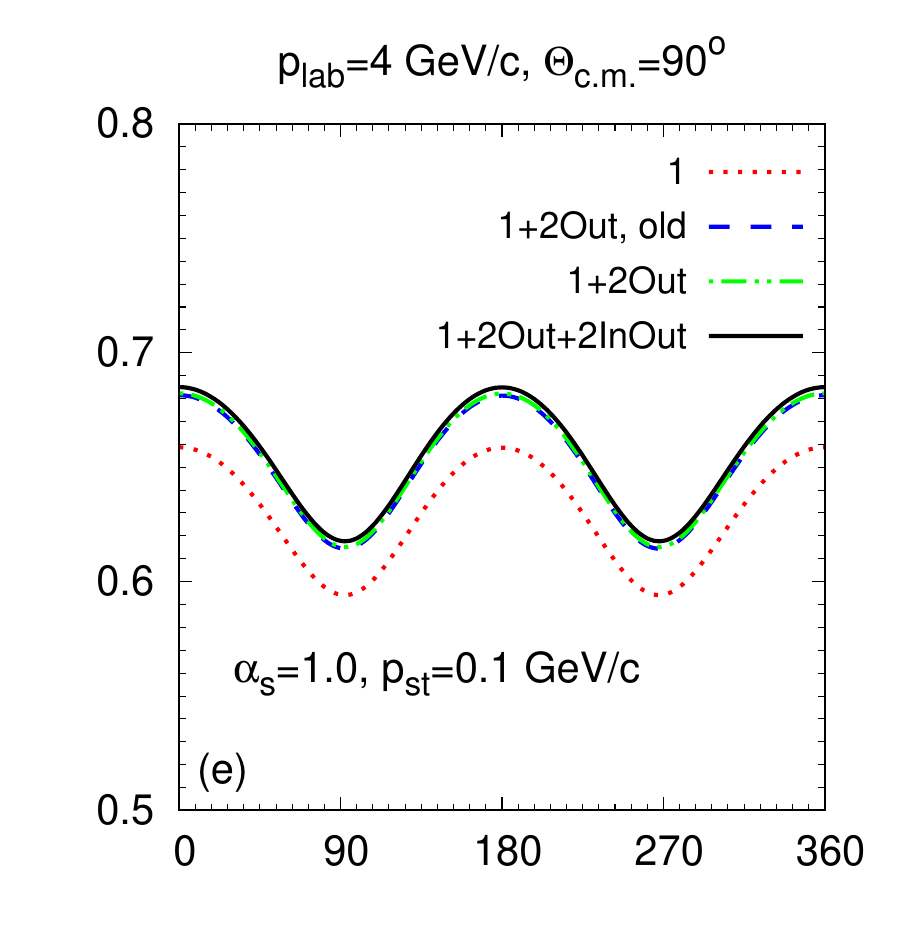} &
   \includegraphics[scale = 0.3]{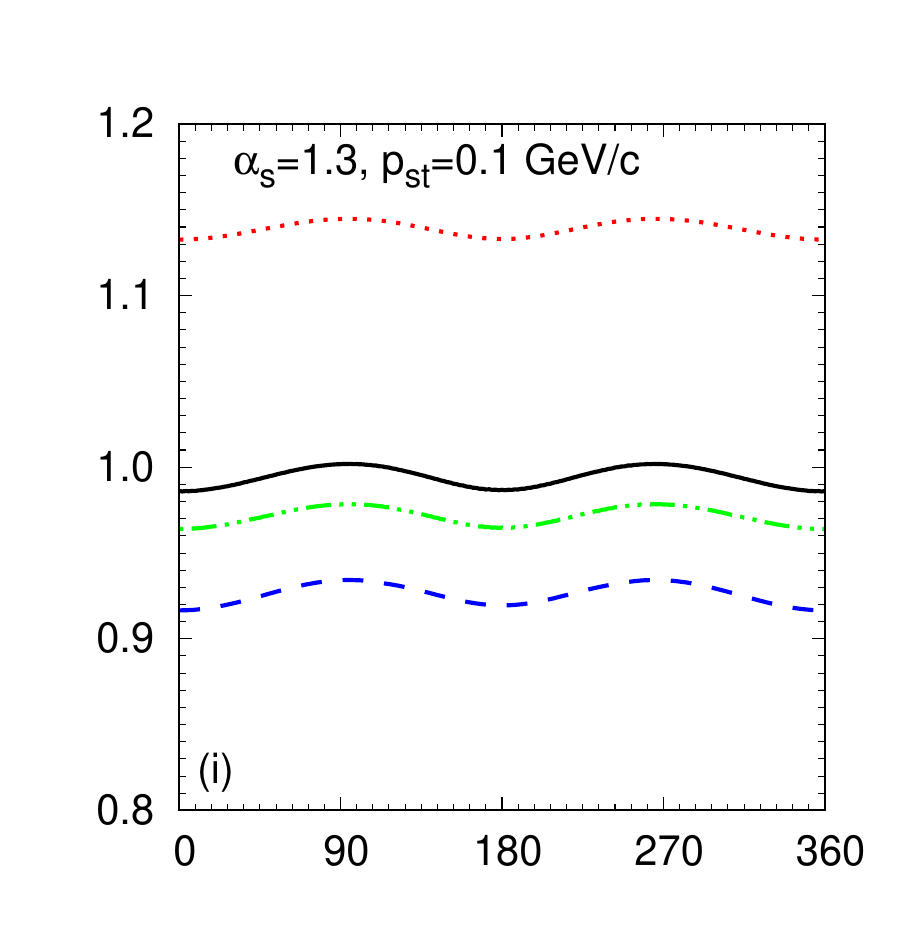} \\
   \vspace{-1cm}
   \includegraphics[scale = 0.3]{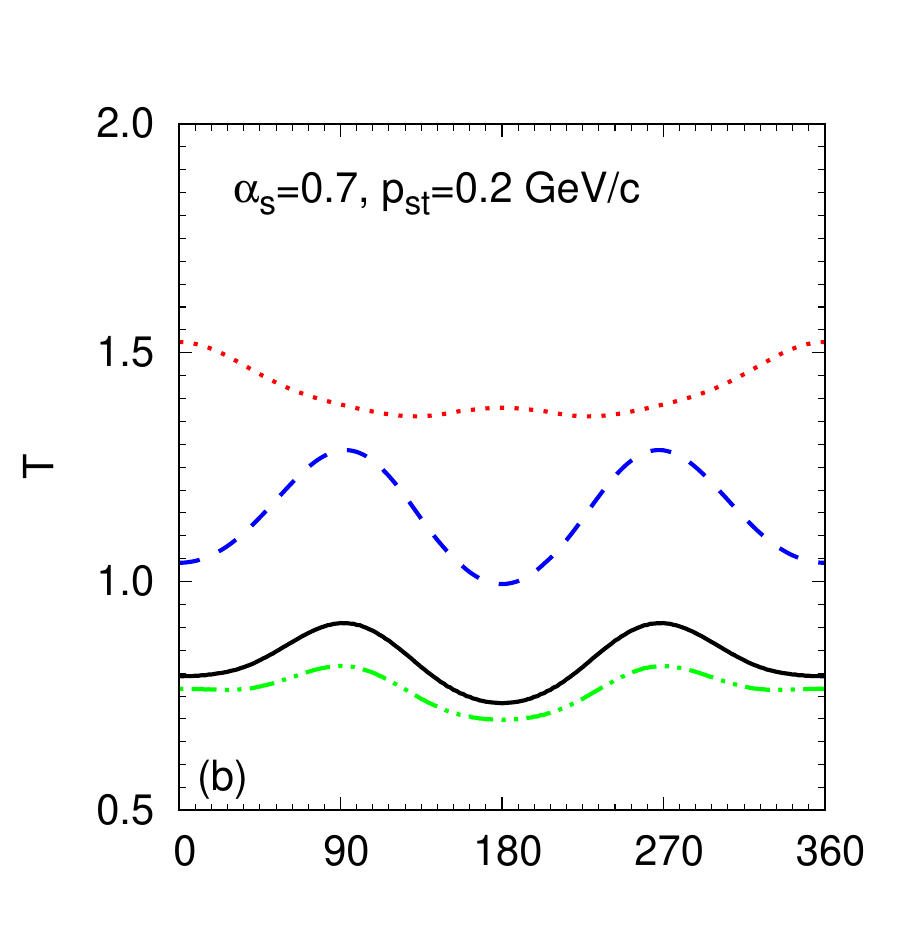} &
   \includegraphics[scale = 0.3]{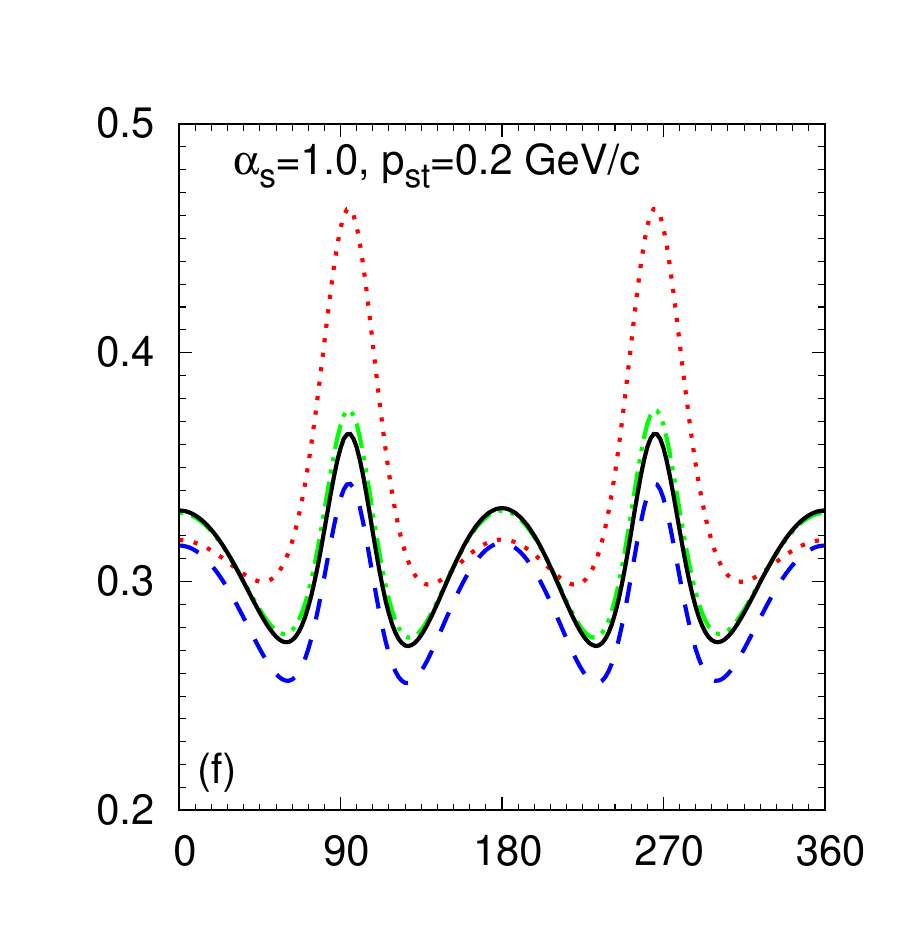} &
   \includegraphics[scale = 0.3]{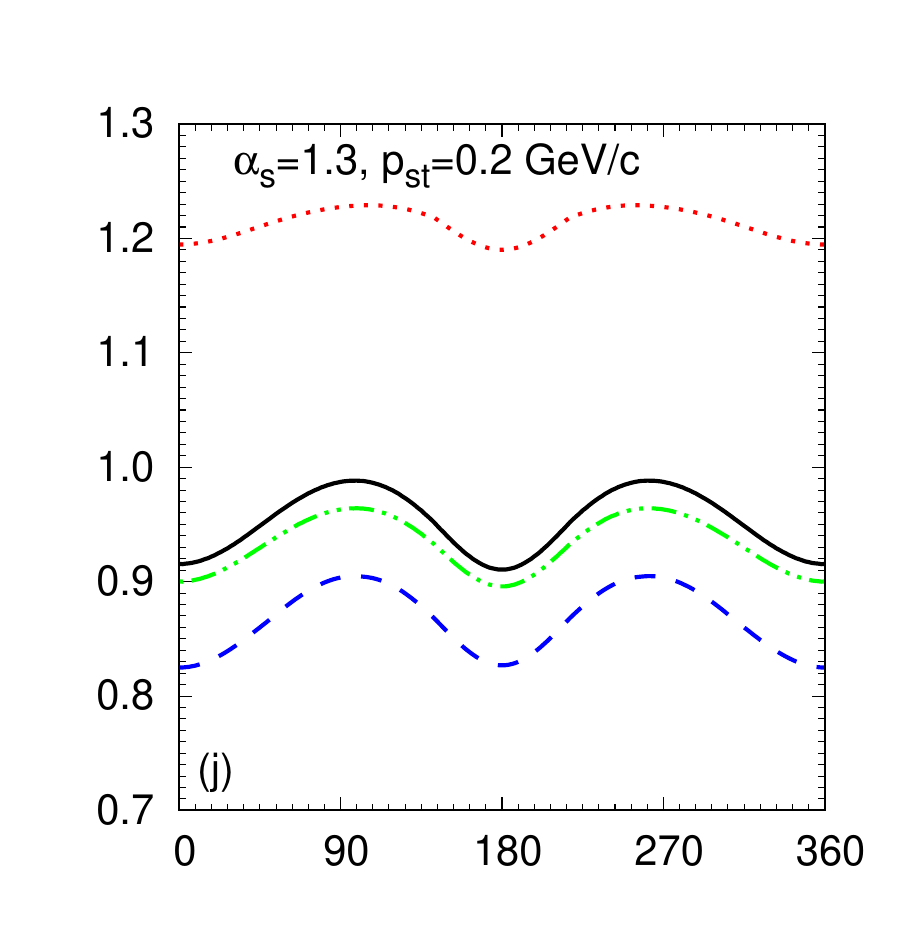} \\
   \vspace{-1cm}
   \includegraphics[scale = 0.3]{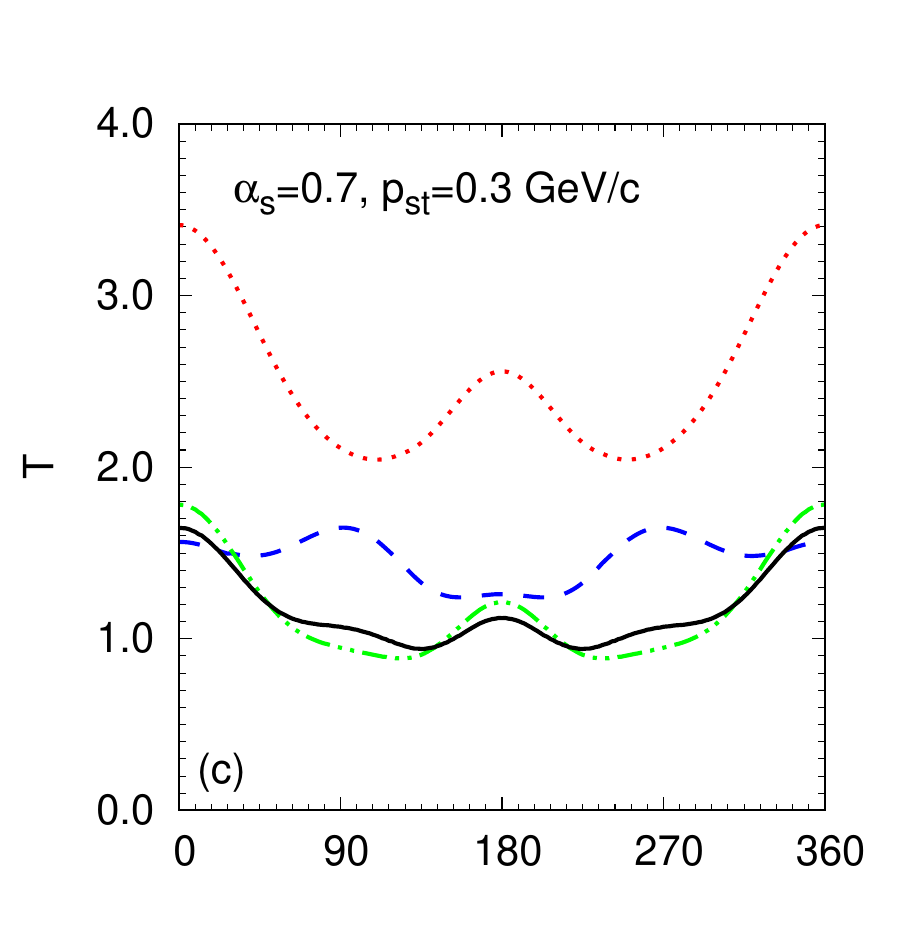} &
   \includegraphics[scale = 0.3]{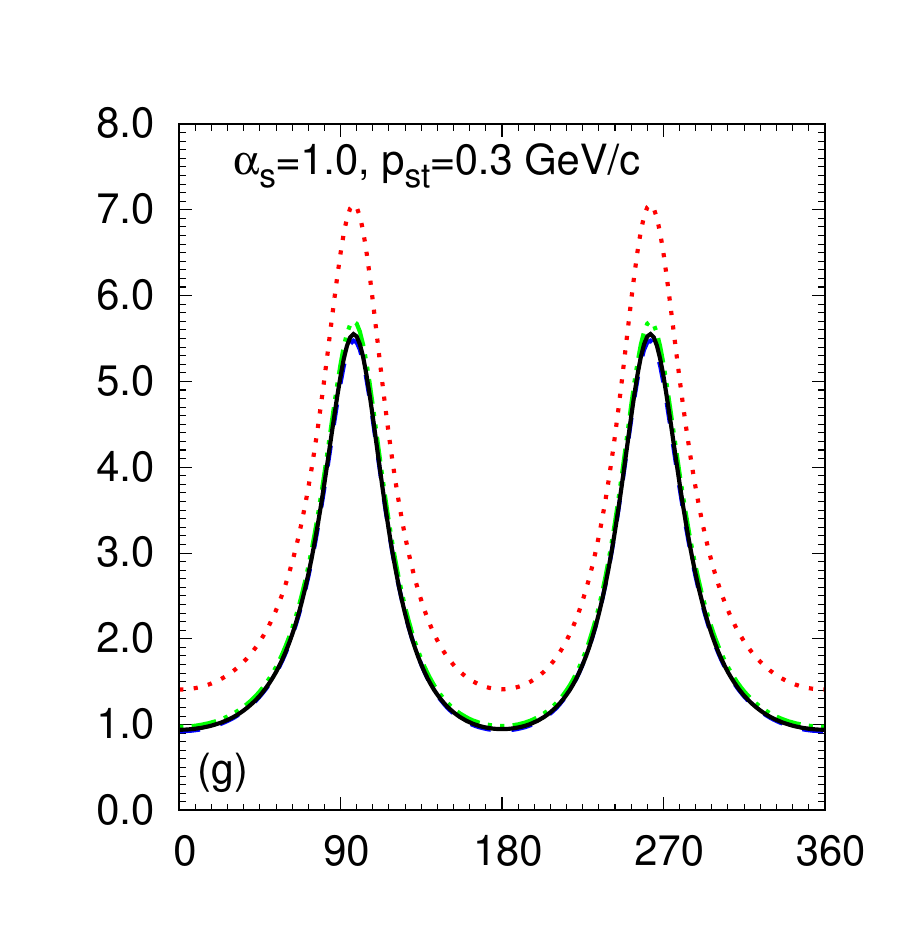} &
   \includegraphics[scale = 0.3]{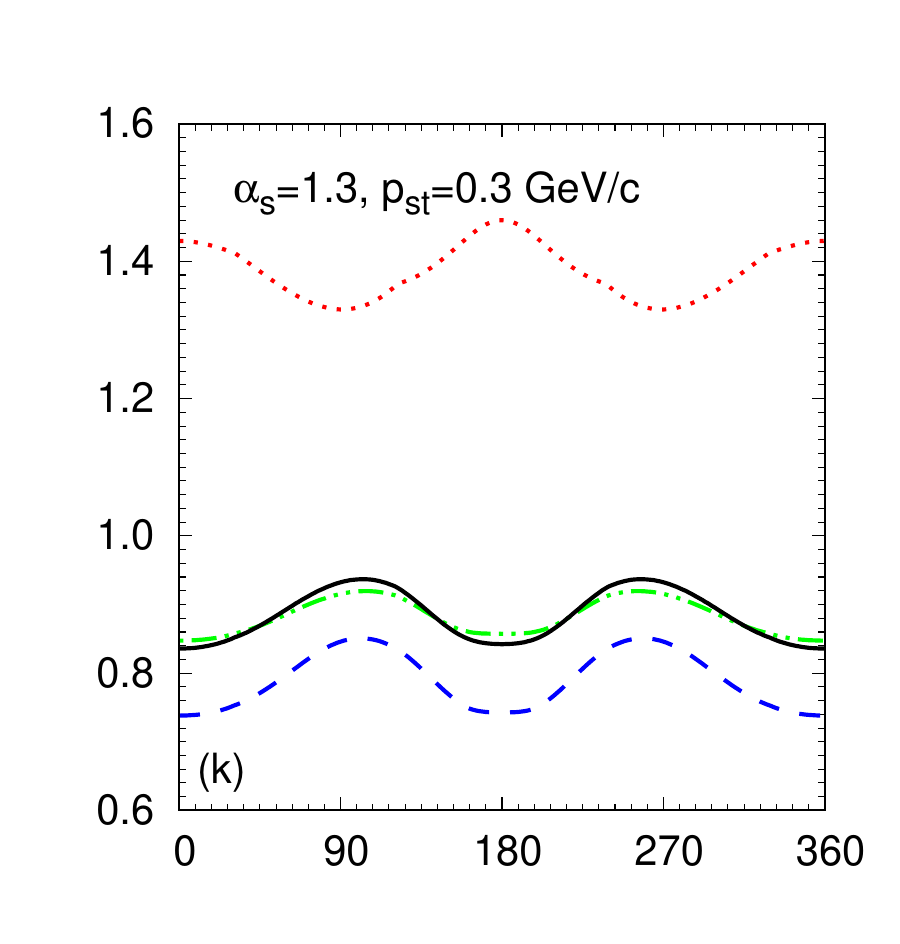} \\
   \includegraphics[scale = 0.3]{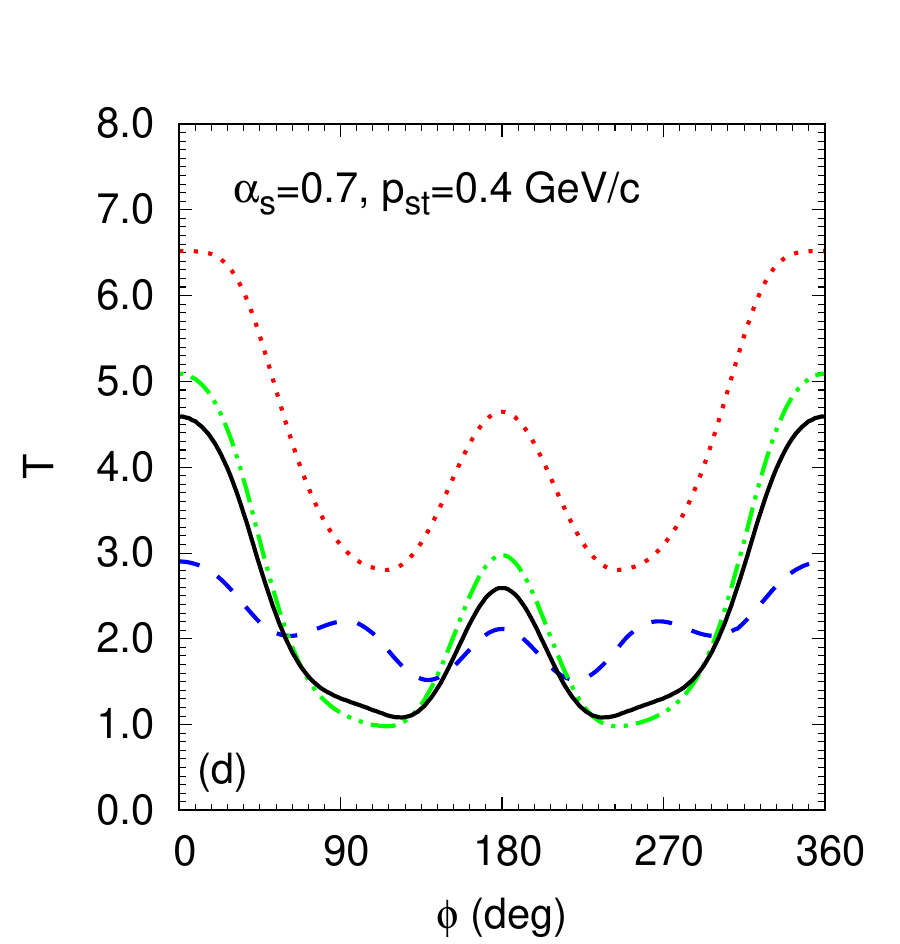} &
   \includegraphics[scale = 0.3]{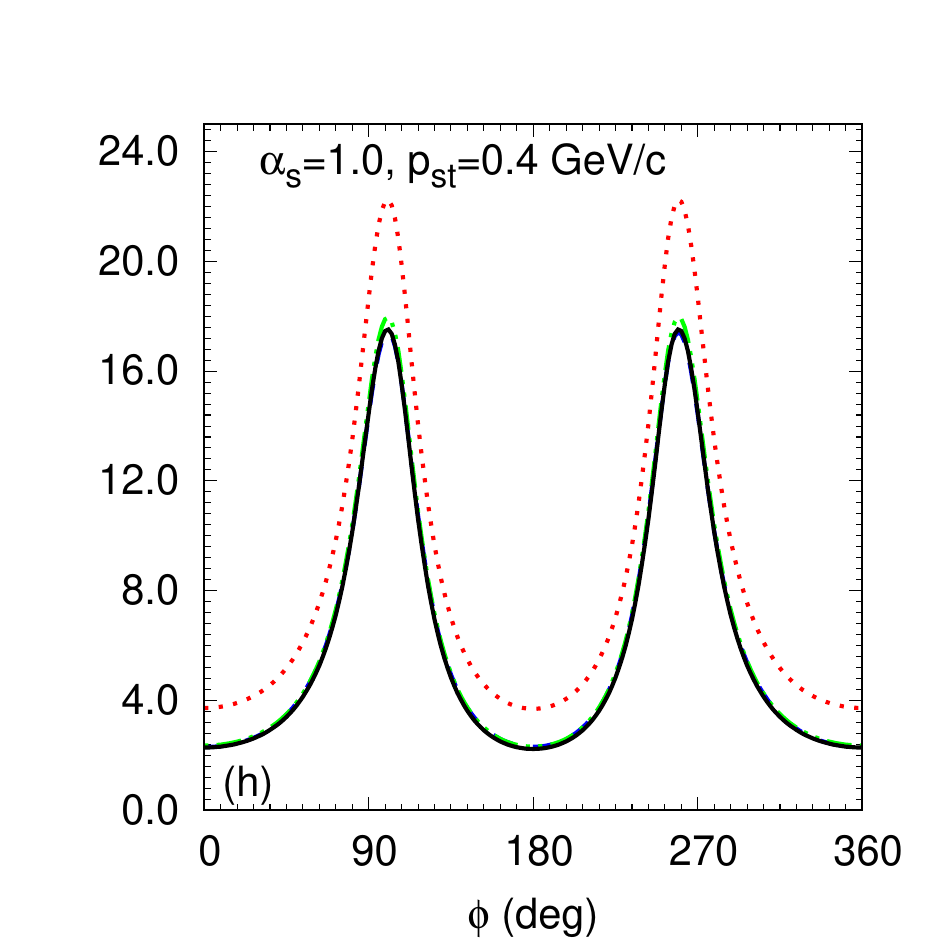} &
   \includegraphics[scale = 0.3]{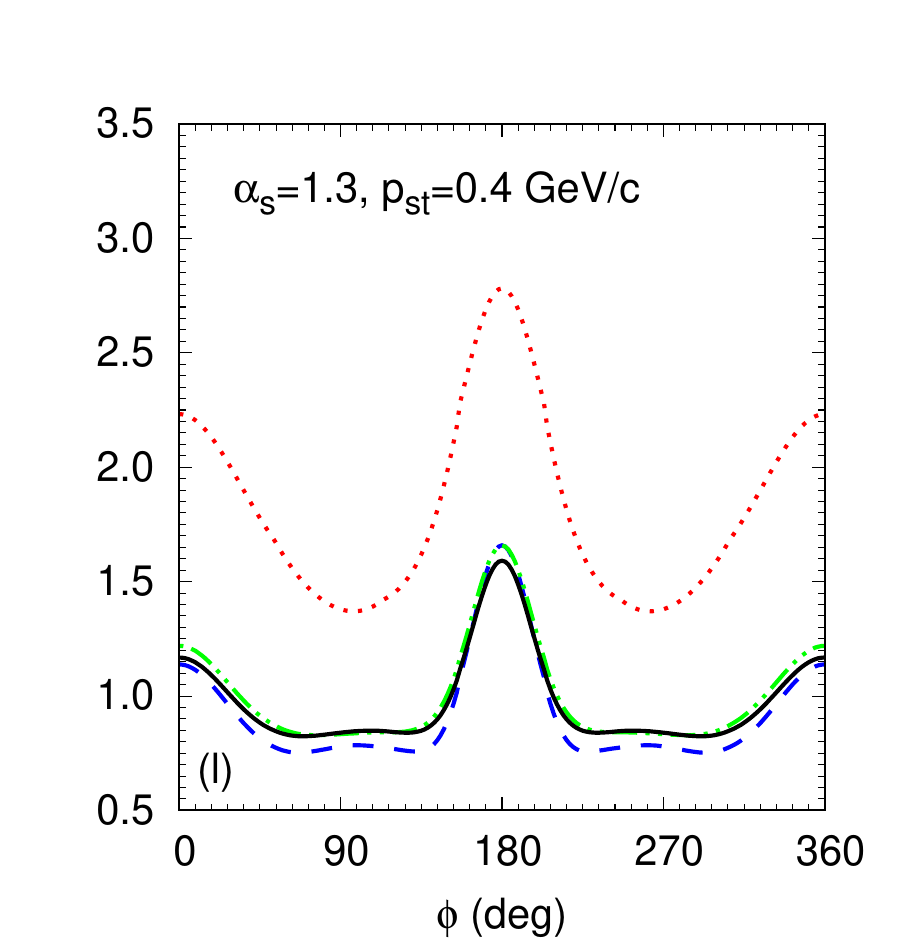} \\
   \end{tabular}
 \end{center}   
  \caption{\label{fig:T_4gevc_t_fix} The azimuthal angle dependence of the transparency ratio $T$ for $pd \to ppn$ at $p_{\rm lab}=4$ GeV/c
    and $\Theta_{\rm c.m.}=90\degree$.
    Different panels correspond to different values of $\alpha_s$ and $p_{st}$ as indicated.
    Shown are the results of GEA calculations for different sets of amplitudes from Fig.~\ref{fig:diagr}.
    Dotted (red) lines -- the IA amplitude (a) and the single rescattering amplitudes (b),(c), and (d).
    Dashed (blue) lines -- same but with the addition of the amplitudes (g) and (h) of the double rescattering
    of the outgoing protons calculated according to Ref.~\cite{Larionov:2022gvn}.
    Dash-double-dotted (green) lines -- same but with the renormalized amplitudes (g) and (h) of the present work.
    Solid (black) lines -- same but with the addition of the amplitudes  (e) and (f) of the double rescattering 
    of the incoming and outgoing protons (full calculation).
  }
\end{figure}

Fig.~\ref{fig:T_4gevc_t_fix} shows the transparency ratio at $p_{\rm lab}=4$ GeV/c and $\Theta_{\rm c.m.}=90\degree$.
In agreement with previous results of Refs.~\cite{Frankfurt:1996uz,Larionov:2022gvn}, for transverse spectator kinematics, the effect of double rescattering grows with
increasing transverse momentum of spectator. The different treatments of the double rescattering lead to practically identical results at $\alpha_s=1$
(except the case of $p_{st}=0.2$ GeV/c where some small differences are still visible).
In the case of forward ($\alpha_s=0.7$) and backward ($\alpha_s=1.3$) spectator kinematics, the effects of amplitudes with rescattering are overall weaker
compared to the case of transverse kinematics. However, some interesting new features appear which deserve to be discussed.

At $\alpha_s=0.7$, the struck proton longitudinal momentum is negative, i.e. $p_2^z < 0$, and, thus, the opening angle between momenta of the outgoing protons is
larger compared to the case of $\alpha_s=1$. This leads to significant difference between the different treatments of double rescattering.
The new formula, Eq.(\ref{M^(g)_GEA}), remains correct when the transverse and longitudinal momenta of particles become comparable.
At small transverse momenta of the spectator ($p_{st} \ltsim 0.3$ GeV/c), the new treatment of the double rescattering of the outgoing protons
gives smaller transparency ratio, while at $p_{st} = 0.4$ GeV/c -- stronger variation with $\phi$ compared to the old treatment (i.e. that according
to Eq.(32) of Ref.\cite{Larionov:2022gvn}). 
Including the double rescattering of the incoming and outgoing protons pushes $T$ slightly closer to the old results.  

At $\alpha_s=1.3$, the opening angle between momenta of the outgoing protons is smaller than at $\alpha_s=1$ which is favorable for the old treatment of the
double rescattering. Thus, the difference between results with various treatments of the double rescattering becomes small.

The choice of large enough c.m. scattering angle is needed for keeping the $pp$ elastic scattering in a 'hard region' ($|t| > 1~\mbox{GeV}^2$).
This is important for CT studies. However, according to Eq.(\ref{dsigma_pp^QC/dt_par}), the differential elastic $pp$ cross section
$d\sigma_{pp}^{\rm QC}/dt$ at fixed $\Theta_{\rm c.m.}$ drops with increasing collision energy as $s^{-10}$. Thus, the cross section at $t=(4m^2-s)/2$, i.e. at $\Theta_{\rm c.m.}=90\degree$,
becomes extremely small and difficult to study experimentally. Selecting smaller $|t|$ would strongly increase the cross section.
For example, at $t=0.4(4m^2-s)/2$, that corresponds to $\Theta_{\rm c.m.}=53\degree$, the $pp$ differential cross section is two orders of magnitude larger
than at $90\degree$. Even at $p_{\rm lab}$ as small as 4 GeV/c the $pp$ scattering at $53\degree$ is still hard ($|t|=1.18~\mbox{GeV}^2$).

\begin{figure}
  \begin{center} 
    \begin{tabular}{ccc}
   \vspace{-1cm}   
   \includegraphics[scale = 0.3]{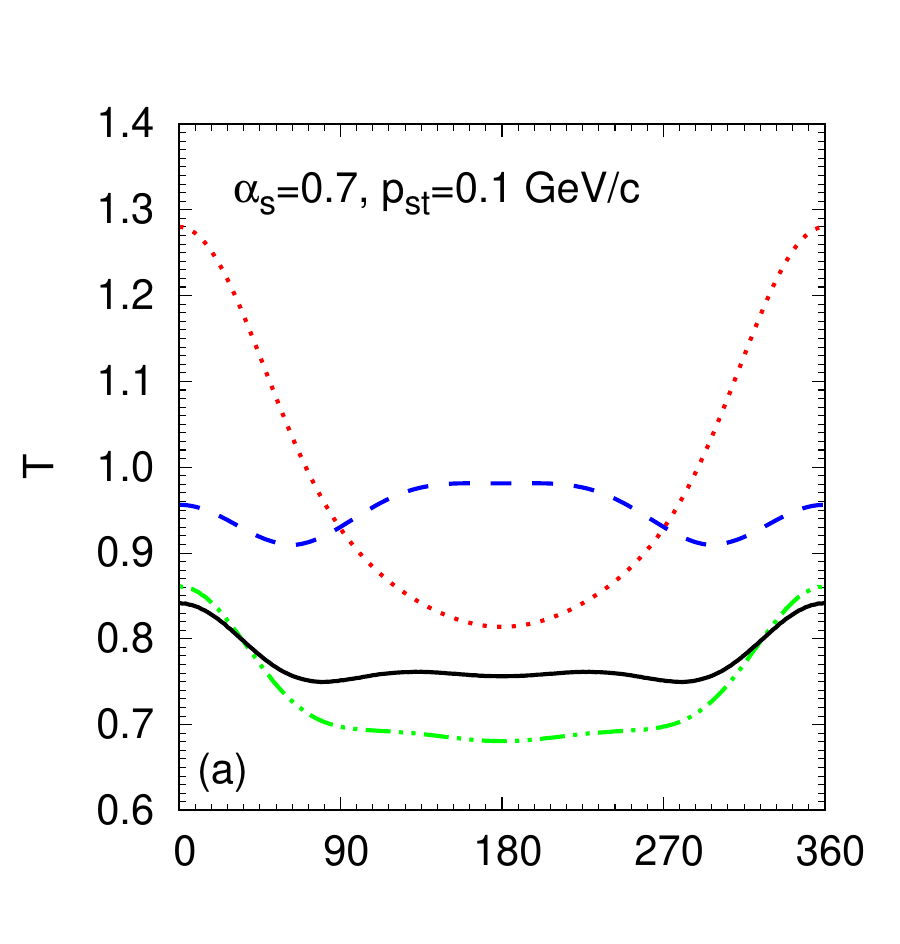} &
   \includegraphics[scale = 0.3]{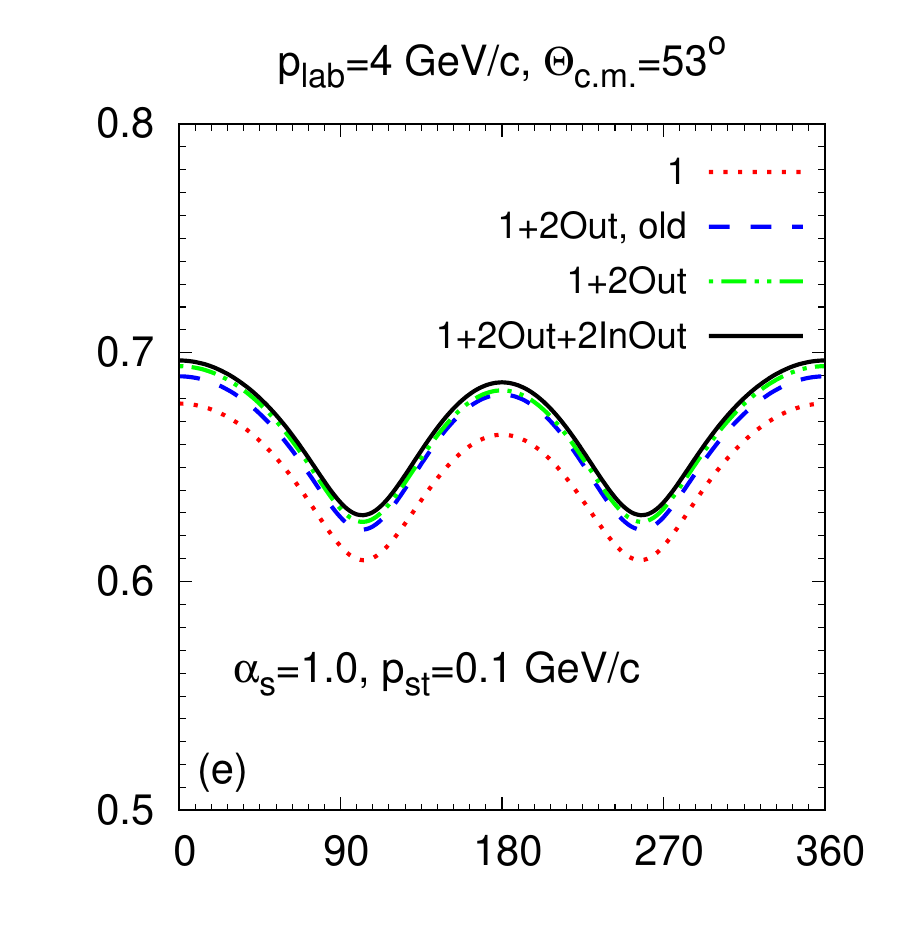} &
   \includegraphics[scale = 0.3]{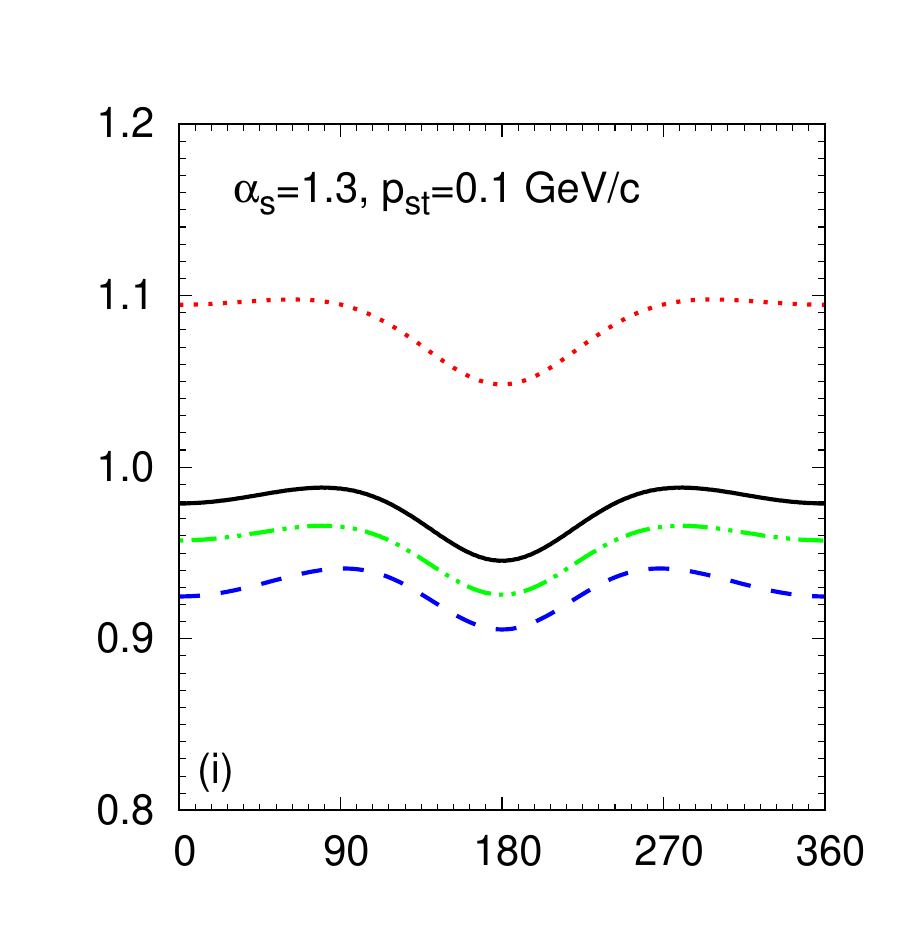} \\
   \vspace{-1cm}
   \includegraphics[scale = 0.3]{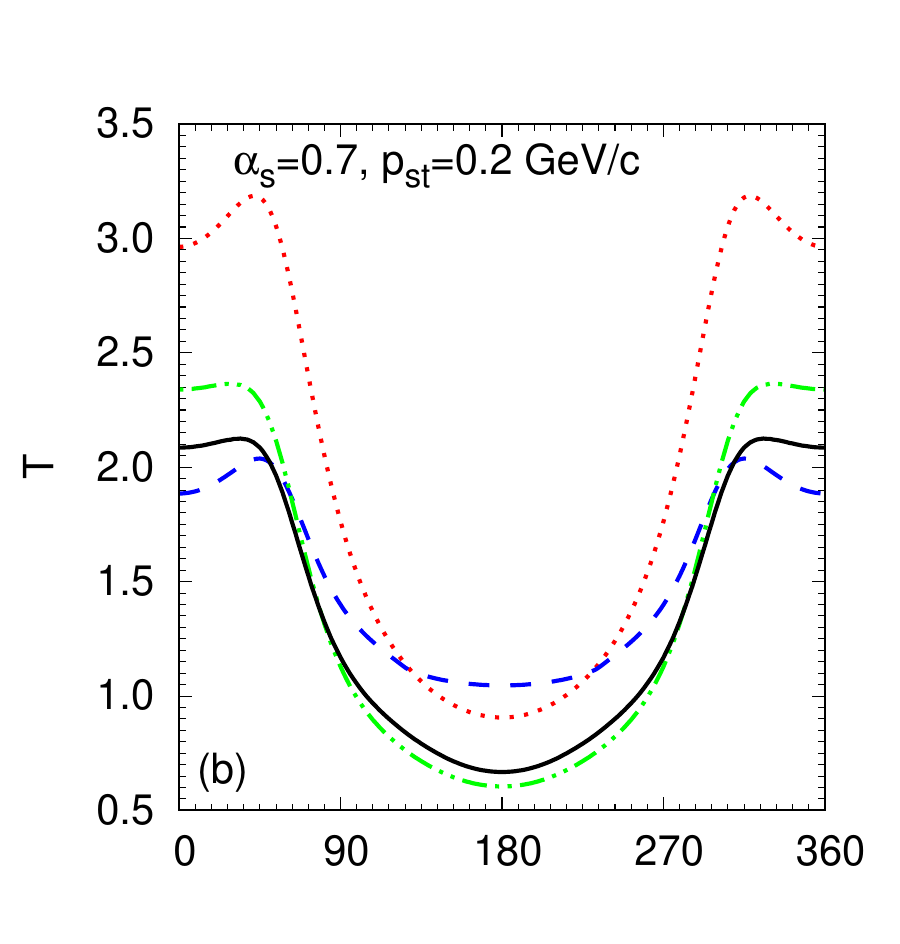} &
   \includegraphics[scale = 0.3]{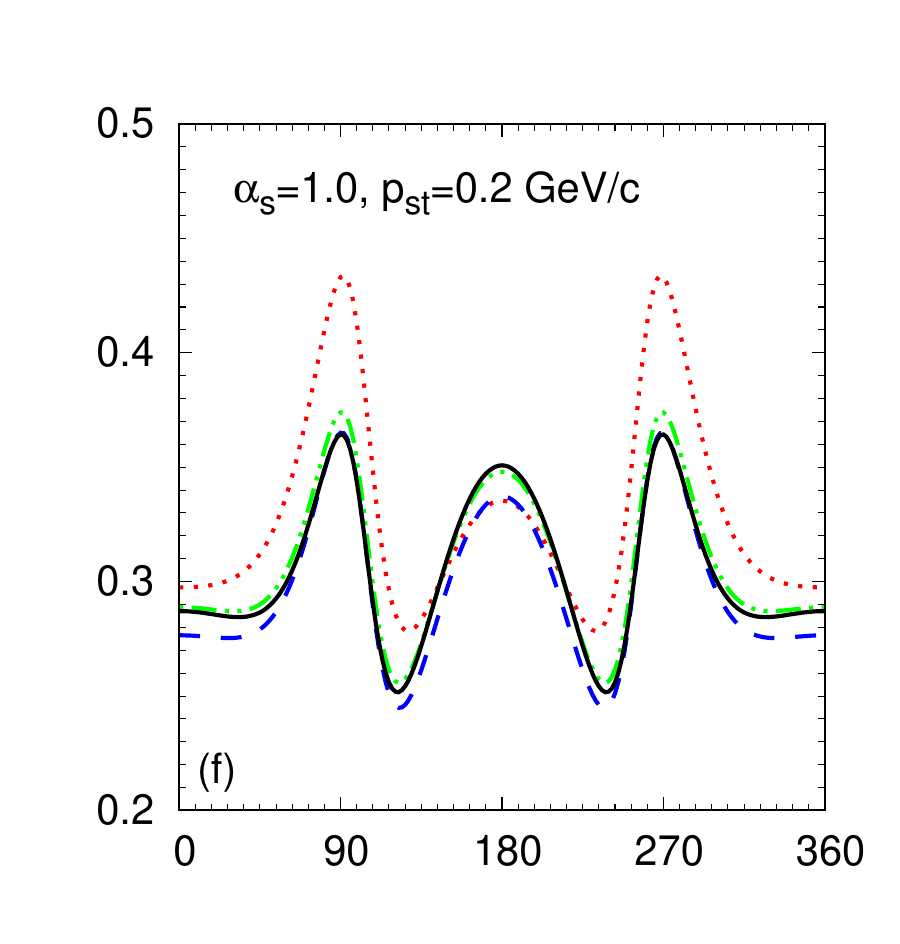} &
   \includegraphics[scale = 0.3]{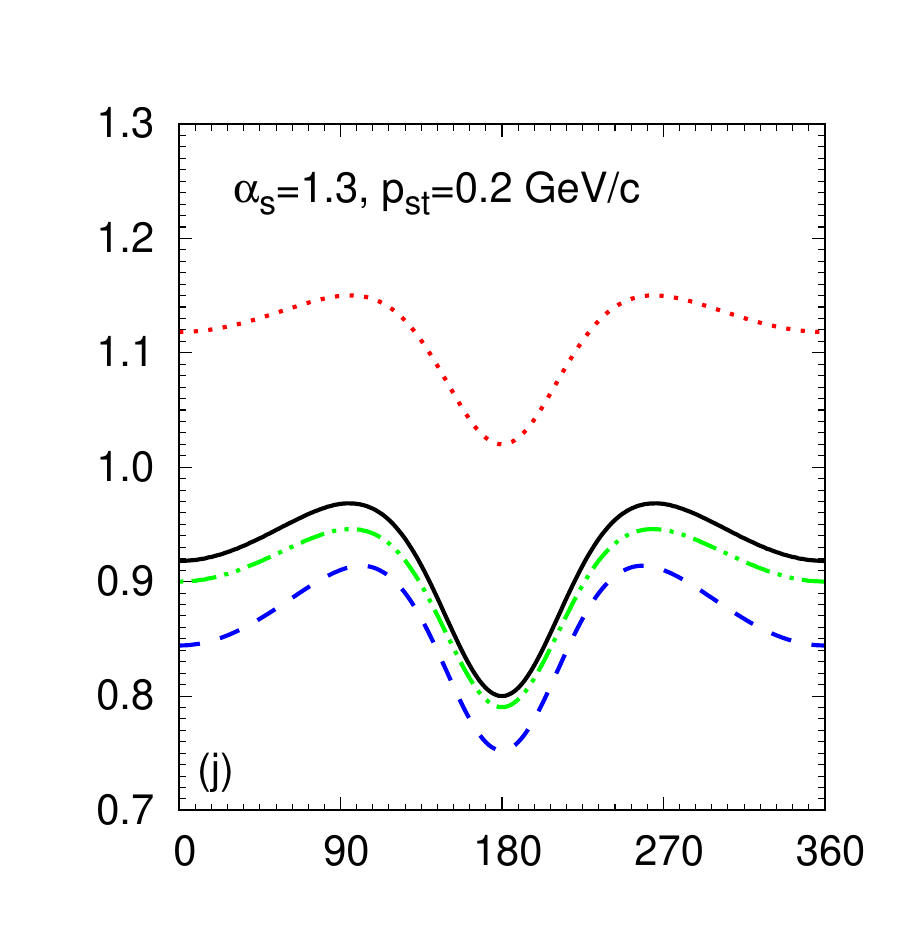} \\
   \vspace{-1cm}
   \includegraphics[scale = 0.3]{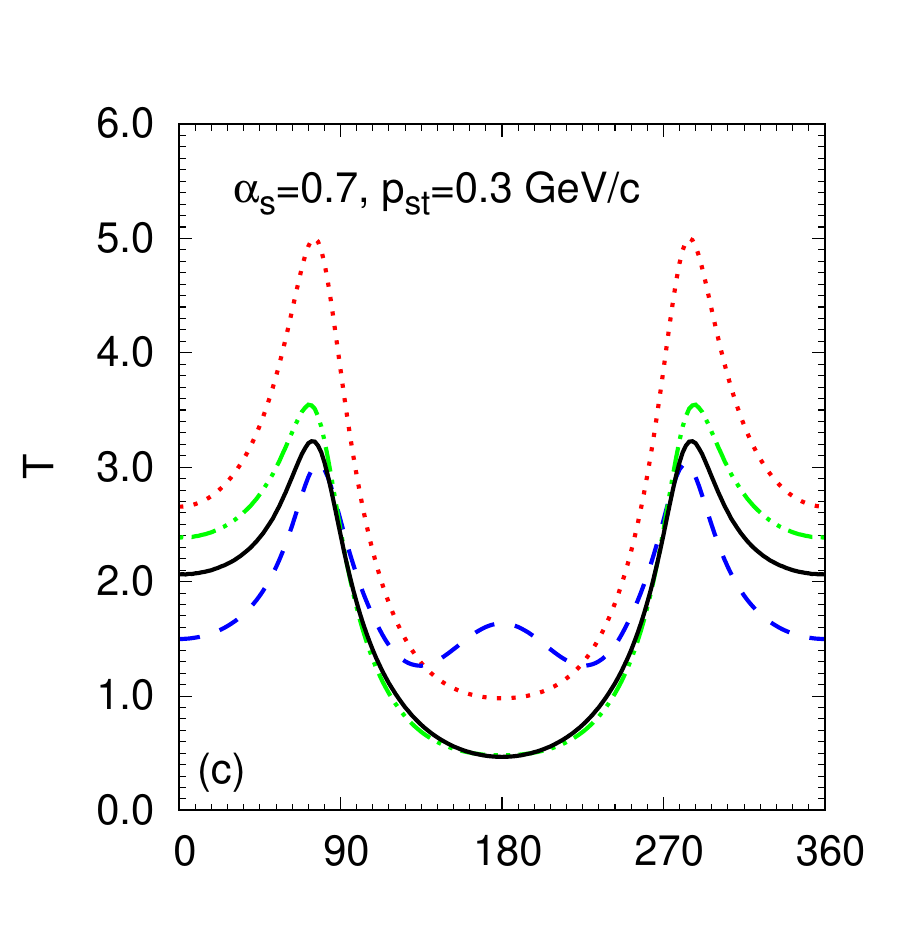} &
   \includegraphics[scale = 0.3]{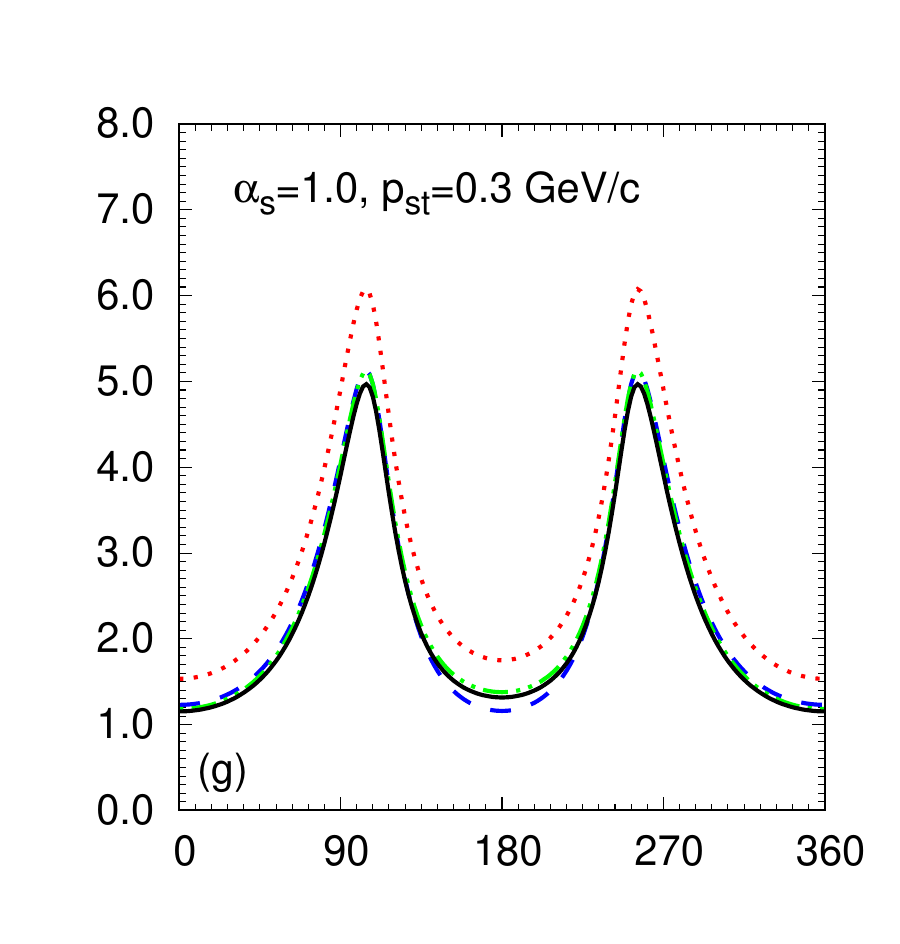} &
   \includegraphics[scale = 0.3]{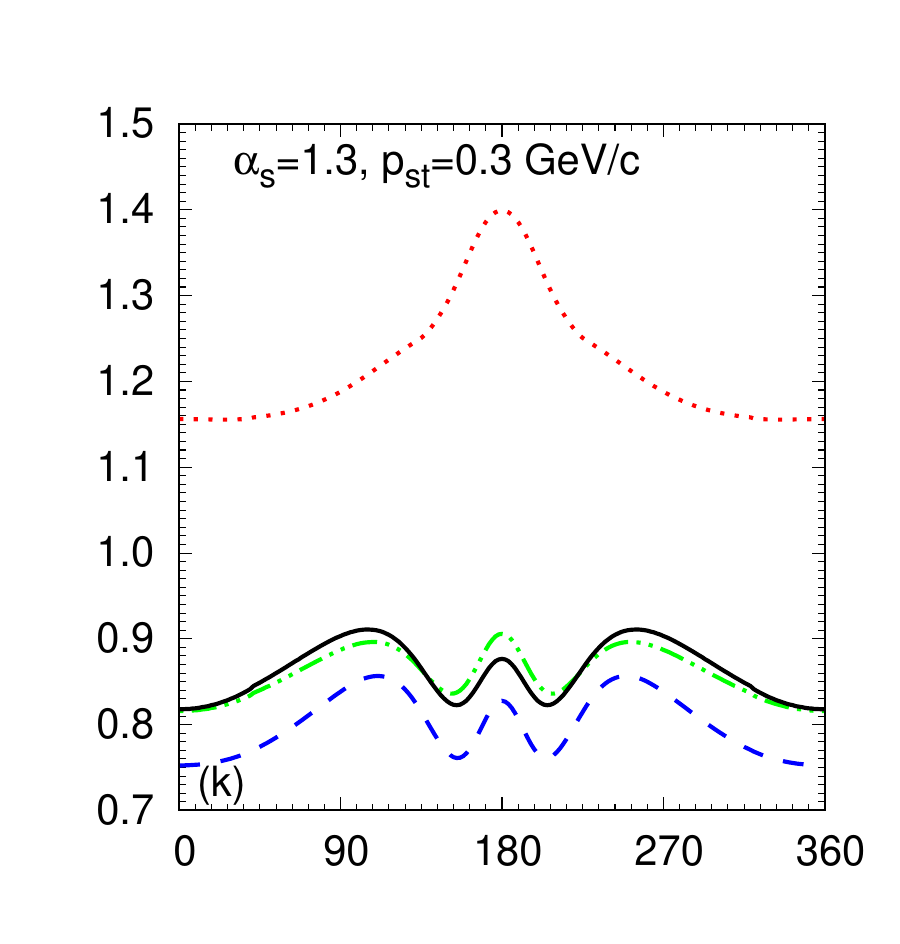} \\
   \includegraphics[scale = 0.3]{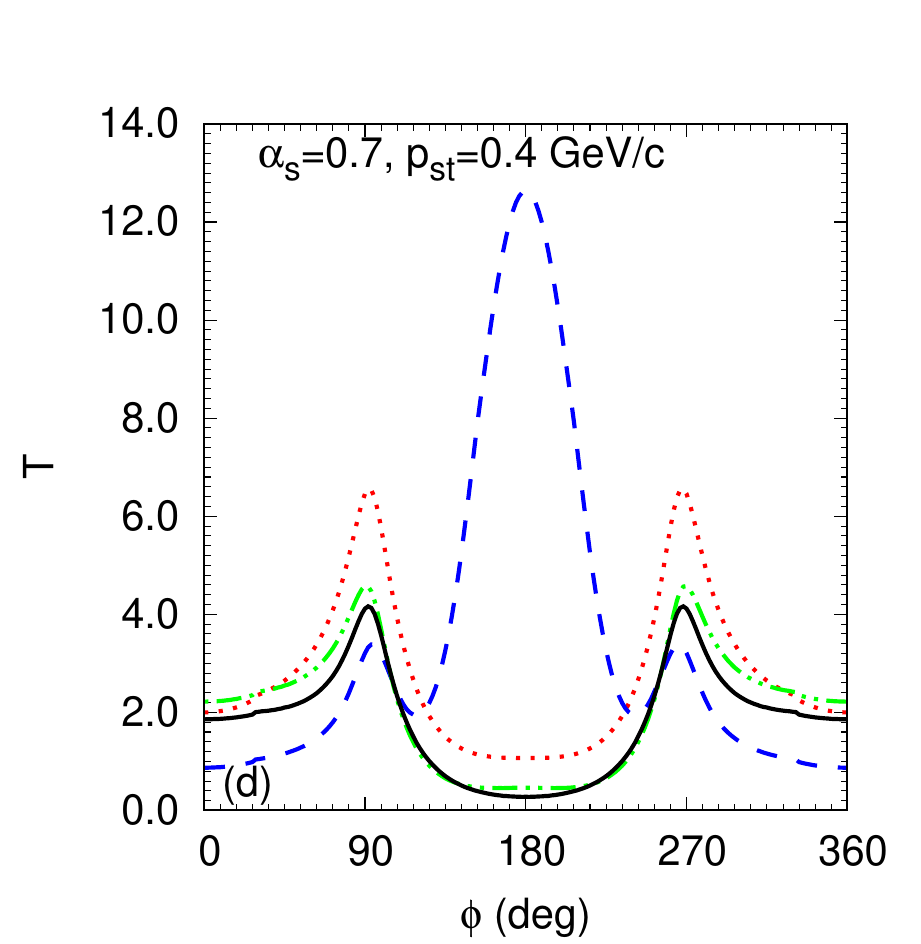} &
   \includegraphics[scale = 0.3]{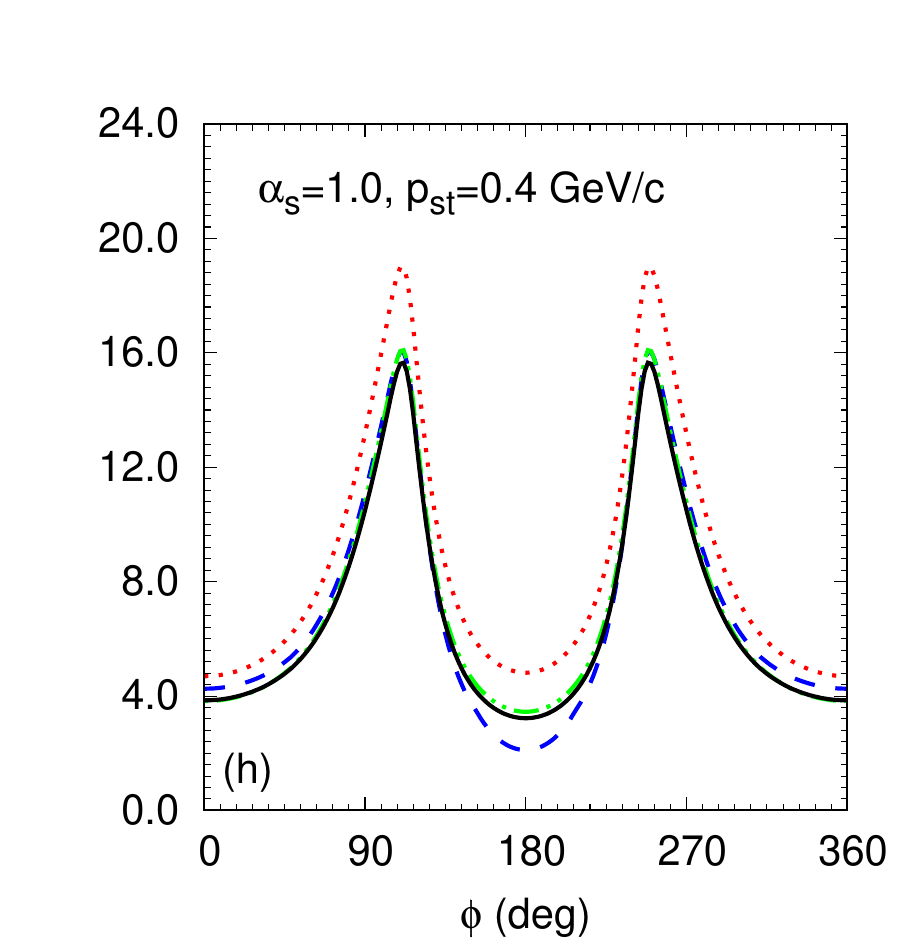} &
   \includegraphics[scale = 0.3]{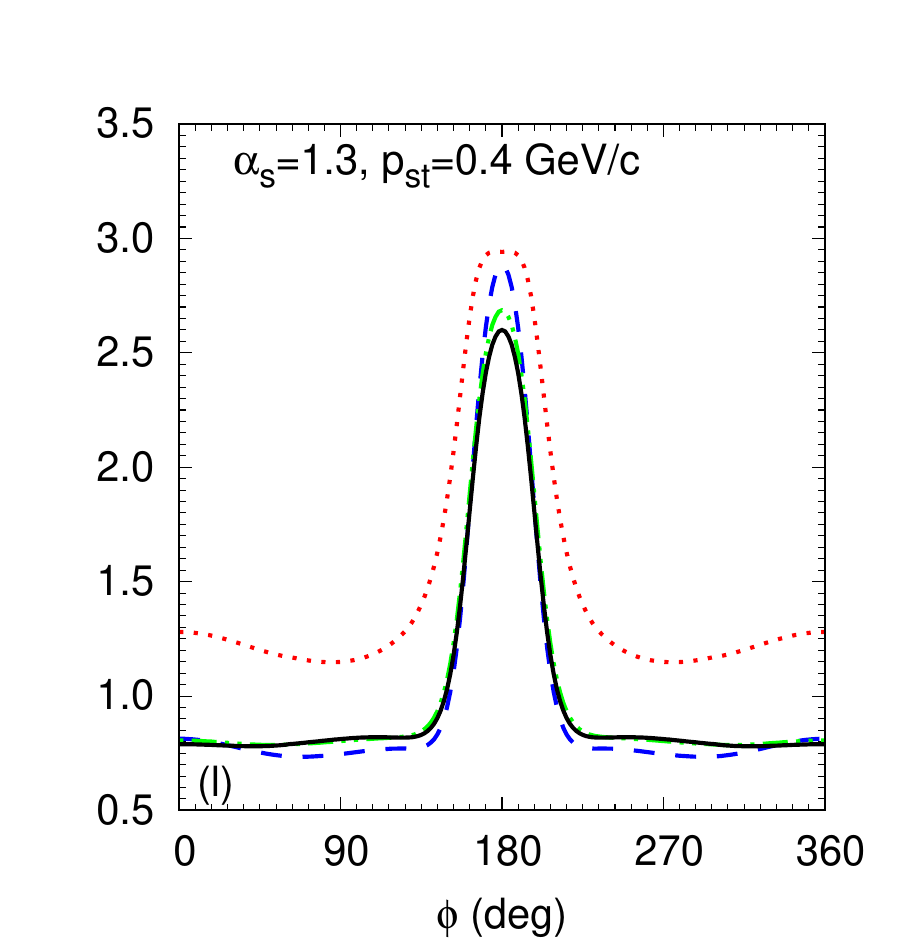} \\
   \end{tabular}
 \end{center}   
  \caption{\label{fig:T_4gevc_t_40per} Same as Fig.~\ref{fig:T_4gevc_t_fix}, but for $\Theta_{\rm c.m.}=53\degree$.}
\end{figure}

Fig.~\ref{fig:T_4gevc_t_40per} shows the transparency ratio at $p_{\rm lab}=4$ GeV/c and $\Theta_{\rm c.m.}=53\degree$. 
At $\alpha_s=1$ and 1.3, the azimuthal dependencies of $T$ at various spectator transverse momenta are quite similar to those at $\Theta_{\rm c.m.}=90\degree$.
However, at $\alpha_s=0.7$, the $\phi$-dependence becomes substantially different from that at $\Theta_{\rm c.m.}=90\degree$.
This is explained by the fact that the slowest outgoing proton has the transverse momentum component larger than the longitudinal one.
The peak positions of the $\phi$-dependence of $T$ approximately correspond to the kinematics when the momentum of the slowest proton is orthogonal to the momentum of the spectator
neutron which enhances the partial amplitudes with soft scattering of that proton.
\footnote{At $\alpha_s=0.7$, $p_{st}=0.4$ GeV/c, the sharp peak at $\phi=180\degree$ in the calculation with old treatment of double rescattering
is spurious because therein the double rescattering amplitude was proportional to $(p_4^zp_3^z)^{-1}$ instead of $(|\bvec{p}_3||\bvec{p}_4|)^{-1}$
in the new treatment of Eq.(\ref{M^(g)_GEA}). This is crucial since the longitudinal momentum of the slowest proton is minimal at this kinematics
and is as small as 0.18 GeV/c.}  

\begin{figure}
  \begin{center} 
    \begin{tabular}{ccc}
   \vspace{-1cm}   
   \includegraphics[scale = 0.3]{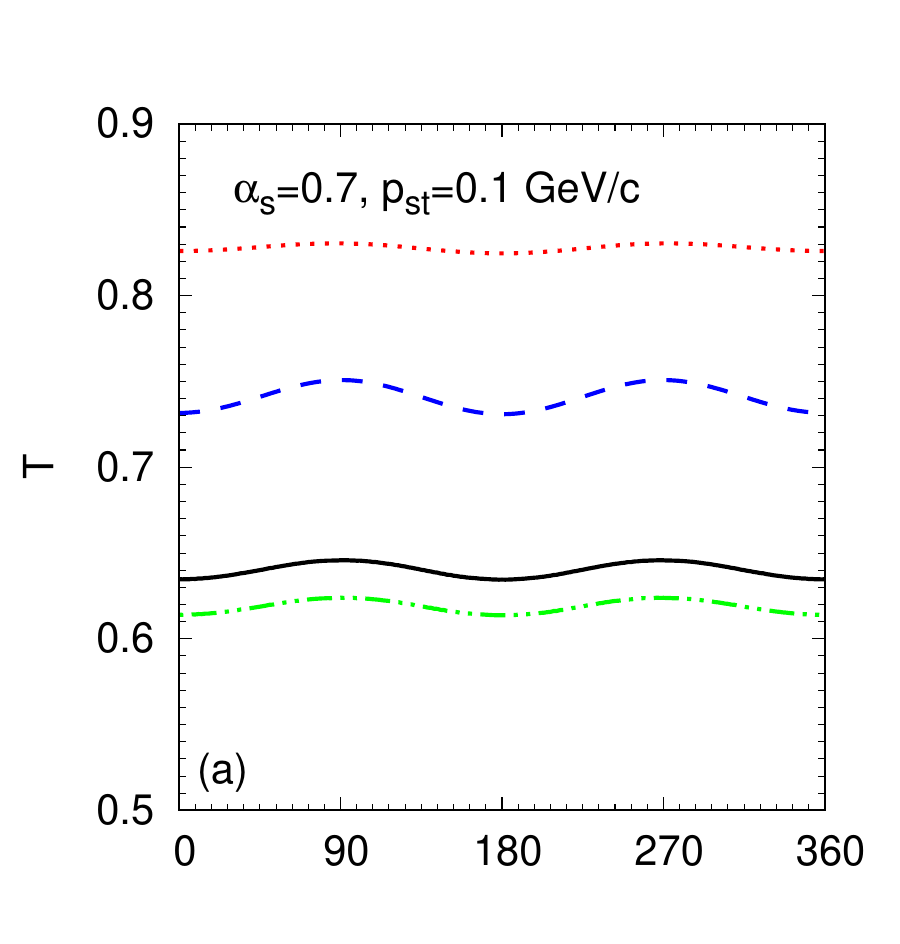} &
   \includegraphics[scale = 0.3]{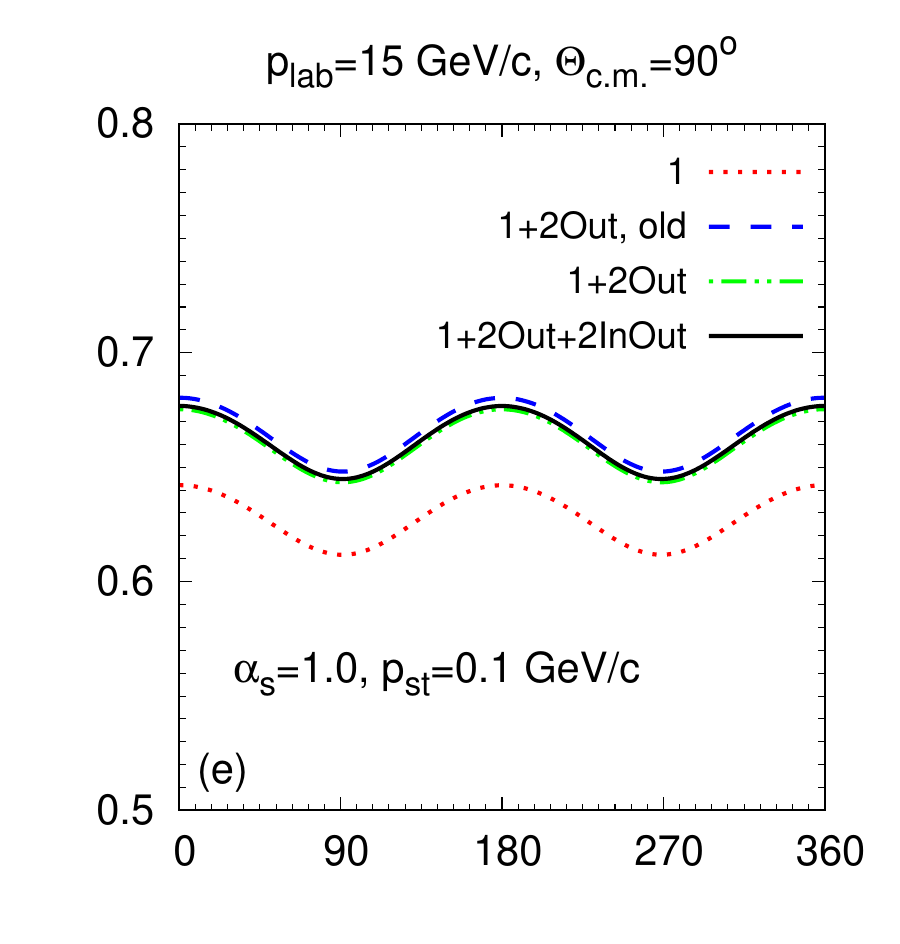} &
   \includegraphics[scale = 0.3]{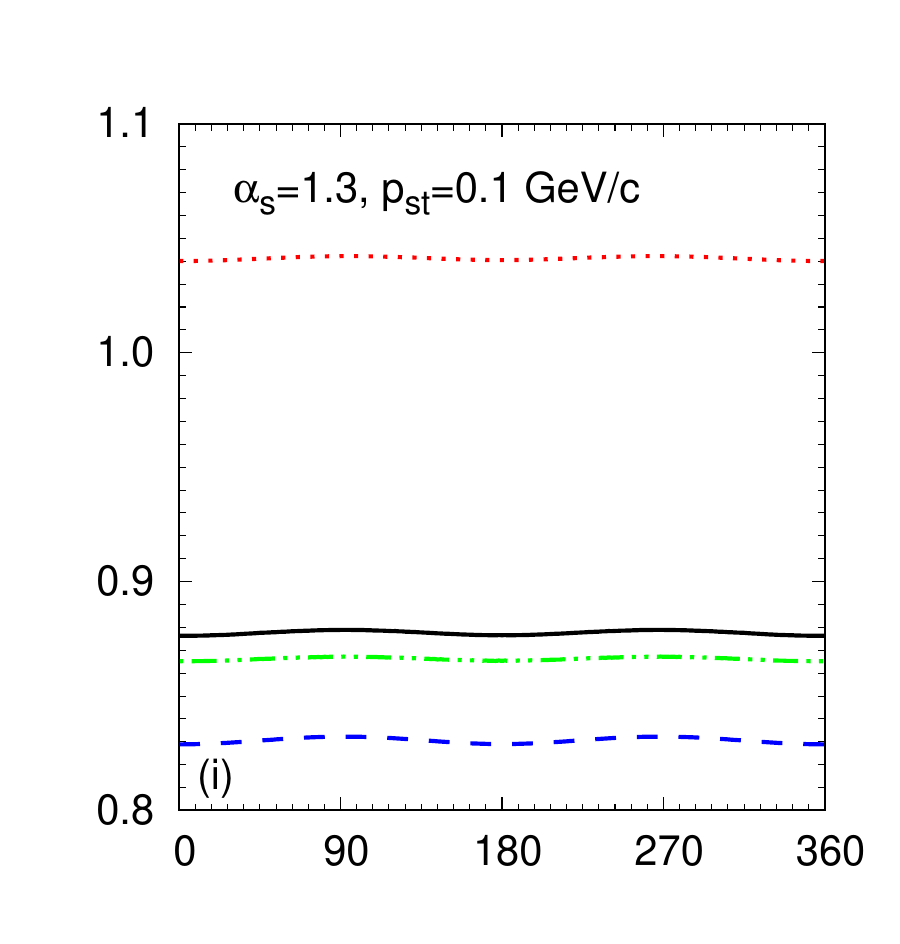} \\
   \vspace{-1cm}
   \includegraphics[scale = 0.3]{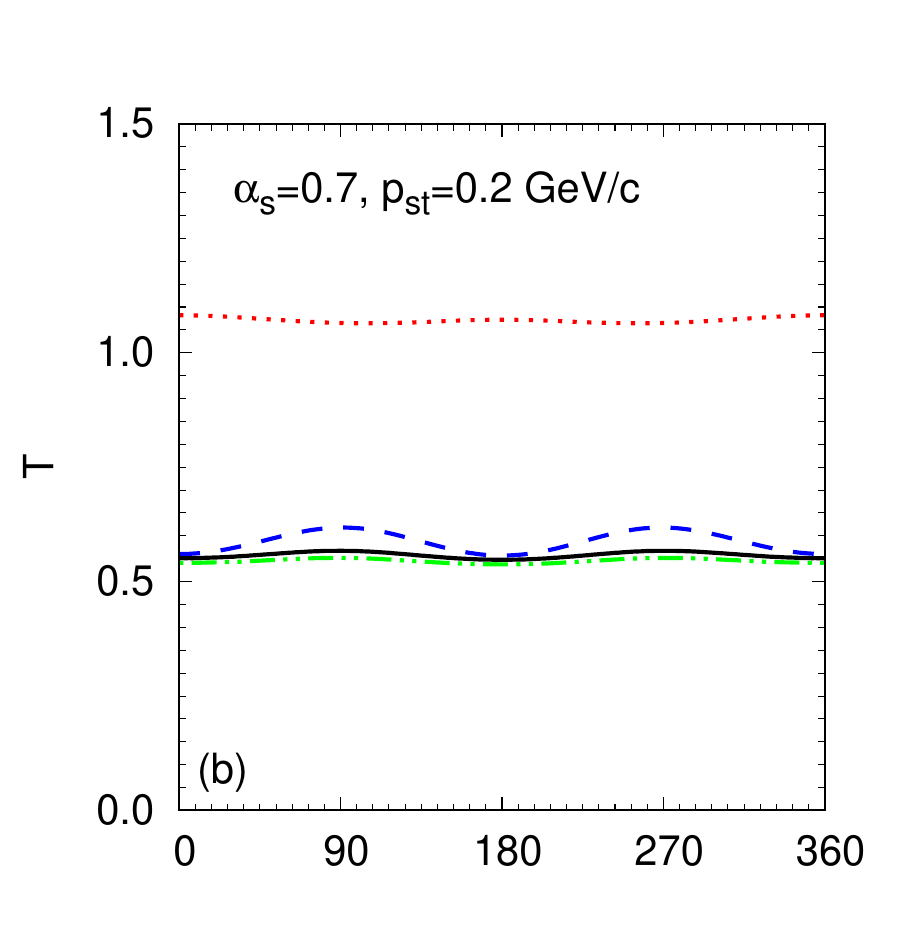} &
   \includegraphics[scale = 0.3]{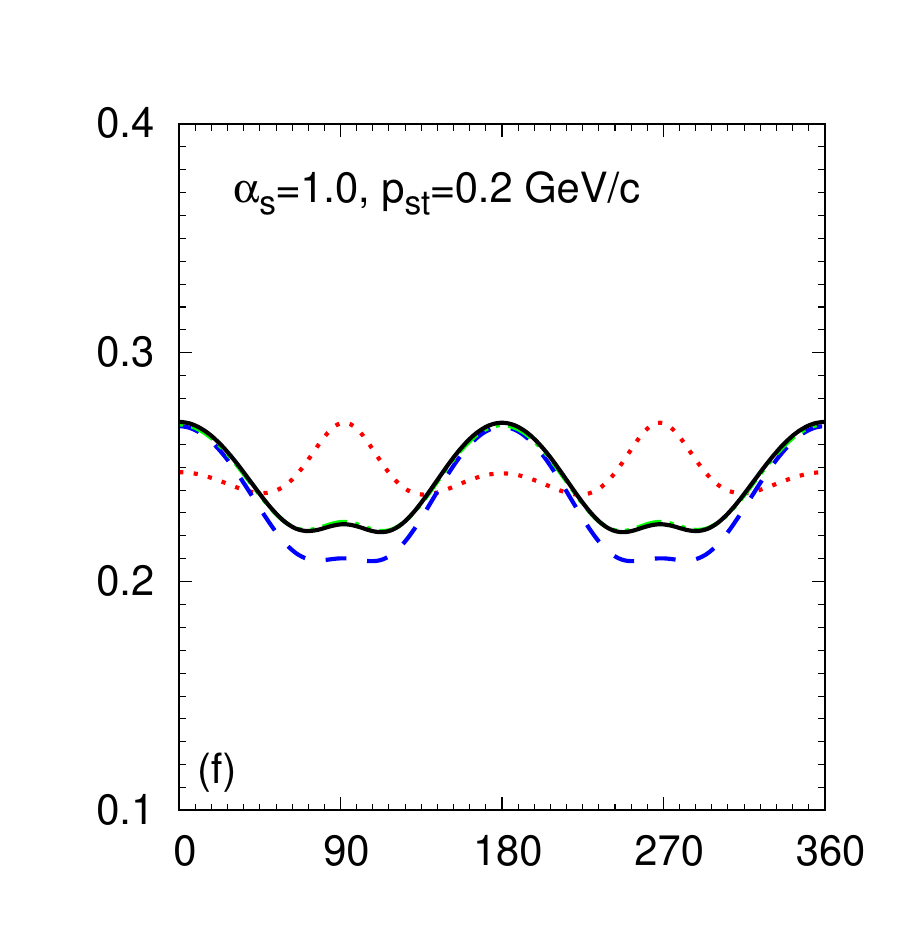} &
   \includegraphics[scale = 0.3]{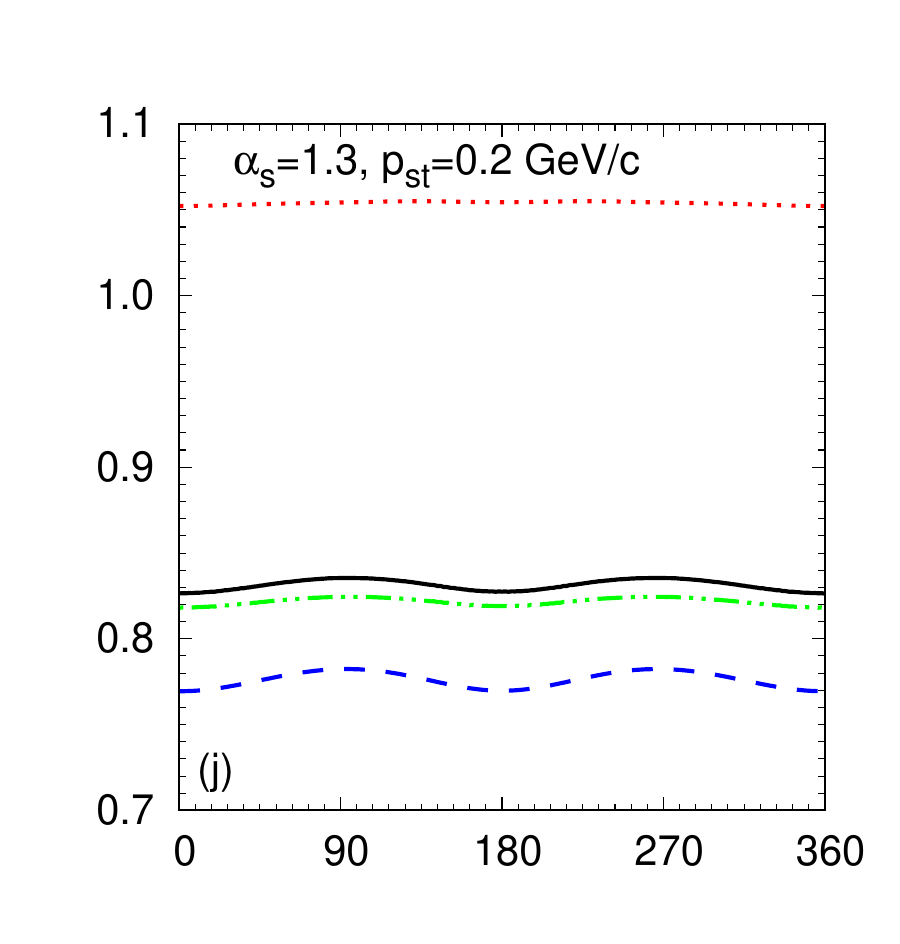} \\
   \vspace{-1cm}
   \includegraphics[scale = 0.3]{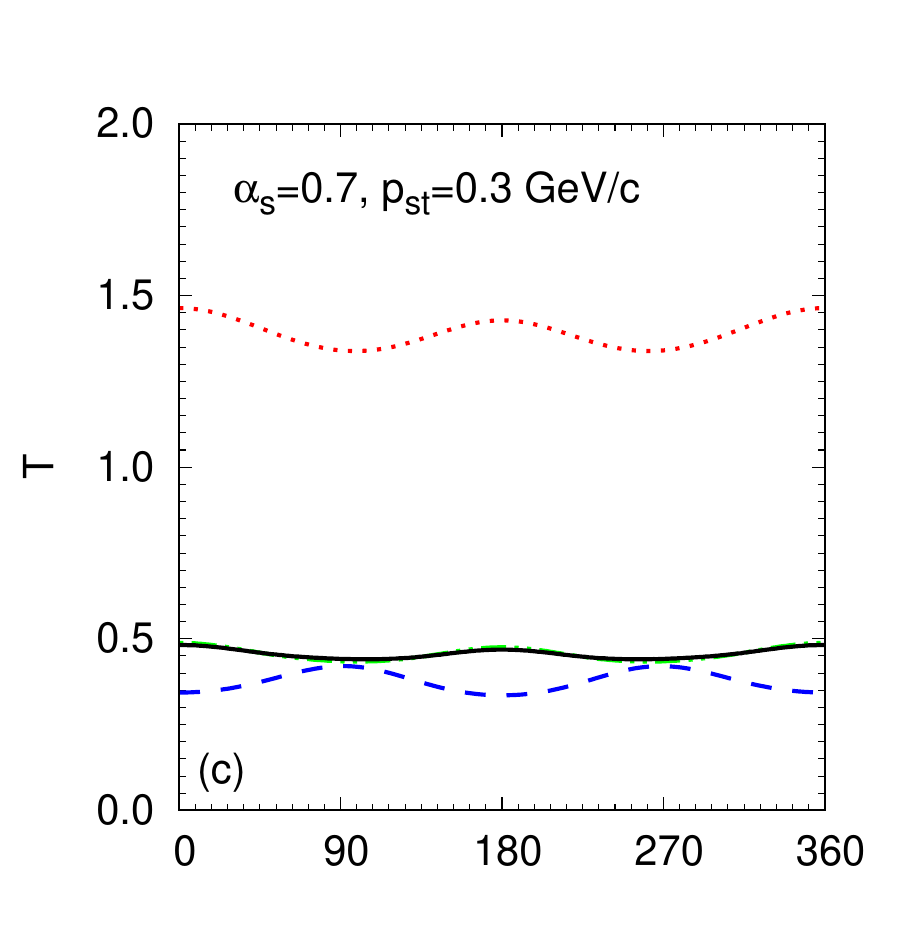} &
   \includegraphics[scale = 0.3]{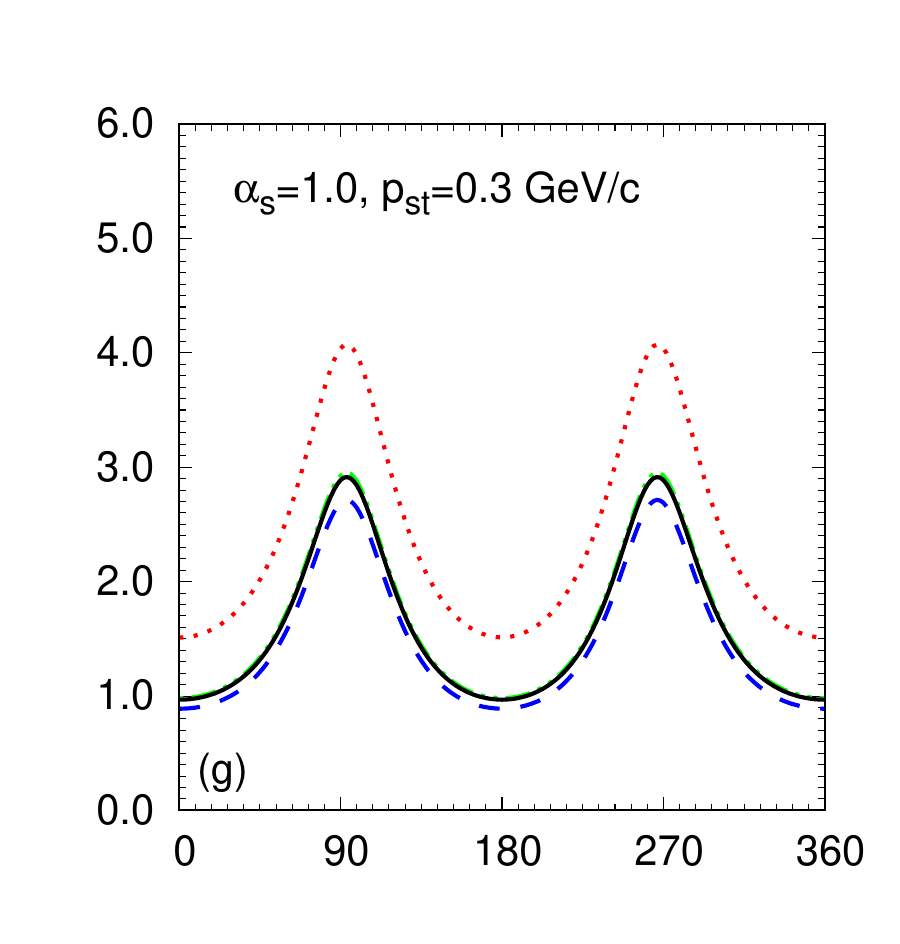} &
   \includegraphics[scale = 0.3]{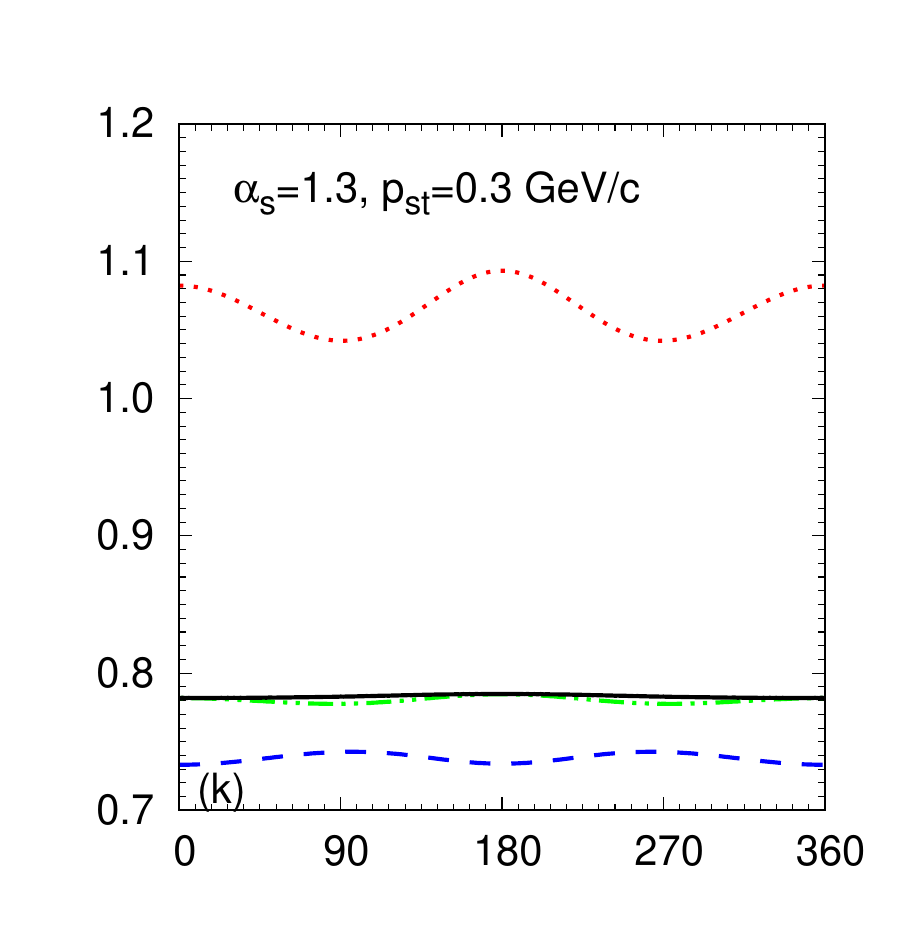} \\
   \includegraphics[scale = 0.3]{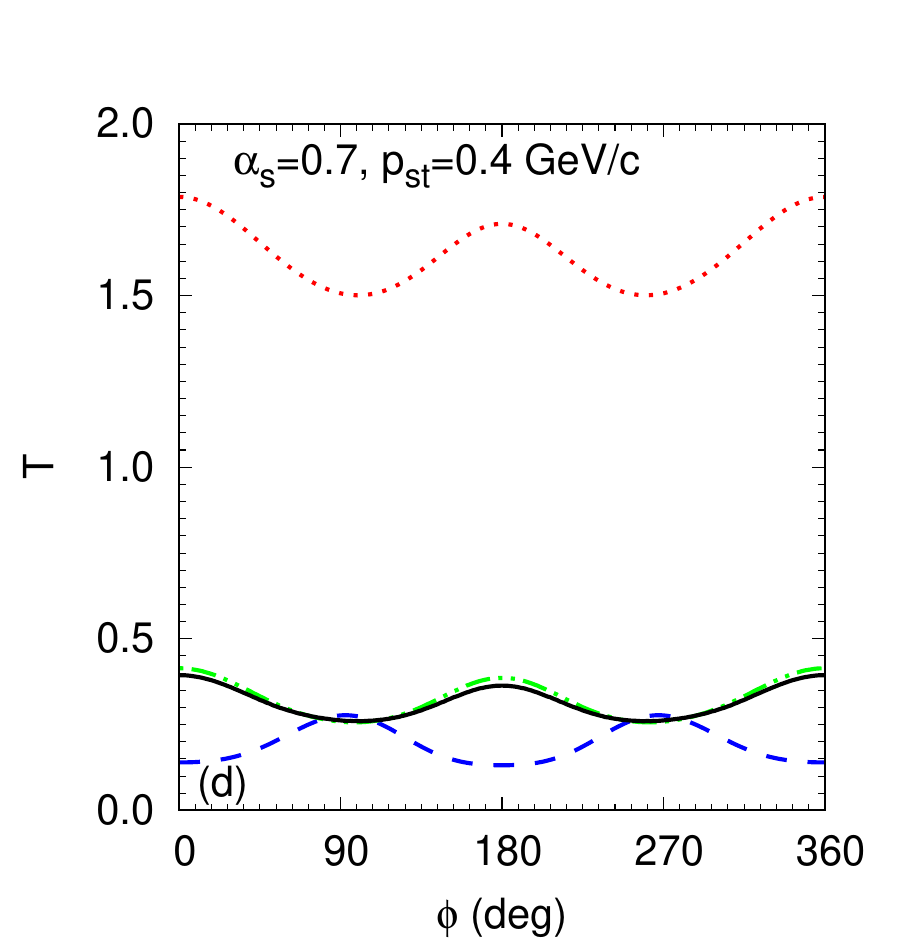} &
   \includegraphics[scale = 0.3]{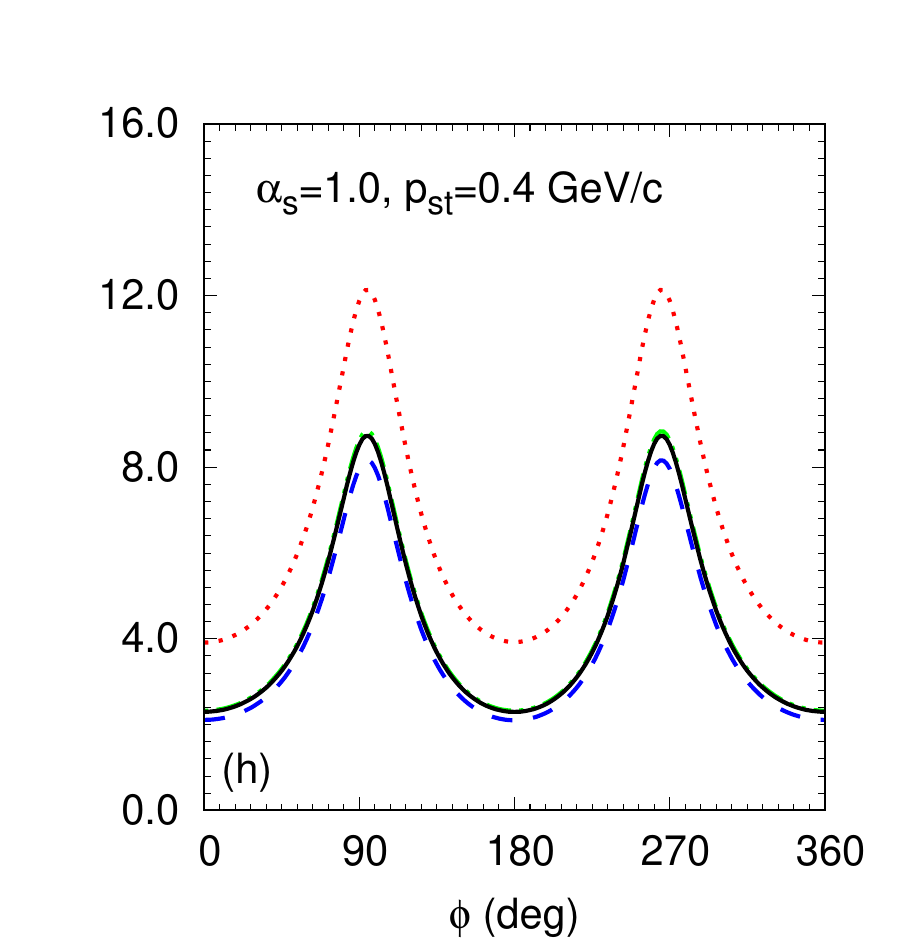} &
   \includegraphics[scale = 0.3]{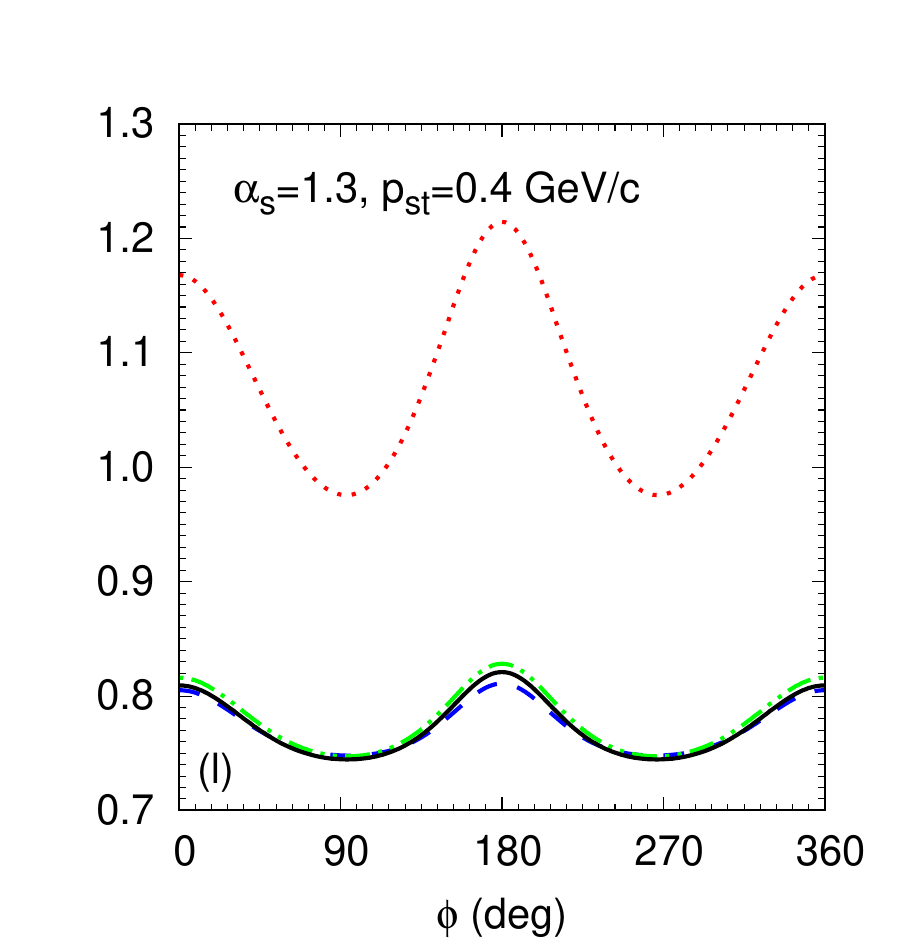} \\
   \end{tabular}
 \end{center}   
  \caption{\label{fig:T_15gevc_t_fix} Same as Fig.~\ref{fig:T_4gevc_t_fix}, but for $p_{\rm lab}=15$ GeV/c.}
\end{figure}

\begin{figure}
  \begin{center} 
    \begin{tabular}{ccc}
   \vspace{-1cm}   
   \includegraphics[scale = 0.3]{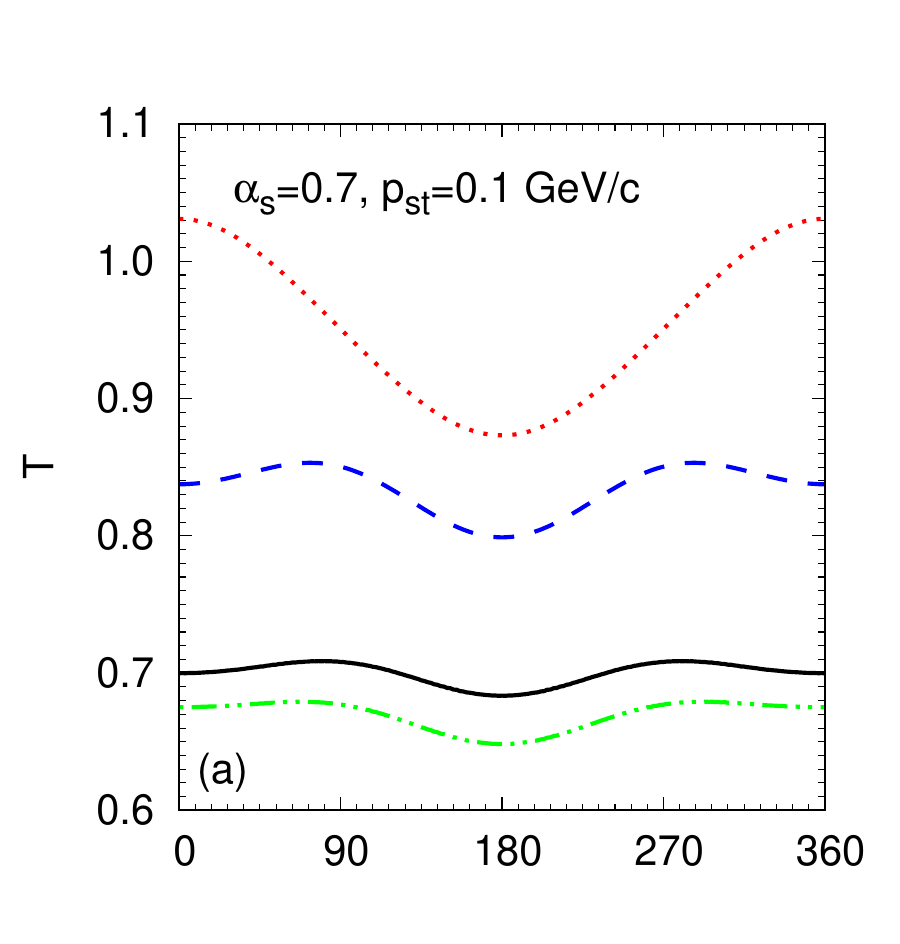} &
   \includegraphics[scale = 0.3]{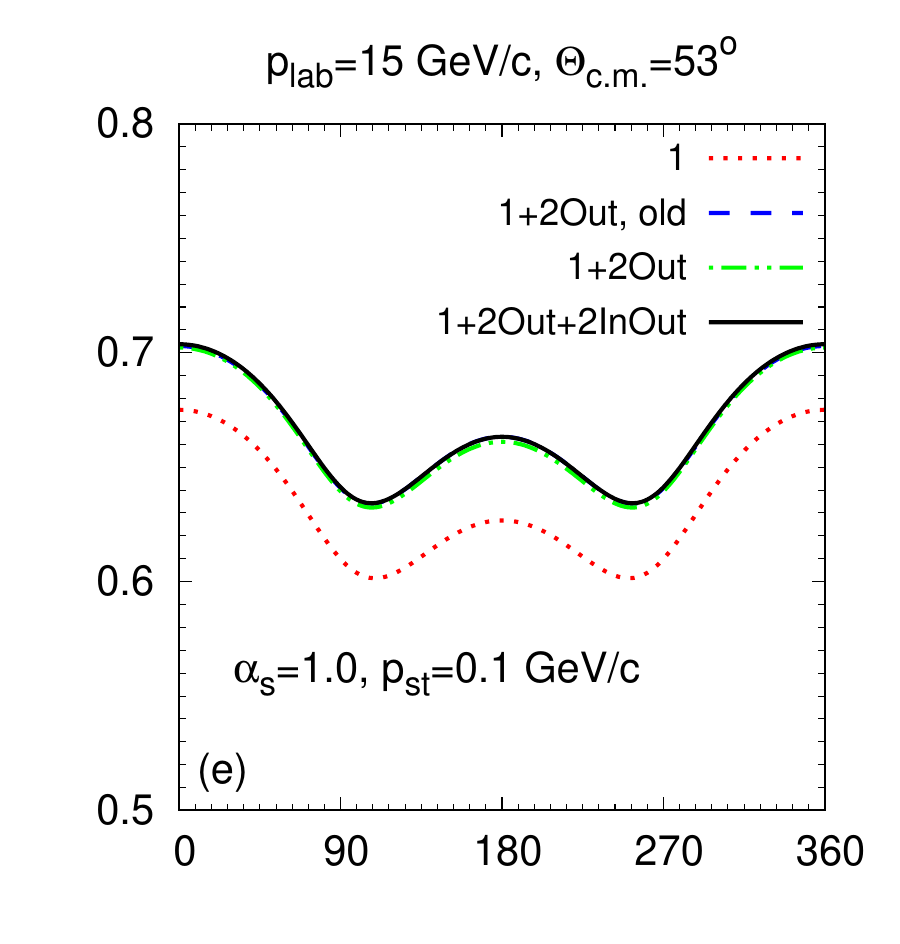} &
   \includegraphics[scale = 0.3]{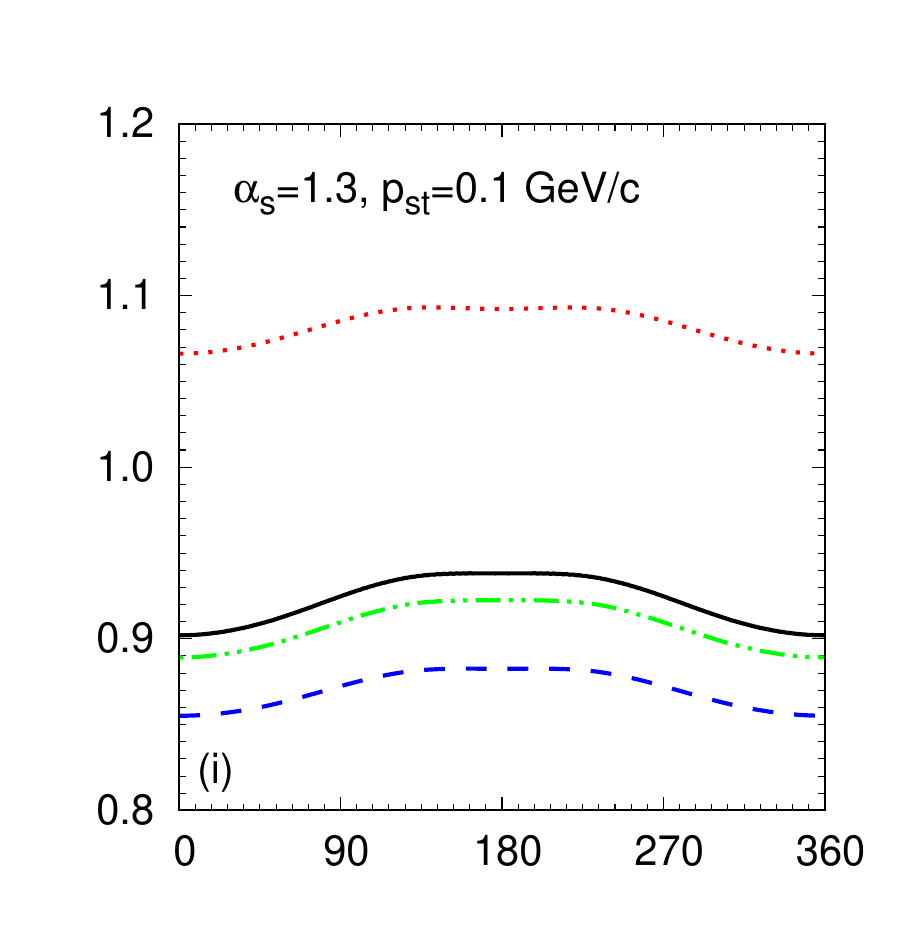} \\
   \vspace{-1cm}
   \includegraphics[scale = 0.3]{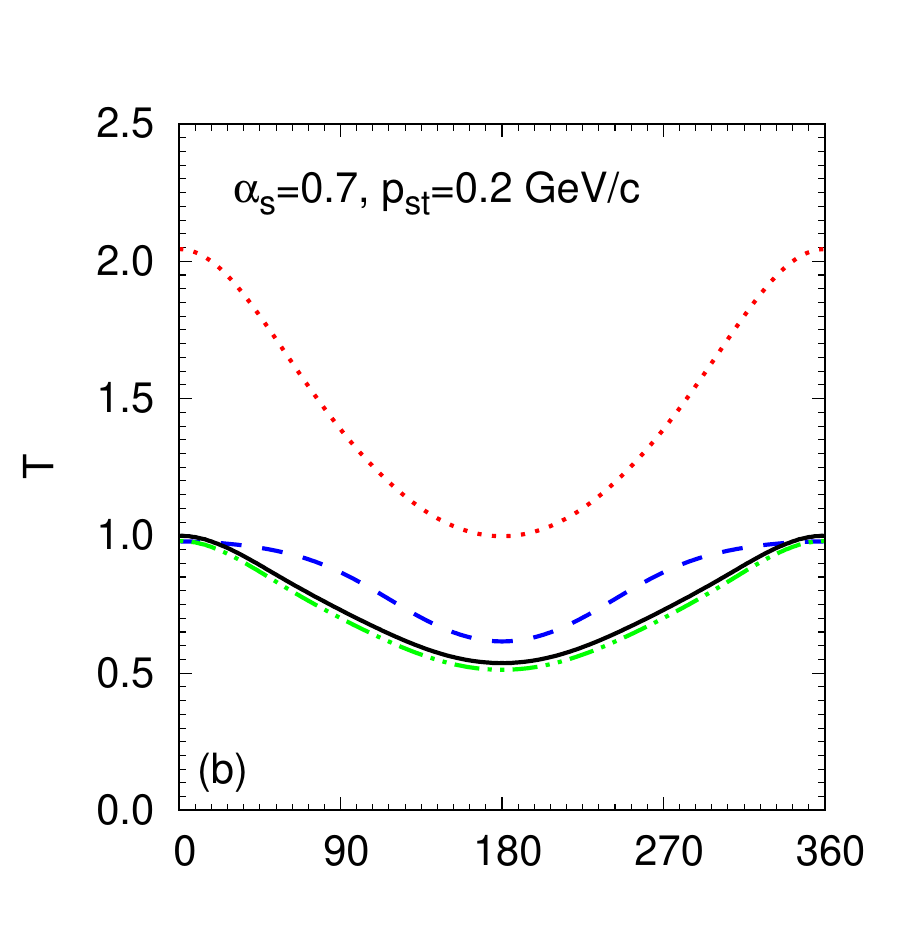} &
   \includegraphics[scale = 0.3]{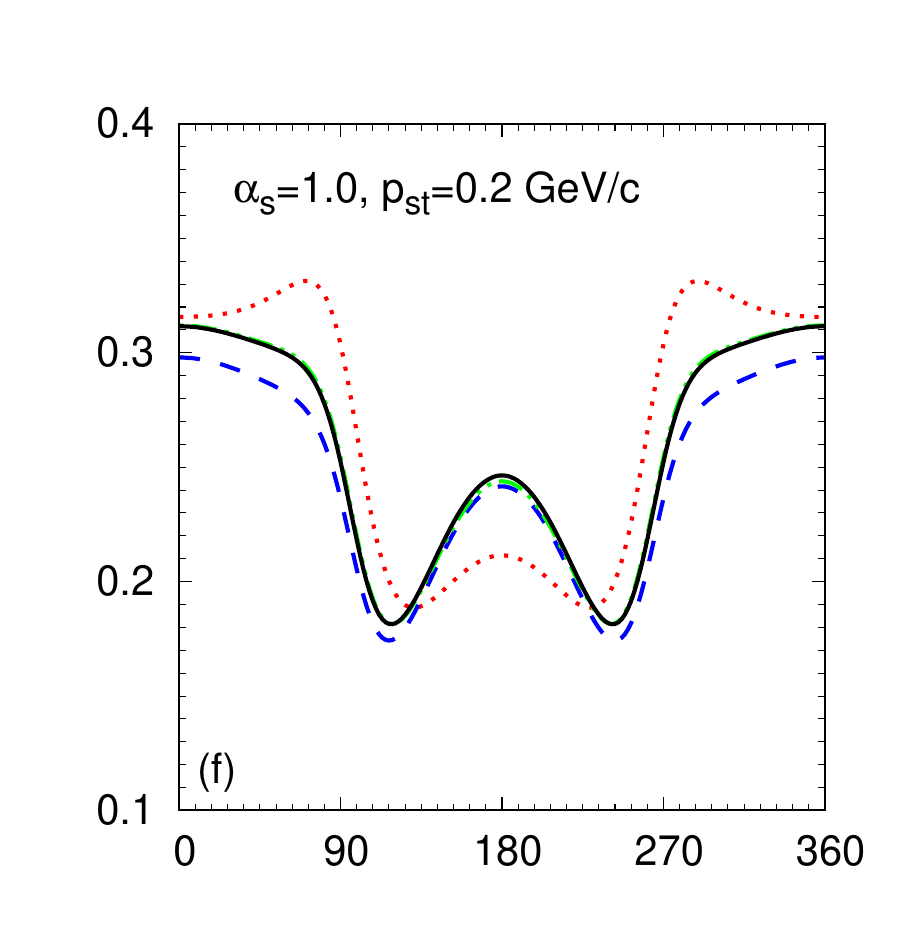} &
   \includegraphics[scale = 0.3]{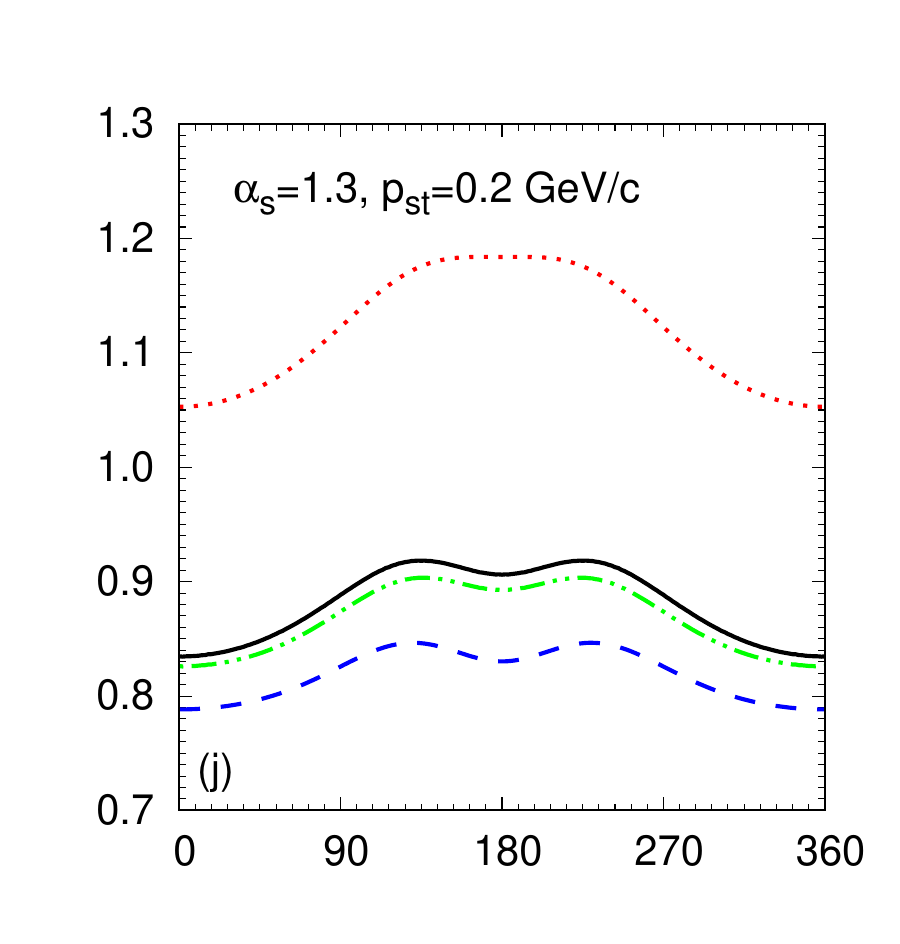} \\
   \vspace{-1cm}
   \includegraphics[scale = 0.3]{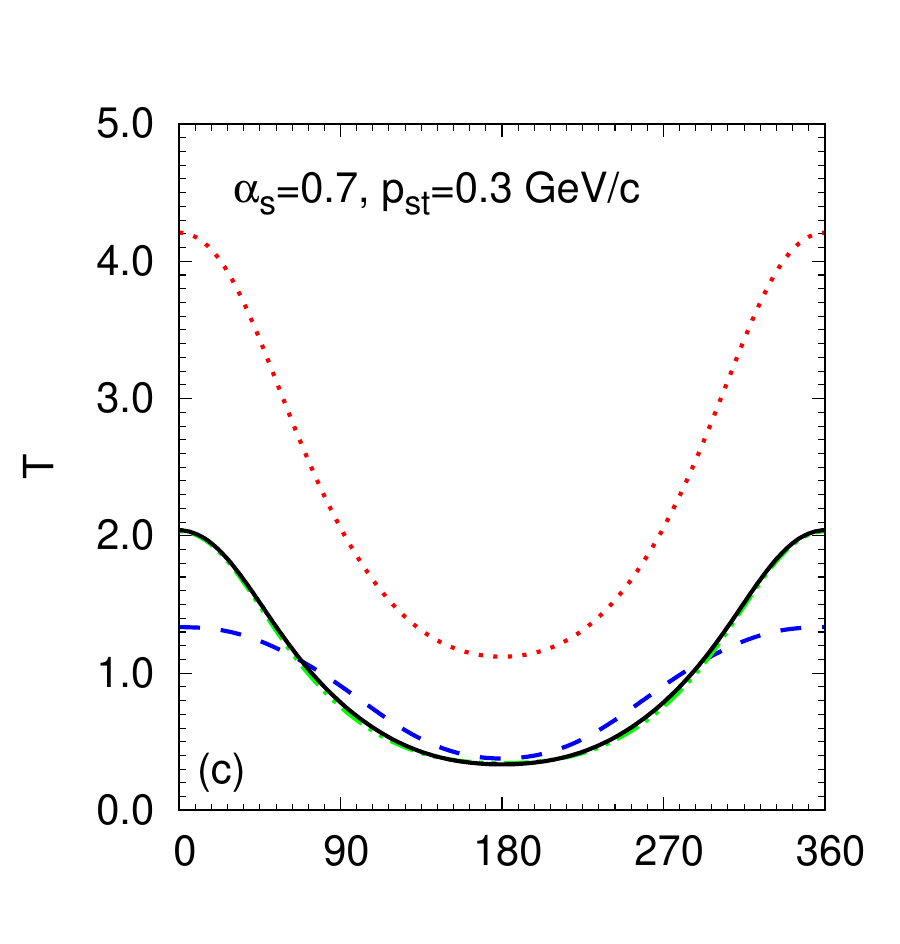} &
   \includegraphics[scale = 0.3]{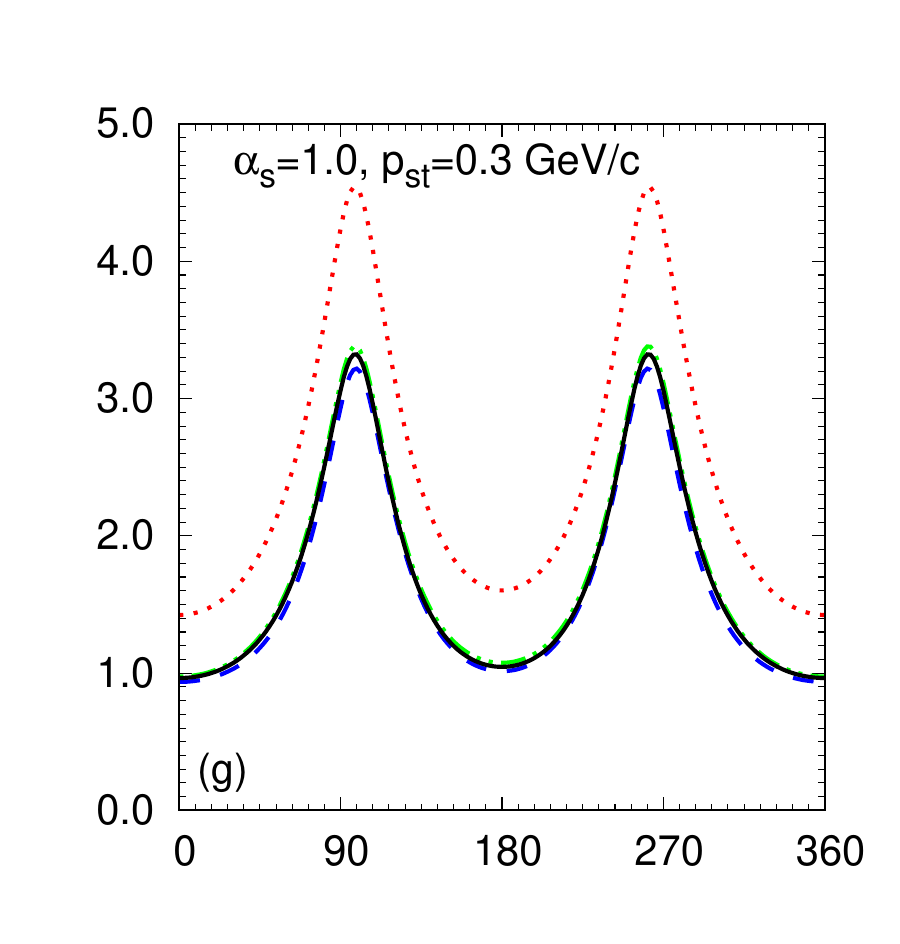} &
   \includegraphics[scale = 0.3]{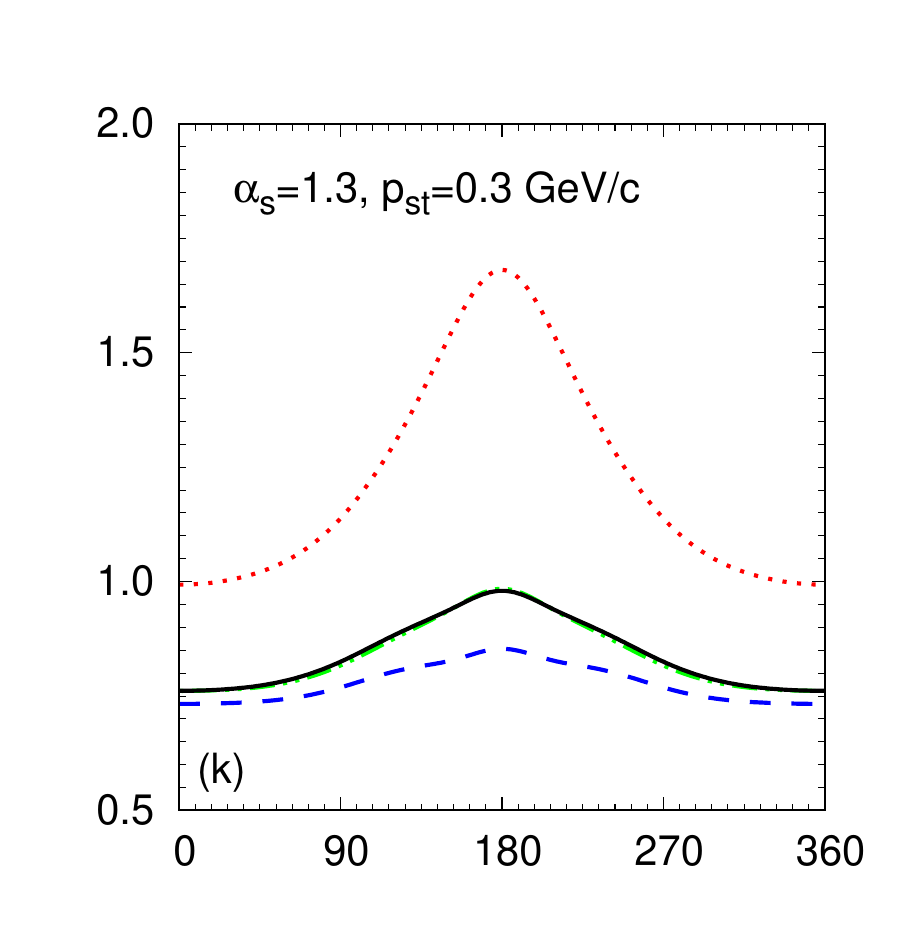} \\
   \includegraphics[scale = 0.3]{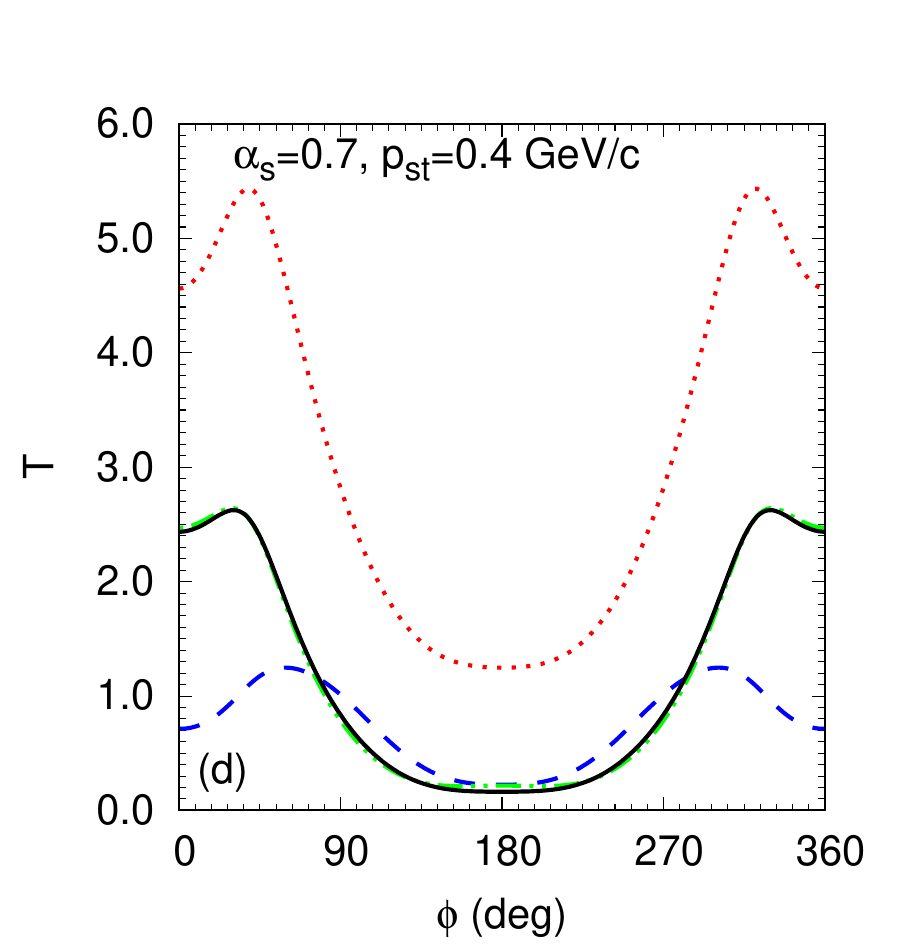} &
   \includegraphics[scale = 0.3]{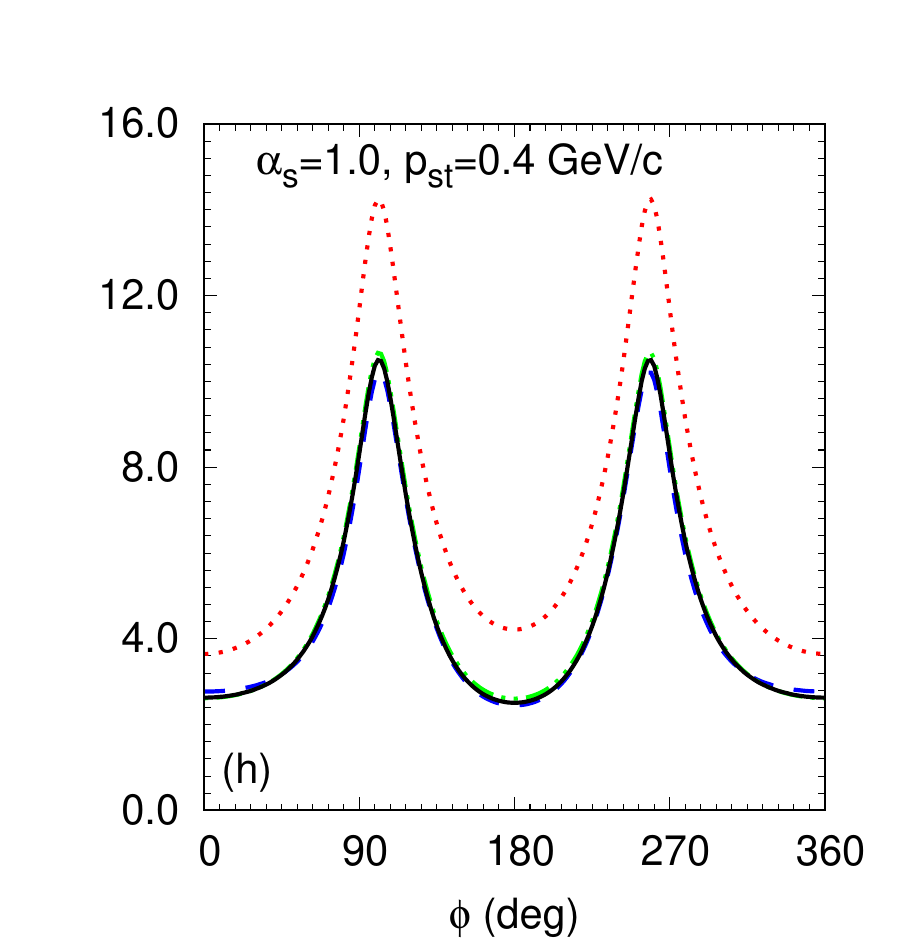} &
   \includegraphics[scale = 0.3]{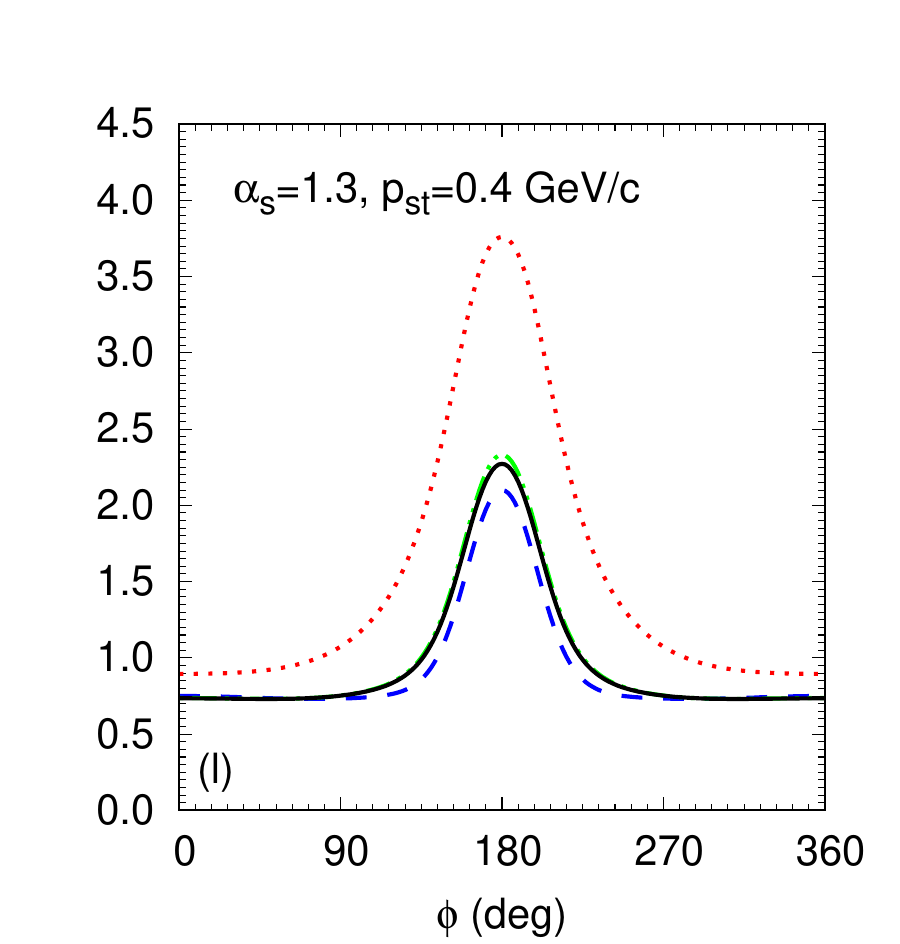} \\
   \end{tabular}
 \end{center}   
  \caption{\label{fig:T_15gevc_t_40per} Same as Fig.~\ref{fig:T_4gevc_t_fix}, but for $p_{\rm lab}=15$ GeV/c
    and $\Theta_{\rm c.m.}=53\degree$.}
\end{figure}

It is expected that with increasing beam momentum the geometrical details of the double rescattering amplitudes become less important.
This is, indeed, visible from Figs.~\ref{fig:T_15gevc_t_fix} and \ref{fig:T_15gevc_t_40per} which show the results at $p_{\rm lab}=15$ GeV/c
for $\Theta_{\rm c.m.}=90\degree$ and $53\degree$, respectively.
At $\alpha_s=1$ and 1.3, the difference between various treatments of the double rescattering is not exceeding 10\%.
At $\alpha_s=0.7$ the difference is greater and, in addition, with an increase in the transverse momentum, the shape of the $\phi$-dependence of $T$
becomes different for the old and new treatment.
  
Interestingly, at $\Theta_{\rm c.m.}=53\degree$, $\alpha_s=1.3$, $p_{st}=0.4$ GeV/c, the sharp peak at $\phi=180\degree$ is visible for both
$p_{\rm lab}=4$ and $15$ GeV/c.
This feature persists regardless of whether double rescattering is enabled or not, and is due to the orthogonality of the momenta of the spectator neutron
and the slowest outgoing proton.

\section{Summary}
\label{summary}

Within the GEA framework, ladder diagrams with all possible soft elastic rescatterings were calculated for the process $d(p,pp)n$ at large $pp$ c.m. angles 
for proton beam momenta from several GeV/c to several tens of GeV/c.
The main purpose was to accurately account for the deviations of the trajectories of the outgoing fast protons from the incoming beam direction.
Accordingly, the propagators of the incoming and outgoing protons in the diagrams with soft rescattering were linearized with respect to the momentum transfers
to the spectator nucleon along the asymptotic momenta of the protons.
In the coordinate representation, this led to an ordering of the positions of the proton and neutron in the deuteron along the asymptotic momentum of the rescattered proton,
which is expressed by the $\Theta$-function as a multiplicative factor.
As a result, when at least one outgoing proton is produced at large angle
in the deuteron rest frame, the amplitudes with double soft rescattering of the outgoing protons are corrected
as compared to those of Refs.~\cite{Frankfurt:1996uz,Larionov:2022gvn}
Moreover, the amplitudes with rescattering of both the incoming and outgoing
protons become finite.
Finally, it is shown that the amplitudes with rescattering of both outgoing protons are effectively renormalized by an infinite series
of successive soft rescatterings.

The improved description has practically no effect in the transverse ($\alpha_s=1$) and backward  ($\alpha_s > 1$)
spectator kinematics, however, becomes quite important in the forward ($\alpha_s < 1$) one.

At the lowest beam momenta, $p_{\rm lab} \simeq 4-6$ GeV/c, where the search for CT effects is still possible,
the effect of the improvement is comparable to the expected effect of CT on nuclear transparency ratio. 
This range of momenta (per nucleon) is accessible at JINR nuclotron with deuteron beam and J-PARC, before
the future FAIR, HIAF, and NICA facilities will come into operation and give an opportunity to study CT effects
in a broader beam momentum range.

\bibliography{mscat}

\begin{thebibliography}{27}%
\makeatletter
\providecommand \@ifxundefined [1]{%
 \@ifx{#1\undefined}
}%
\providecommand \@ifnum [1]{%
 \ifnum #1\expandafter \@firstoftwo
 \else \expandafter \@secondoftwo
 \fi
}%
\providecommand \@ifx [1]{%
 \ifx #1\expandafter \@firstoftwo
 \else \expandafter \@secondoftwo
 \fi
}%
\providecommand \natexlab [1]{#1}%
\providecommand \enquote  [1]{``#1''}%
\providecommand \bibnamefont  [1]{#1}%
\providecommand \bibfnamefont [1]{#1}%
\providecommand \citenamefont [1]{#1}%
\providecommand \href@noop [0]{\@secondoftwo}%
\providecommand \href [0]{\begingroup \@sanitize@url \@href}%
\providecommand \@href[1]{\@@startlink{#1}\@@href}%
\providecommand \@@href[1]{\endgroup#1\@@endlink}%
\providecommand \@sanitize@url [0]{\catcode `\\12\catcode `\$12\catcode
  `\&12\catcode `\#12\catcode `\^12\catcode `\_12\catcode `\%12\relax}%
\providecommand \@@startlink[1]{}%
\providecommand \@@endlink[0]{}%
\providecommand \url  [0]{\begingroup\@sanitize@url \@url }%
\providecommand \@url [1]{\endgroup\@href {#1}{\urlprefix }}%
\providecommand \urlprefix  [0]{URL }%
\providecommand \Eprint [0]{\href }%
\providecommand \doibase [0]{http://dx.doi.org/}%
\providecommand \selectlanguage [0]{\@gobble}%
\providecommand \bibinfo  [0]{\@secondoftwo}%
\providecommand \bibfield  [0]{\@secondoftwo}%
\providecommand \translation [1]{[#1]}%
\providecommand \BibitemOpen [0]{}%
\providecommand \bibitemStop [0]{}%
\providecommand \bibitemNoStop [0]{.\EOS\space}%
\providecommand \EOS [0]{\spacefactor3000\relax}%
\providecommand \BibitemShut  [1]{\csname bibitem#1\endcsname}%
\let\auto@bib@innerbib\@empty
\bibitem [{\citenamefont {Glauber}(1959)}]{Glauber}%
  \BibitemOpen
  \bibfield  {author} {\bibinfo {author} {\bibfnamefont {R.~J.}\ \bibnamefont
  {Glauber}},\ }in\ \href@noop {} {\emph {\bibinfo {booktitle} {Lectures in
  Theoretical Physics, Vol. 1}}},\ \bibinfo {editor} {edited by\ \bibinfo
  {editor} {\bibfnamefont {W.~E.}\ \bibnamefont {Brittin}}\ and\ \bibinfo
  {editor} {\bibfnamefont {L.~G.}\ \bibnamefont {Dunham}}}\ (\bibinfo
  {publisher} {Interscience Publishers, Inc.},\ \bibinfo {address} {New York},\
  \bibinfo {year} {1959})\ p.\ \bibinfo {pages} {315}\BibitemShut {NoStop}%
\bibitem [{\citenamefont {Glauber}\ and\ \citenamefont
  {Matthiae}(1970)}]{Glauber:1970jm}%
  \BibitemOpen
  \bibfield  {author} {\bibinfo {author} {\bibfnamefont {R.~J.}\ \bibnamefont
  {Glauber}}\ and\ \bibinfo {author} {\bibfnamefont {G.}~\bibnamefont
  {Matthiae}},\ }\href {\doibase 10.1016/0550-3213(70)90511-0} {\bibfield
  {journal} {\bibinfo  {journal} {Nucl. Phys.}\ }\textbf {\bibinfo {volume}
  {B21}},\ \bibinfo {pages} {135} (\bibinfo {year} {1970})}\BibitemShut
  {NoStop}%
\bibitem [{\citenamefont {Gribov}(1970)}]{Gribov:1968gs}%
  \BibitemOpen
  \bibfield  {author} {\bibinfo {author} {\bibfnamefont {V.~N.}\ \bibnamefont
  {Gribov}},\ }\href {http://jetp.ras.ru/cgi-bin/e/index/e/30/4/p709?a=list}
  {\bibfield  {journal} {\bibinfo  {journal} {Sov. Phys. JETP}\ }\textbf
  {\bibinfo {volume} {30}},\ \bibinfo {pages} {709} (\bibinfo {year}
  {1970})}\BibitemShut {NoStop}%
\bibitem [{\citenamefont {Bertocchi}(1972)}]{Bertocchi:1972cj}%
  \BibitemOpen
  \bibfield  {author} {\bibinfo {author} {\bibfnamefont {L.}~\bibnamefont
  {Bertocchi}},\ }\href {\doibase 10.1007/BF02722777} {\bibfield  {journal}
  {\bibinfo  {journal} {Nuovo Cim. A}\ }\textbf {\bibinfo {volume} {11}},\
  \bibinfo {pages} {45} (\bibinfo {year} {1972})}\BibitemShut {NoStop}%
\bibitem [{\citenamefont {Frankfurt}\ \emph
  {et~al.}(1997{\natexlab{a}})\citenamefont {Frankfurt}, \citenamefont
  {Sargsian},\ and\ \citenamefont {Strikman}}]{Frankfurt:1996xx}%
  \BibitemOpen
  \bibfield  {author} {\bibinfo {author} {\bibfnamefont {L.~L.}\ \bibnamefont
  {Frankfurt}}, \bibinfo {author} {\bibfnamefont {M.~M.}\ \bibnamefont
  {Sargsian}}, \ and\ \bibinfo {author} {\bibfnamefont {M.~I.}\ \bibnamefont
  {Strikman}},\ }\href {\doibase 10.1103/PhysRevC.56.1124} {\bibfield
  {journal} {\bibinfo  {journal} {Phys. Rev. C}\ }\textbf {\bibinfo {volume}
  {56}},\ \bibinfo {pages} {1124} (\bibinfo {year} {1997}{\natexlab{a}})},\
  \Eprint {http://arxiv.org/abs/nucl-th/9603018} {arXiv:nucl-th/9603018}
  \BibitemShut {NoStop}%
\bibitem [{\citenamefont {Frankfurt}\ \emph
  {et~al.}(1997{\natexlab{b}})\citenamefont {Frankfurt}, \citenamefont
  {Piasetzky}, \citenamefont {Sargsian},\ and\ \citenamefont
  {Strikman}}]{Frankfurt:1996uz}%
  \BibitemOpen
  \bibfield  {author} {\bibinfo {author} {\bibfnamefont {L.~L.}\ \bibnamefont
  {Frankfurt}}, \bibinfo {author} {\bibfnamefont {E.}~\bibnamefont
  {Piasetzky}}, \bibinfo {author} {\bibfnamefont {M.~M.}\ \bibnamefont
  {Sargsian}}, \ and\ \bibinfo {author} {\bibfnamefont {M.~I.}\ \bibnamefont
  {Strikman}},\ }\href {\doibase 10.1103/PhysRevC.56.2752} {\bibfield
  {journal} {\bibinfo  {journal} {Phys. Rev. C}\ }\textbf {\bibinfo {volume}
  {56}},\ \bibinfo {pages} {2752} (\bibinfo {year} {1997}{\natexlab{b}})},\
  \Eprint {http://arxiv.org/abs/hep-ph/9607395} {arXiv:hep-ph/9607395}
  \BibitemShut {NoStop}%
\bibitem [{\citenamefont {Sargsian}(2001)}]{Sargsian:2001ax}%
  \BibitemOpen
  \bibfield  {author} {\bibinfo {author} {\bibfnamefont {M.~M.}\ \bibnamefont
  {Sargsian}},\ }\href {\doibase 10.1142/S0218301301000617} {\bibfield
  {journal} {\bibinfo  {journal} {Int. J. Mod. Phys. E}\ }\textbf {\bibinfo
  {volume} {10}},\ \bibinfo {pages} {405} (\bibinfo {year} {2001})},\ \Eprint
  {http://arxiv.org/abs/nucl-th/0110053} {arXiv:nucl-th/0110053} \BibitemShut
  {NoStop}%
\bibitem [{\citenamefont {Boeglin}\ and\ \citenamefont
  {Sargsian}(2024)}]{Boeglin:2024spd}%
  \BibitemOpen
  \bibfield  {author} {\bibinfo {author} {\bibfnamefont {W.~U.}\ \bibnamefont
  {Boeglin}}\ and\ \bibinfo {author} {\bibfnamefont {M.~M.}\ \bibnamefont
  {Sargsian}},\ }\href {\doibase 10.1016/j.physletb.2024.138742} {\bibfield
  {journal} {\bibinfo  {journal} {Phys. Lett. B}\ }\textbf {\bibinfo {volume}
  {854}},\ \bibinfo {pages} {138742} (\bibinfo {year} {2024})},\ \Eprint
  {http://arxiv.org/abs/2402.13411} {arXiv:2402.13411 [nucl-th]} \BibitemShut
  {NoStop}%
\bibitem [{\citenamefont {Larionov}(2023)}]{Larionov:2022gvn}%
  \BibitemOpen
  \bibfield  {author} {\bibinfo {author} {\bibfnamefont {A.~B.}\ \bibnamefont
  {Larionov}},\ }\href {\doibase 10.1103/PhysRevC.107.014605} {\bibfield
  {journal} {\bibinfo  {journal} {Phys. Rev. C}\ }\textbf {\bibinfo {volume}
  {107}},\ \bibinfo {pages} {014605} (\bibinfo {year} {2023})},\ \Eprint
  {http://arxiv.org/abs/2208.08832} {arXiv:2208.08832 [nucl-th]} \BibitemShut
  {NoStop}%
\bibitem [{\citenamefont {Larionov}\ \emph {et~al.}(2014)\citenamefont
  {Larionov}, \citenamefont {Strikman},\ and\ \citenamefont
  {Bleicher}}]{Larionov:2013nga}%
  \BibitemOpen
  \bibfield  {author} {\bibinfo {author} {\bibfnamefont {A.~B.}\ \bibnamefont
  {Larionov}}, \bibinfo {author} {\bibfnamefont {M.}~\bibnamefont {Strikman}},
  \ and\ \bibinfo {author} {\bibfnamefont {M.}~\bibnamefont {Bleicher}},\
  }\href {\doibase 10.1103/PhysRevC.89.014621} {\bibfield  {journal} {\bibinfo
  {journal} {Phys. Rev. C}\ }\textbf {\bibinfo {volume} {89}},\ \bibinfo
  {pages} {014621} (\bibinfo {year} {2014})},\ \Eprint
  {http://arxiv.org/abs/1312.2150} {arXiv:1312.2150 [nucl-th]} \BibitemShut
  {NoStop}%
\bibitem [{\citenamefont {Larionov}\ \emph {et~al.}(2019)\citenamefont
  {Larionov}, \citenamefont {Gillitzer},\ and\ \citenamefont
  {Strikman}}]{Larionov:2019mwa}%
  \BibitemOpen
  \bibfield  {author} {\bibinfo {author} {\bibfnamefont {A.~B.}\ \bibnamefont
  {Larionov}}, \bibinfo {author} {\bibfnamefont {A.}~\bibnamefont {Gillitzer}},
  \ and\ \bibinfo {author} {\bibfnamefont {M.}~\bibnamefont {Strikman}},\
  }\href {\doibase 10.1140/epja/i2019-12849-4} {\bibfield  {journal} {\bibinfo
  {journal} {Eur. Phys. J. A}\ }\textbf {\bibinfo {volume} {55}},\ \bibinfo
  {pages} {154} (\bibinfo {year} {2019})},\ \Eprint
  {http://arxiv.org/abs/1905.10419} {arXiv:1905.10419 [nucl-th]} \BibitemShut
  {NoStop}%
\bibitem [{\citenamefont {Larionov}\ and\ \citenamefont
  {Strikman}(2020)}]{Larionov:2019xdn}%
  \BibitemOpen
  \bibfield  {author} {\bibinfo {author} {\bibfnamefont {A.~B.}\ \bibnamefont
  {Larionov}}\ and\ \bibinfo {author} {\bibfnamefont {M.}~\bibnamefont
  {Strikman}},\ }\href {\doibase 10.1140/epja/s10050-020-00022-1} {\bibfield
  {journal} {\bibinfo  {journal} {Eur. Phys. J. A}\ }\textbf {\bibinfo {volume}
  {56}},\ \bibinfo {pages} {21} (\bibinfo {year} {2020})},\ \Eprint
  {http://arxiv.org/abs/1909.00379} {arXiv:1909.00379 [nucl-th]} \BibitemShut
  {NoStop}%
\bibitem [{\citenamefont {Larionov}(2022)}]{Larionov:2022uri}%
  \BibitemOpen
  \bibfield  {author} {\bibinfo {author} {\bibfnamefont {A.~B.}\ \bibnamefont
  {Larionov}},\ }\href {\doibase 10.3390/physics4010020} {\bibfield  {journal}
  {\bibinfo  {journal} {MDPI Physics}\ }\textbf {\bibinfo {volume} {4}},\
  \bibinfo {pages} {294} (\bibinfo {year} {2022})}\BibitemShut {NoStop}%
\bibitem [{\citenamefont {Larionov}(2025)}]{Larionov:2025equ}%
  \BibitemOpen
  \bibfield  {author} {\bibinfo {author} {\bibfnamefont {A.~B.}\ \bibnamefont
  {Larionov}},\ }\href {\doibase 10.1134/S1063779624701673} {\bibfield
  {journal} {\bibinfo  {journal} {Phys. Part. Nucl.}\ }\textbf {\bibinfo
  {volume} {56}},\ \bibinfo {pages} {381} (\bibinfo {year} {2025})},\ \Eprint
  {http://arxiv.org/abs/2409.07845} {arXiv:2409.07845 [nucl-th]} \BibitemShut
  {NoStop}%
\bibitem [{\citenamefont {Larionov}\ \emph {et~al.}(2018)\citenamefont
  {Larionov}, \citenamefont {Gillitzer}, \citenamefont {Haidenbauer},\ and\
  \citenamefont {Strikman}}]{Larionov:2018lpk}%
  \BibitemOpen
  \bibfield  {author} {\bibinfo {author} {\bibfnamefont {A.~B.}\ \bibnamefont
  {Larionov}}, \bibinfo {author} {\bibfnamefont {A.}~\bibnamefont {Gillitzer}},
  \bibinfo {author} {\bibfnamefont {J.}~\bibnamefont {Haidenbauer}}, \ and\
  \bibinfo {author} {\bibfnamefont {M.}~\bibnamefont {Strikman}},\ }\href
  {\doibase 10.1103/PhysRevC.98.054611} {\bibfield  {journal} {\bibinfo
  {journal} {Phys. Rev.}\ }\textbf {\bibinfo {volume} {C98}},\ \bibinfo {pages}
  {054611} (\bibinfo {year} {2018})},\ \Eprint
  {http://arxiv.org/abs/1807.05105} {arXiv:1807.05105 [nucl-th]} \BibitemShut
  {NoStop}%
\bibitem [{\citenamefont {Glauber}(1955)}]{Glauber:1955qq}%
  \BibitemOpen
  \bibfield  {author} {\bibinfo {author} {\bibfnamefont {R.~J.}\ \bibnamefont
  {Glauber}},\ }\href {\doibase 10.1103/PhysRev.100.242} {\bibfield  {journal}
  {\bibinfo  {journal} {Phys. Rev.}\ }\textbf {\bibinfo {volume} {100}},\
  \bibinfo {pages} {242} (\bibinfo {year} {1955})}\BibitemShut {NoStop}%
\bibitem [{\citenamefont {Bassel}\ and\ \citenamefont
  {Wilkin}(1968)}]{Bassel:1968stc}%
  \BibitemOpen
  \bibfield  {author} {\bibinfo {author} {\bibfnamefont {R.~H.}\ \bibnamefont
  {Bassel}}\ and\ \bibinfo {author} {\bibfnamefont {C.}~\bibnamefont
  {Wilkin}},\ }\href {\doibase 10.1103/PhysRev.174.1179} {\bibfield  {journal}
  {\bibinfo  {journal} {Phys. Rev.}\ }\textbf {\bibinfo {volume} {174}},\
  \bibinfo {pages} {1179} (\bibinfo {year} {1968})}\BibitemShut {NoStop}%
\bibitem [{\citenamefont {Platonova}\ and\ \citenamefont
  {Kukulin}(2010)}]{Platonova:2010wjt}%
  \BibitemOpen
  \bibfield  {author} {\bibinfo {author} {\bibfnamefont {M.~N.}\ \bibnamefont
  {Platonova}}\ and\ \bibinfo {author} {\bibfnamefont {V.~I.}\ \bibnamefont
  {Kukulin}},\ }\href {\doibase 10.1103/PhysRevC.81.014004} {\bibfield
  {journal} {\bibinfo  {journal} {Phys. Rev. C}\ }\textbf {\bibinfo {volume}
  {81}},\ \bibinfo {pages} {014004} (\bibinfo {year} {2010})},\ \bibinfo {note}
  {[Erratum: Phys.Rev.C 94, 069902 (2016)]},\ \Eprint
  {http://arxiv.org/abs/1612.08694} {arXiv:1612.08694 [nucl-th]} \BibitemShut
  {NoStop}%
\bibitem [{\citenamefont {Uzikov}\ and\ \citenamefont
  {Haidenbauer}(2016)}]{Uzikov:2016lsc}%
  \BibitemOpen
  \bibfield  {author} {\bibinfo {author} {\bibfnamefont {Y.~N.}\ \bibnamefont
  {Uzikov}}\ and\ \bibinfo {author} {\bibfnamefont {J.}~\bibnamefont
  {Haidenbauer}},\ }\href {\doibase 10.1103/PhysRevC.94.035501} {\bibfield
  {journal} {\bibinfo  {journal} {Phys. Rev. C}\ }\textbf {\bibinfo {volume}
  {94}},\ \bibinfo {pages} {035501} (\bibinfo {year} {2016})},\ \Eprint
  {http://arxiv.org/abs/1607.04409} {arXiv:1607.04409 [nucl-th]} \BibitemShut
  {NoStop}%
\bibitem [{\citenamefont {Lacombe}\ \emph {et~al.}(1981)\citenamefont
  {Lacombe}, \citenamefont {Loiseau}, \citenamefont {Vinh~Mau}, \citenamefont
  {Cote}, \citenamefont {Pires},\ and\ \citenamefont
  {de~Tourreil}}]{Lacombe:1981eg}%
  \BibitemOpen
  \bibfield  {author} {\bibinfo {author} {\bibfnamefont {M.}~\bibnamefont
  {Lacombe}}, \bibinfo {author} {\bibfnamefont {B.}~\bibnamefont {Loiseau}},
  \bibinfo {author} {\bibfnamefont {R.}~\bibnamefont {Vinh~Mau}}, \bibinfo
  {author} {\bibfnamefont {J.}~\bibnamefont {Cote}}, \bibinfo {author}
  {\bibfnamefont {P.}~\bibnamefont {Pires}}, \ and\ \bibinfo {author}
  {\bibfnamefont {R.}~\bibnamefont {de~Tourreil}},\ }\href {\doibase
  10.1016/0370-2693(81)90659-6} {\bibfield  {journal} {\bibinfo  {journal}
  {Phys. Lett.}\ }\textbf {\bibinfo {volume} {101B}},\ \bibinfo {pages} {139}
  (\bibinfo {year} {1981})}\BibitemShut {NoStop}%
\bibitem [{\citenamefont {Ralston}\ and\ \citenamefont
  {Pire}(1988)}]{Ralston:1988rb}%
  \BibitemOpen
  \bibfield  {author} {\bibinfo {author} {\bibfnamefont {J.~P.}\ \bibnamefont
  {Ralston}}\ and\ \bibinfo {author} {\bibfnamefont {B.}~\bibnamefont {Pire}},\
  }\href {\doibase 10.1103/PhysRevLett.61.1823} {\bibfield  {journal} {\bibinfo
   {journal} {Phys. Rev. Lett.}\ }\textbf {\bibinfo {volume} {61}},\ \bibinfo
  {pages} {1823} (\bibinfo {year} {1988})}\BibitemShut {NoStop}%
\bibitem [{\citenamefont {Frankfurt}\ \emph {et~al.}(1995)\citenamefont
  {Frankfurt}, \citenamefont {Piasetsky}, \citenamefont {Sargsian},\ and\
  \citenamefont {Strikman}}]{Frankfurt:1994nw}%
  \BibitemOpen
  \bibfield  {author} {\bibinfo {author} {\bibfnamefont {L.}~\bibnamefont
  {Frankfurt}}, \bibinfo {author} {\bibfnamefont {E.}~\bibnamefont
  {Piasetsky}}, \bibinfo {author} {\bibfnamefont {M.}~\bibnamefont {Sargsian}},
  \ and\ \bibinfo {author} {\bibfnamefont {M.}~\bibnamefont {Strikman}},\
  }\href {\doibase 10.1103/PhysRevC.51.890} {\bibfield  {journal} {\bibinfo
  {journal} {Phys. Rev. C}\ }\textbf {\bibinfo {volume} {51}},\ \bibinfo
  {pages} {890} (\bibinfo {year} {1995})},\ \Eprint
  {http://arxiv.org/abs/nucl-th/9405003} {arXiv:nucl-th/9405003} \BibitemShut
  {NoStop}%
\bibitem [{\citenamefont {Van~Overmeire}\ and\ \citenamefont
  {Ryckebusch}(2007)}]{VanOvermeire:2006tk}%
  \BibitemOpen
  \bibfield  {author} {\bibinfo {author} {\bibfnamefont {B.}~\bibnamefont
  {Van~Overmeire}}\ and\ \bibinfo {author} {\bibfnamefont {J.}~\bibnamefont
  {Ryckebusch}},\ }\href {\doibase 10.1016/j.physletb.2006.11.060} {\bibfield
  {journal} {\bibinfo  {journal} {Phys. Lett. B}\ }\textbf {\bibinfo {volume}
  {644}},\ \bibinfo {pages} {304} (\bibinfo {year} {2007})},\ \Eprint
  {http://arxiv.org/abs/nucl-th/0608040} {arXiv:nucl-th/0608040} \BibitemShut
  {NoStop}%
\bibitem [{\citenamefont {Cugnon}\ \emph {et~al.}(1996)\citenamefont {Cugnon},
  \citenamefont {Vandermeulen},\ and\ \citenamefont {L'Hote}}]{Cugnon:1996kh}%
  \BibitemOpen
  \bibfield  {author} {\bibinfo {author} {\bibfnamefont {J.}~\bibnamefont
  {Cugnon}}, \bibinfo {author} {\bibfnamefont {J.}~\bibnamefont
  {Vandermeulen}}, \ and\ \bibinfo {author} {\bibfnamefont {D.}~\bibnamefont
  {L'Hote}},\ }\href {\doibase 10.1016/0168-583X(95)01384-9} {\bibfield
  {journal} {\bibinfo  {journal} {Nucl. Instrum. Meth. B}\ }\textbf {\bibinfo
  {volume} {111}},\ \bibinfo {pages} {215} (\bibinfo {year}
  {1996})}\BibitemShut {NoStop}%
\bibitem [{\citenamefont {Patrignani}\ \emph {et~al.}(2016)\citenamefont
  {Patrignani} \emph {et~al.}}]{Patrignani:2016xqp}%
  \BibitemOpen
  \bibfield  {author} {\bibinfo {author} {\bibfnamefont {C.}~\bibnamefont
  {Patrignani}} \emph {et~al.} (\bibinfo {collaboration} {Particle Data
  Group}),\ }\href {\doibase 10.1088/1674-1137/40/10/100001} {\bibfield
  {journal} {\bibinfo  {journal} {Chin. Phys.}\ }\textbf {\bibinfo {volume}
  {C40}},\ \bibinfo {pages} {100001} (\bibinfo {year} {2016})}\BibitemShut
  {NoStop}%
\bibitem [{\citenamefont {Workman}\ \emph {et~al.}(2022)\citenamefont {Workman}
  \emph {et~al.}}]{ParticleDataGroup:2022pth}%
  \BibitemOpen
  \bibfield  {author} {\bibinfo {author} {\bibfnamefont {R.~L.}\ \bibnamefont
  {Workman}} \emph {et~al.} (\bibinfo {collaboration} {Particle Data Group}),\
  }\href {\doibase 10.1093/ptep/ptac097} {\bibfield  {journal} {\bibinfo
  {journal} {PTEP}\ }\textbf {\bibinfo {volume} {2022}},\ \bibinfo {pages}
  {083C01} (\bibinfo {year} {2022})}\BibitemShut {NoStop}%
\bibitem [{\citenamefont {Falter}\ \emph {et~al.}(2004)\citenamefont {Falter},
  \citenamefont {Cassing}, \citenamefont {Gallmeister},\ and\ \citenamefont
  {Mosel}}]{Falter:2004uc}%
  \BibitemOpen
  \bibfield  {author} {\bibinfo {author} {\bibfnamefont {T.}~\bibnamefont
  {Falter}}, \bibinfo {author} {\bibfnamefont {W.}~\bibnamefont {Cassing}},
  \bibinfo {author} {\bibfnamefont {K.}~\bibnamefont {Gallmeister}}, \ and\
  \bibinfo {author} {\bibfnamefont {U.}~\bibnamefont {Mosel}},\ }\href
  {\doibase 10.1103/PhysRevC.70.054609} {\bibfield  {journal} {\bibinfo
  {journal} {Phys. Rev.}\ }\textbf {\bibinfo {volume} {C70}},\ \bibinfo {pages}
  {054609} (\bibinfo {year} {2004})},\ \Eprint
  {http://arxiv.org/abs/nucl-th/0406023} {arXiv:nucl-th/0406023 [nucl-th]}
  \BibitemShut {NoStop}%
\end{thebibliography}%

\newpage

\appendix

\section{Forward elastic $pd$ amplitude}
\label{pdForw}

In this Appendix, the Glauber model results are reproduced by using the Feynman diagrams approach for the case of $pd$ forward elastic scattering amplitude.
\begin{figure}
  \begin{center}
    \begin{tabular}{cc}
  \includegraphics[scale = 0.4]{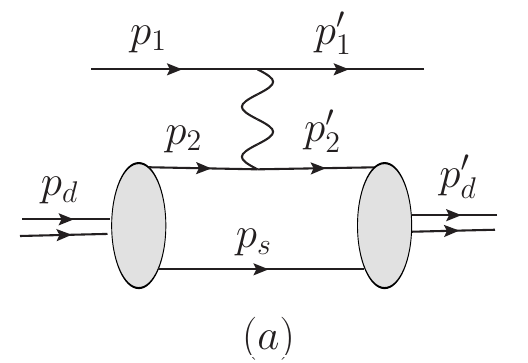} &
  \includegraphics[scale = 0.4]{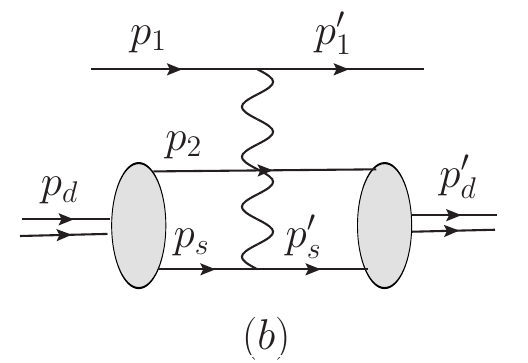}\\
  \includegraphics[scale = 0.4]{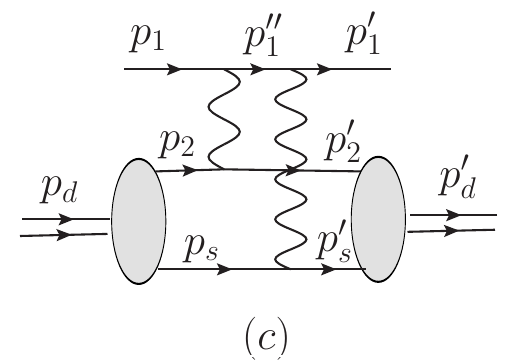} &
  \includegraphics[scale = 0.4]{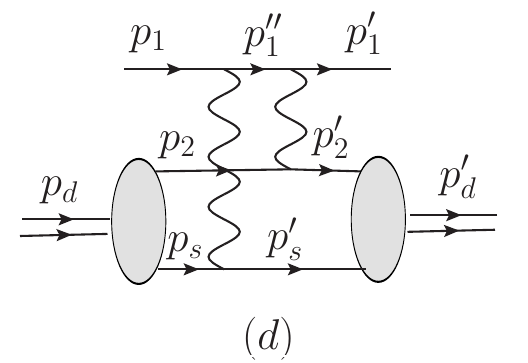}\\
  \includegraphics[scale = 0.4]{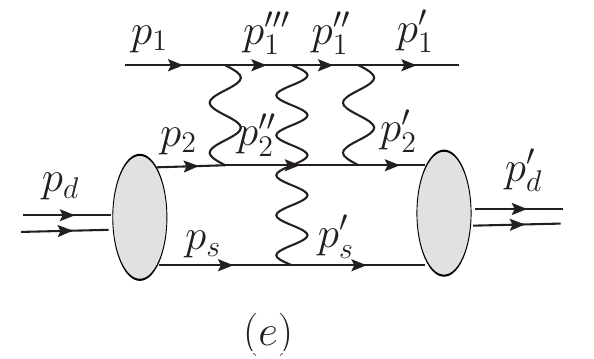} &  \\
    \end{tabular}
    \end{center}
  \caption{\label{fig:pd} Feynman diagrams for the elastic scattering process $p d  \to p d$.
    Notations are the same as in Fig.~\ref{fig:diagr}.}
\end{figure}
The possible single-, double-, and triple scattering amplitudes are shown in Fig.~\ref{fig:pd}.
The single scattering amplitude (a) on the proton is
\begin{eqnarray}
 iM_{pd}^{(a)} &=& \int \frac{d^4p_s}{(2\pi)^4}  \frac{-i\Gamma_{d \to pn}^*(p_d^\prime,p_s)}{p_2^{\prime 2}-m^2+i\epsilon} iM_{\rm el}^{pp}(p_1^\prime,p_2^\prime,p_1)
    \frac{i}{p_s^2-m^2+i\epsilon} \frac{i\Gamma_{d \to pn}(p_d,p_s)}{p_2^2-m^2+i\epsilon}  \nonumber \\
  &=& \int \frac{d^3p_s}{(2\pi)^3 2E_s} \left(\frac{2E_s m_d}{p_2^{\prime 0}}\right)^{1/2} (2\pi)^{3/2} \phi^*(\bvec{p}_2^\prime-\frac{\bvec{p}_d^\prime}{2})
    iM_{\rm el}^{pp}(p_1^\prime,p_2^\prime,p_1)    \nonumber \\
  && \times \left(\frac{2E_s m_d}{p_2^0}\right)^{1/2} (2\pi)^{3/2} \phi(\bvec{p}_2-\frac{\bvec{p}_d}{2})~,  \label{M_{pd}^(a)}
\end{eqnarray}
where the integration over $dp_s^0$ was done according to Eq.(\ref{contIntegral}) and
the relation (\ref{Gamma_d}) between the $d \to pn$ on-shell vertex function and the DWF was used on the second step;
the DWF's depend on the struck proton momentum in the deuteron rest frame calculated non-relativistically.
For the forward scattering amplitude, by replacing the energies of slow particles (i.e. the deuteron and its constituent nucleons) by their masses, we, thus, have
\begin{equation}
  M_{pd}^{(a)}(0) = 2 \int d^3p_s |\phi(\frac{\bvec{p}_d}{2}-\bvec{p}_s)|^2 M_{\rm el}^{pp}(p_1,p_d-p_s,p_1)~,     \label{M_pd^(a)_fin}
\end{equation}
where the momentum conservation in vertices, $\bvec{p}_2+\bvec{p}_s=\bvec{p}_d$, has been used.
Similar expression can be obtained for the forward amplitude (b):
\begin{equation}
  M_{pd}^{(b)}(0) = 2 \int d^3p_2  |\phi(\bvec{p}_2-\frac{\bvec{p}_d}{2})|^2 M_{\rm el}^{pn}(p_1,p_d-p_2,p_1)~.     \label{M_pd^(b)_fin}
\end{equation}

The amplitude (c) can be written as
\begin{eqnarray}
  iM_{pd}^{(c)} &=& \int \frac{d^4p_s}{(2\pi)^4} \int \frac{d^4p_s^\prime}{(2\pi)^4}  \frac{-i\Gamma_{d \to pn}^*(p_d^\prime,p_s^\prime)}{p_2^{\prime 2}-m^2+i\epsilon}
  \frac{i}{p_s^{\prime 2}-m^2+i\epsilon} iM_{\rm el}^{pn}(p_1^\prime,p_s^\prime,p_1^{\prime\prime}) \nonumber \\
  && \times \frac{i}{p_1^{\prime\prime 2}-m^2+i\epsilon} iM_{\rm el}^{pp}(p_1^{\prime\prime},p_2^\prime,p_1) \frac{i}{p_s^2-m^2+i\epsilon}
  \frac{i\Gamma_{d \to pn}(p_d,p_s)}{p_2^2-m^2+i\epsilon} \nonumber \\
  &=& \int \frac{d^3p_s}{(2\pi)^3 2E_s} \int \frac{d^3p_s^\prime}{(2\pi)^3 2E_s^\prime} \left(\frac{2E_s^\prime m_d}{p_2^{\prime 0}}\right)^{1/2} (2\pi)^{3/2} \phi^*(\bvec{p}_2^\prime-\frac{\bvec{p}_d^\prime}{2})
      iM_{\rm el}^{pn}(p_1^\prime,p_s^\prime,p_1^{\prime\prime}) \nonumber \\
  &&  \times \frac{i}{p_1^{\prime\prime 2}-m^2+i\epsilon} iM_{\rm el}^{pp}(p_1^{\prime\prime},p_2^\prime,p_1) 
      \left(\frac{2E_s m_d}{p_2^0}\right)^{1/2} (2\pi)^{3/2} \phi(\bvec{p}_2-\frac{\bvec{p}_d}{2})~,  \label{M_{pd}^(c)}
\end{eqnarray}
where  the contour integrations over $dp_s^0$ and $dp_s^{\prime 0}$ (see Eq.(\ref{contIntegral}) have been performed in the second step.
By introducing the partial momentum transfer, $k=p_1^{\prime\prime}-p_1^\prime$, and replacing the energies of slow particles
by their masses, we have for the forward amplitude (c):
\begin{eqnarray}
  M_{pd}^{(c)}(0) &=& i \int \frac{d^3p_s}{(2\pi)^3 m} \int d^3k \phi^*(\frac{\bvec{p}_d}{2}-\bvec{p}_s-\bvec{k})
        M_{\rm el}^{pn}(p_1,p_s+k,p_1+k)  \frac{i}{(p_1+k)^2-m^2+i\epsilon} \nonumber \\
   && \times M_{\rm el}^{pp}(p_1+k,p_d-p_s-k,p_1) \phi(\frac{\bvec{p}_d}{2}-\bvec{p}_s)~.    \label{M_{pd}^(c)_1}
\end{eqnarray}
In a similar way, the forward amplitude (d) can be expressed as
\begin{eqnarray}
  M_{pd}^{(d)}(0) &=& i \int \frac{d^3p_s}{(2\pi)^3 m} \int d^3k \phi^*(\frac{\bvec{p}_d}{2}-\bvec{p}_s+\bvec{k})
        M_{\rm el}^{pp}(p_1,p_d-p_s+k,p_1+k) \frac{i}{(p_1+k)^2-m^2+i\epsilon} \nonumber \\
   && \times M_{\rm el}^{pn}(p_1+k,p_s-k,p_1) \phi(\frac{\bvec{p}_d}{2}-\bvec{p}_s)~.    \label{M_{pd}^(d)_1} 
\end{eqnarray}
The inverse propagator of the fast intermediate proton  can be now expressed in the eikonal form as
\begin{equation}
  (p_1+k)^2-m^2+i\epsilon = 2p_1k + k^2 + i\epsilon = 2|\bvec{p}_1|(-k^z+\Delta+i\epsilon)~,   \label{invPropEik}
\end{equation}
with
\begin{equation}
  \Delta = \frac{E_1k^0}{|\bvec{p}_1|} + \frac{k^2}{2|\bvec{p}_1|}~.
\end{equation}
Note that $k^0$ is different for the amplitudes (c) and (d) for given $\bvec{k}$. However, since large $|k^0|$ and/or $|\bvec{k}|$
values are anyway suppressed by the DWF's, we can approximately set $\Delta=0$ in the both amplitudes. 

Taking into account the fact that the scale of momentum transfer on which the $pp$ and $pn$ elastic amplitude change is much
larger than the momentum scale on which the DWF change, the $pp$ and $pn$ elastic amplitudes in Eqs.(\ref{M_{pd}^(c)_1}),(\ref{M_{pd}^(d)_1})
can be replaced by the forward ones. Then, after changing $\bvec{k} \to -\bvec{k}$ in the integrand of Eq.(\ref{M_{pd}^(d)_1}),
the sum of the (c) and (d) forward amplitudes can be rewritten as
\begin{eqnarray}
  M_{pd}^{(c)}(0) + M_{pd}^{(d)}(0) &=& -\frac{1}{2|\bvec{p}_1|m} \int \frac{d^3p_s}{(2\pi)^3} \int d^3k \phi^*(\frac{\bvec{p}_d}{2}-\bvec{p}_s-\bvec{k})
        M_{\rm el}^{pn}(p_1,p_s,p_1) M_{\rm el}^{pp}(p_1,p_d-p_s,p_1)    \nonumber \\
   && \times \phi(\frac{\bvec{p}_d}{2}-\bvec{p}_s) \left(\frac{1}{-k^z+i\epsilon} + \frac{1}{k^z+i\epsilon} \right)~.    \label{M_{pd}^(c+d)}
\end{eqnarray}
By using the rule (\ref{LandauRule}), the integral over $dk^z$ can be taken that leads to the following result:
\begin{eqnarray}
  M_{pd}^{(c)}(0) + M_{pd}^{(d)}(0) &=& \frac{i\pi}{|\bvec{p}_1|m} \int \frac{d^3p_s}{(2\pi)^3} \int d^2k_t \phi^*(\frac{\bvec{p}_d}{2}-\bvec{p}_s-\bvec{k}_t)
        M_{\rm el}^{pn}(p_1,p_s,p_1) M_{\rm el}^{pp}(p_1,p_d-p_s,p_1)    \nonumber \\
   && \times \phi(\frac{\bvec{p}_d}{2}-\bvec{p}_s)~.    \label{M_{pd}^(c+d)_fin}
\end{eqnarray}

According to the optical theorem,
\begin{eqnarray}
   && \mbox{Im} M_{pd}(0) = 2 |\bvec{p}_1| m_d \sigma_{pd}~, \label{optTheor_pN} \\
   && \mbox{Im} M_{\rm el}^{pN}(p_1,p,p_1) = 2 I_{NN}(p_1,p) \sigma_{pN}~,     \label{optTheor_pN} 
\end{eqnarray}
where $\sigma_{pd}$ and $\sigma_{pN}$ are the total $pd$ and $pN$ cross sections, and
\begin{equation}
  I_{NN}(p_1,p) = \sqrt{(p_1p)^2-m^4} = |\bvec{p}_1|m - E_1p^z + O[(p^z)^2] 
\end{equation}
is the M\"oller flux factor, where the decomposition with accuracy up to the linear terms in the momentum, $\bvec{p}$,
of the struck nucleon is performed in the second step.

Therefore, by considering the calculated forward $pd$ amplitudes, Eqs.(\ref{M_pd^(a)_fin}),(\ref{M_pd^(b)_fin}),(\ref{M_{pd}^(c+d)_fin}),
in the deuteron rest frame, one sees that
the amplitudes (a) and (b) provide, respectively, the contributions $\sigma_{pp}$ and $\sigma_{pn}$
to the total $pd$ cross section.
The sum of the amplitudes (c) and (d) can be rewritten as
\begin{eqnarray}
  M_{pd}^{(c)}(0) + M_{pd}^{(d)}(0) &=& -\frac{i \pi 4 |\bvec{p}_1|^2 m^2 \sigma_{pp} \sigma_{pn}}{|\bvec{p}_1|m}
  \int \frac{d^3p_s}{(2\pi)^3} \int d^2k_t \phi^*(-\bvec{p}_s-\bvec{k}_t) \phi(-\bvec{p}_s)   \nonumber \\
    &=& -2 i |\bvec{p}_1| m \sigma_{pp} \sigma_{pn} \int \frac{d^3p_s}{(2\pi)^3} \int \frac{d^2k_t}{(2\pi)^2} \int d^3r \mbox{e}^{i(-\bvec{p}_s-\bvec{k}_t)\bvec{r}} \phi^*(\bvec{r})
         \int d^3r^\prime \mbox{e}^{i\bvec{p}_s\bvec{r}^\prime} \phi(\bvec{r}^\prime)  \nonumber \\
    &=& -2 i |\bvec{p}_1| m \sigma_{pp} \sigma_{pn} \int d^3r |\phi(\bvec{r})|^2 \delta^{(2)}(\bvec{r}_t)~,    \label{M_{pd}^(c+d)_coord}
\end{eqnarray}
where the small real parts of the elastic $pN$ amplitudes were neglected, and, in the second step, the relation between DWF's in the momentum and coordinate
representations, Eq.(\ref{phi(p_2)}), was used. Since the DWF modulus squared does not depend on the direction of $\bvec{r}$, the last integral in Eq.(\ref{M_{pd}^(c+d)_coord})
is equal to $\langle r^{-2} \rangle/2\pi$. Thus, we come to the following formula:
\begin{equation}
  \sigma_{pd} = \frac{\mbox{Im}(M_{pd}^{(a)}(0)+M_{pd}^{(b)}(0)+M_{pd}^{(c)}(0) + M_{pd}^{(d)}(0))}{2 |\bvec{p}_1| m_d}
             = \sigma_{pp} + \sigma_{pn} - \frac{\sigma_{pp} \sigma_{pn}}{4\pi} \langle r^{-2} \rangle~,      \label{sigma_pd}
\end{equation}
which is in full agreement with Glauber result \cite{Glauber:1955qq}. 

Let us now evaluate the contribution of triple scattering amplitude (e):
\begin{eqnarray}
  iM_{pd}^{(e)} &=& \int \frac{d^4p_s}{(2\pi)^4} \int \frac{d^4p_2^{\prime\prime}}{(2\pi)^4}  \int \frac{d^4p_s^\prime}{(2\pi)^4}  \frac{-i\Gamma_{d \to pn}^*(p_d^\prime,p_s^\prime)}{p_2^{\prime 2}-m^2+i\epsilon}
  \frac{i}{p_s^{\prime 2}-m^2+i\epsilon} iM_{\rm el}^{pp}(p_1^\prime,p_2^\prime,p_1^{\prime\prime}) \nonumber \\
  && \times \frac{i}{p_1^{\prime\prime 2}-m^2+i\epsilon} \frac{i}{p_2^{\prime\prime 2}-m^2+i\epsilon}  iM_{\rm el}^{pn}(p_1^{\prime\prime},p_s^\prime,p_1^{\prime\prime\prime})
             \frac{i}{p_1^{\prime\prime\prime 2}-m^2+i\epsilon} \nonumber \\
  && \times \frac{i}{p_s^2-m^2+i\epsilon}  iM_{\rm el}^{pp}(p_1^{\prime\prime\prime},p_2^{\prime\prime},p_1)
  \frac{i\Gamma_{d \to pn}(p_d,p_s)}{p_2^2-m^2+i\epsilon} \nonumber \\
  &=& \int \frac{d^3p_s}{(2\pi)^3 2E_s} \int \frac{d^3p_2^{\prime\prime}}{(2\pi)^32E_2^{\prime\prime}} \int \frac{d^3p_s^\prime}{(2\pi)^3 2E_s^\prime}
       \left(\frac{2E_s^\prime m_d}{p_2^{\prime 0}}\right)^{1/2} (2\pi)^{3/2} \phi^*(\bvec{p}_2^\prime-\frac{\bvec{p}_d^\prime}{2})
       iM_{\rm el}^{pp}(p_1^\prime,p_2^\prime,p_1^{\prime\prime}) \nonumber \\
  &&  \times \frac{i}{p_1^{\prime\prime 2}-m^2+i\epsilon} iM_{\rm el}^{pn}(p_1^{\prime\prime},p_s^\prime,p_1^{\prime\prime\prime})
            \frac{i}{p_1^{\prime\prime\prime 2}-m^2+i\epsilon} iM_{\rm el}^{pp}(p_1^{\prime\prime\prime},p_2^{\prime\prime},p_1) \nonumber \\
   &&  \times \left(\frac{2E_s m_d}{p_2^0}\right)^{1/2} (2\pi)^{3/2} \phi(\bvec{p}_2-\frac{\bvec{p}_d}{2})~,  \label{M_{pd}^(e)}
\end{eqnarray}
where, in the second step, the contour integrations over $dp_s^0, dp_2^{\prime\prime 0}$ and $dp_s^{\prime 0}$ are performed using Eq.(\ref{contIntegral}). 
By introducing the partial four-momentum transfers to the deuteron, $k^\prime = p_1-p_1^{\prime\prime\prime},k^{\prime\prime} = p_1^{\prime\prime\prime}- p_1^{\prime\prime}$,
$k^{\prime\prime\prime} = p_1^{\prime\prime}- p_1^\prime$,
and $q=k^\prime+k^{\prime\prime} = p_1-p_1^{\prime\prime}$,
the inverse propagators of the intermediate fast proton in Eq.(\ref{M_{pd}^(e)}) can be represented in the eikonal form as
\begin{eqnarray}
  \lefteqn{p_1^{\prime\prime\prime 2}-m^2+i\epsilon = (p_1-k^\prime)^2 - m^2 + i\epsilon = -2p_1k^\prime + k^{\prime 2} + i\epsilon} && \nonumber \\
  &=& 2|\bvec{p}_1|(k^{\prime z}-\Delta^{\prime\prime\prime} + i\epsilon)~,
  ~~~\Delta^{\prime\prime\prime}=\frac{E_1 k^{\prime 0}}{|\bvec{p}_1|} -\frac{k^{\prime 2}}{2|\bvec{p}_1|}~,         \label{invProp_p1ppp}  \\
  \lefteqn{p_1^{\prime\prime 2}-m^2+i\epsilon = (p_1-q)^2 - m^2 + i\epsilon = -2p_1q + q^2 + i\epsilon} && \nonumber \\
  &=& 2|\bvec{p}_1|(q^z-\Delta^{\prime\prime} + i\epsilon)~,
  ~~~\Delta^{\prime\prime}=\frac{E_1q^0}{|\bvec{p}_1|}-\frac{q^2}{2|\bvec{p}_1|}~.    \label{invProp_p1pp}
\end{eqnarray}
While the large four-momentum transfer $k^{\prime\prime}$ is suppressed by the DWF, $k^\prime$ (and, thus, also $q$) is not. However, large negative $k^{\prime 2}$ values are anyway suppressed
by the factor $\propto \exp(B k^{\prime 2}/2) \simeq \exp(-B k_t^{\prime 2}/2)$ which enters the $pp$ elastic scattering amplitude. Thus, we will assume $\Delta^{\prime\prime\prime}$ and
$\Delta^{\prime\prime}$ in Eqs.(\ref{invProp_p1ppp}),(\ref{invProp_p1pp}) to be small constant values.

The integrations over $d^3p_s d^3p_2^{\prime\prime} d^3p_s^\prime$ can be equivalently replaced by the integrations over $d^3k^{\prime\prime} d^3k^{\prime} d^3p_2$.
This allows us to rewrite Eq.(\ref{M_{pd}^(e)}) as
\begin{eqnarray}
  iM_{pd}^{(e)} &=& \frac{2m_d}{(2m)^3 4|\bvec{p}_1|^2} \int \frac{d^3k^{\prime\prime}}{(2\pi)^3} \int \frac{d^3k^\prime}{(2\pi)^3} \int d^3p_2
  \phi^*(\bvec{p}_2-\bvec{k}^{\prime\prime}-\bvec{p}_d+\frac{\bvec{p}_d^\prime}{2})  \phi(\bvec{p}_2-\frac{\bvec{p}_d}{2})  iM_{\rm el}^{pp}(|\bvec{p}_1|,k_t^{\prime\prime\prime}) \nonumber \\
  && \times  iM_{\rm el}^{pn}(|\bvec{p}_1|,k_t^{\prime\prime}) iM_{\rm el}^{pp}(|\bvec{p}_1|,k_t^\prime)
     \frac{i}{k^{\prime z} - \Delta^{\prime\prime\prime} + i\epsilon} \frac{i}{k^{\prime z} + k^{\prime\prime z} -\Delta^{\prime\prime} + i\epsilon}~,     \label{M_{pd}^(e)_fin} 
\end{eqnarray}
where the $pp$ and $pn$ elastic scattering amplitudes are supposed to depend on the transverse momentum transfers only (and not on the momentum $\bvec{p}_2$
of the struck proton in the deuteron). The integral over $dk^{\prime z}$ can be factorized-out and is equal to zero:
\begin{equation}
   \int \frac{dk^{\prime z}}{(k^{\prime z} - \Delta^{\prime\prime\prime} + i\epsilon)(k^{\prime z}+k^{\prime\prime z} -\Delta^{\prime\prime} + i\epsilon)} = 0~,   \label{contInt1} 
\end{equation}
where the integration contour is closed in the upper part of the complex $k^{\prime z}$ plane, where no singularities present.
Thus, the triple scattering term (e) of Fig.~\ref{fig:pd} with repeated scattering of the fast proton on the proton of the deuteron is kinematically suppressed.
The same of course applies to the other term (not shown) of the triple scattering with repeated scattering on the neutron.

\end{document}